\documentclass[12pt]{article}
\newif\iffigs\figstrue
\usepackage{latexsym,amssymb}
\iffigs
  \input{epsf}
\else 
\fi
\def\lcm{{\rm lcm}}
\def\m1{M^{1,1,1}}
\def\zz{\mathbb Z}
\def\ra#1{\buildrel #1 \over \longrightarrow}
\def\dim{{\rm dim}}
\font\tenmsbm=msbm10 scaled 1200
\font\sevenmsbm=msbm9
\newfam\msbmfam
\textfont\msbmfam=\tenmsbm
\scriptfont\msbmfam=\sevenmsbm
\def\msbm{\fam\msbmfam\tenmsbm}

\newtheorem{definizione}{Definition}[section]
\newcommand{\bdefi}{\begin{definizione}}
\newcommand{\edefi}{\end{definizione}}
\makeatletter
\@addtoreset{equation}{section}
\makeatother

\newcommand{\ft}[2]{{\textstyle\frac{#1}{#2}}}
\newsavebox{\uuunit}
\sbox{\uuunit}
    {\setlength{\unitlength}{0.825em}
     \begin{picture}(0.6,0.7)
        \thinlines
        \put(0,0){\line(1,0){0.5}}
        \put(0.15,0){\line(0,1){0.7}}
        \put(0.35,0){\line(0,1){0.8}}
       \multiput(0.3,0.8)(-0.04,-0.02){10}{\rule{0.5pt}{0.5pt}}
     \end {picture}}
\newcommand {\unity}{\mathord{\!\usebox{\uuunit}}}

\def\IP{\relax{\rm I\kern-.18em P}}

\font\tenmsbm=msbm10 scaled 1200
\font\sevenmsbm=msbm9
\newfam\msbmfam
\textfont\msbmfam=\tenmsbm
\scriptfont\msbmfam=\sevenmsbm
\def\msbm{\fam\msbmfam\tenmsbm}

\def\inbar{\vrule height1.5ex width.4pt depth0pt}
\def\IC{\relax\,\hbox{$\inbar\kern-.3em{\rm C}$}}
\def\IG{\relax\,\hbox{$\inbar\kern-.3em{\rm G}$}}
\def\IB{\relax{\rm I\kern-.18em B}}
\def\ID{\relax{\rm I\kern-.18em D}}
\def\IL{\relax{\rm I\kern-.18em L}}
\def\IF{\relax{\rm I\kern-.18em F}}
\def\IH{\relax{\rm I\kern-.18em H}}
\def\II{\relax{\rm I\kern-.17em I}}
\def\IN{\relax{\rm I\kern-.18em N}}
\def\IP{\relax{\rm I\kern-.18em P}}
\def\IQ{\relax\,\hbox{$\inbar\kern-.3em{\rm Q}$}}
\def\bfzero{\relax\,\hbox{$\inbar\kern-.3em{\rm 0}$}}
\def\IK{\relax{\rm I\kern-.18em K}}
\def\IG{\relax\,\hbox{$\inbar\kern-.3em{\rm G}$}}
 \font\cmss=cmss10 \font\cmsss=cmss10 at 7pt
\def\IR{\relax{\rm I\kern-.18em R}}
\def\IGam{\relax{{\rm I}\kern-.18em \Gamma}}
\def\ZZ{\relax\ifmmode\mathchoice
{\hbox{\cmss Z\kern-.4em Z}}{\hbox{\cmss Z\kern-.4em Z}}
{\lower.9pt\hbox{\cmsss Z\kern-.4em Z}}
{\lower1.2pt\hbox{\cmsss Z\kern-.4em Z}}\else{\cmss Z\kern-.4em
Z}\fi}
\def\bfone{\relax{\rm 1\kern-.35em 1}}

\def\a{\alpha}
\def\b{\beta}

\font\cmss=cmss10 \font\cmsss=cmss10 at 7pt

\def\inbar{\vrule height1.5ex width.4pt depth0pt}
\def\IC{\relax\,\hbox{$\inbar\kern-.3em{\rm C}$}}
\def\bfzero{\relax\,\hbox{$\inbar\kern-.3em{\rm 0}$}}
\def\bfone{\relax{\rm 1\kern-.35em 1}}

\def\diag{{\rm diag}}

\def\tilde{\widetilde}

\def\IE{\relax{{\rm I\kern-.18em E}}}

\def\IGam{\relax{{\rm I}\kern-.18em \Gamma}}

\def\bet{\begin{tabular}}
\def\eet{\end{tabular}}

\def\a{\alpha}
\def\b{\beta}

\begin{document}
\begin{titlepage}
\begin{flushright}
hep-th/9907219\\
July 1999\\
SISSA/113/99/FM, CERN-TH/99-231\\
\end{flushright}
\vskip 2cm
\begin{center}
{\Large \bf   3D superconformal theories\\
\vskip  0.3 cm
 from Sasakian seven-manifolds:\\
\vskip  0.5 cm
 new nontrivial evidences for $AdS_4/CFT_3$  $^*$}\\
\vfill
{\large    Davide Fabbri$^1$, Pietro Fr\'e$^1$, Leonardo Gualtieri$^1$, Cesare Reina$^2$,\\
\vskip  0.5 cm
 Alessandro Tomasiello$^2$, Alberto Zaffaroni$^3$ and Alessandro Zampa$^2$} \\
\vfill
{\small
$^1$  Dipartimento di Fisica Teorica, Universit\'a di Torino, via P.
Giuria 1,
I-10125 Torino, \\
 Istituto Nazionale di Fisica Nucleare (INFN) - Sezione di Torino,
Italy \\
\vspace{6pt}
$^2$ International School for Advanced Studies (ISAS), via Beirut 2-4,
I-34100 Trieste\\
\vspace{6pt}
$^3$ CERN, Theoretical Division, CH 1211 Geneva,
Switzerland,\\
}
\end{center}
\vfill
\begin{abstract}
In this paper we discuss candidate superconformal ${\cal N}=2$ gauge theories
that realize the AdS/CFT correspondence with M--theory compactified
on the homogeneous Sasakian $7$-manifolds $M^7$ that were classified long ago. In
particular we focus on the two cases $M^7=Q^{1,1,1}$ and $M^7=M^{1,1,1}$, for
the latter the Kaluza Klein spectrum being completely known. We show
how the toric description of $M^7$ suggests the gauge group and the
supersingleton fields. The conformal dimensions of the latter can be
independently calculated by comparison with the mass of baryonic
operators that correspond to  $5$--branes wrapped on supersymmetric
$5$--cycles and are charged with respect to the Betti multiplets.
The entire Kaluza Klein spectrum of short multiplets
agrees with these dimensions. Furthermore, the metric cone over the
Sasakian manifold is a conifold algebraically embedded in some
${\mathbb C}^p$.
The ring of chiral primary fields is defined as the coordinate ring
of ${\mathbb C}^p$ modded by the ideal generated by the embedding
equations; this ideal has a nice characterization by means of
representation theory. The entire Kaluza Klein spectrum is
explained in terms of these vanishing relations. We give the
superfield interpretation of all short multiplets and we point out
the existence of many long multiplets with rational protected
dimensions, whose presence and pattern were already noticed in other 
compactifications and seem to be universal.
\end{abstract}
\vspace{2mm} \vfill \hrule width 3.cm
{\footnotesize
 $^*$ Supported in part by   EEC  under TMR contract
 ERBFMRX-CT96-0045 and by GNFM.}
\end{titlepage}
\tableofcontents
%
%
\section{Synopsis}
In this paper we consider
M--theory compactified on anti de Sitter four dimensional space $AdS_4$  times
a homogeneous Sasakian $7$--manifold $M^7$  and we study the
correspondence with
the infrared conformal point of
suitable $D=3,{\cal N}=2$ gauge theories describing the appropriate M2--brane
dynamics. For the reader's convenience we have divided our paper into
three parts.
\begin{itemize}
  \item Part \ref{gendiscus} contains a general discussion of
the problem we have addressed and a summary of all our results.
  \item Part \ref{comparo} presents the superconformal gauge--theory interpretation
  of the Kaluza Klein multiplet spectra previously obtained from harmonic
  analysis and illustrates the non--trivial predictions one obtains
  from such a comparison.
  \item Part \ref{geometry} provides a detailed analysis of the
  algebraic geometry, topology and metric structures of
  homogeneous Sasakian $7$--manifolds. This part contains all the
  geometrical background and the explicit derivations on which our
  results and conclusions are based.
\end{itemize}
\vskip 0.3cm
\part{General Discussion}
\label{gendiscus}

\section{Introduction}
The basic principle of  the AdS/CFT correspondence
\cite{maldapasto,polkleb,witten} states that every consistent M-theory
or type II background with
metric $AdS_{p+2}\times M^{d-p-2}$ in d-dimensions, where $M^{d-p-2}$
is an Einstein manifold, is associated with a conformal quantum field theory
living on the boundary of $AdS_{p+2}$. The background is typically
generated by the near horizon geometry of a set of p-branes and the
boundary conformal field theory is identified with the IR limit of the gauge
theory living on the world-volume of the p-branes. One remarkable example
with ${\cal N}=1$ supersymmetry on the boundary and with a non-trivial smooth
manifold
$M^5=T^{1,1}$ was found in \cite{witkleb} and the associated
superconformal theory was identified. Some general properties and the 
complete spectrum of the $T^{1,1}$ compactification have been discussed
in \cite{gubser,gubserkleb,sergiotorino}, finding complete agreement
between gauge theory expectations and supergravity predictions.  
In this paper we will focus on the case $p=2$ when
$M$ is a coset manifold $G/H$ with ${\cal N} = 2$ supersymmetry.
\par
Backgrounds of the form $AdS_4\times M^7$ arise as the near horizon
geometry of a collection of M2-branes in M-theory. The ${\cal N}=8$
supersymmetric case corresponds to $M^7=S^7$.
 Examples of superconformal theories
with less supersymmetry can be obtained by orbifolding the M2-brane solution
\cite{fkpz,gomis}. Orbifold models have the advantage that the gauge theory can
be directly obtained as a quotient of the ${\cal N}=8$ theory using standard
techniques \cite{douglasmoore}. On the other hand, the internal manifold $M^7$
is $S^7$ divided by some discrete group and it is generically singular.
Smooth manifolds $M^7$ can be obtained
by considering M2-branes sitting at the singular point of the cone over $M^7$,
${\cal C}(M^7)$ \cite{witkleb,fig,morpless}. Many examples where $M^7$ is a 
coset
manifold $G/H$ were studied in the old days of KK theories \cite{freurub}
(see \cite{duffrev,round7a,squash7a,englert,biran,casher,dewit2,osp48,gunawar}
for the cases $S^7$ and squashed $S^7$, see
\cite{kkwitten,noi321,spectfer,spec321,multanna,pagepopeM} for the case
$M^{p,q,r}$, see \cite{dafrepvn,pagepopeQ} for the case $Q^{1,1,1}$,
see \cite{spectfer,univer,bosmass,frenico,gunawar} for general methods 
of harmonic analysis in compactification
and the structure of $Osp(N\vert 4)$ supermultiplets, and finally see
\cite{castromwar} for a complete classification of $G/H$
compactifications.)
\par
For obvious reason, $AdS_4$ was much more investigated in those days
than his simpler cousin $AdS_5$. As a consequence, we have a plethora
of $AdS_4\times G/H$ compactifications for which the dual superconformal
theory is still to be found. If we require supersymmetric solutions,
which are guaranteed to be stable and are simpler to study, and
furthermore we require  $2\le {\cal N} \le 8$, we find
four examples: $N^{0,1,0}$ with ${\cal N}=3$ and
$Q^{1,1,1},M^{1,1,1},V_{5,2}$ with ${\cal N}=2$ supersymmetry.
These are the natural $AdS_4$ counterparts of the $T^{1,1}$ conifold theory
studied in \cite{witkleb}.
\par
In this paper we shall consider in some detail the two cases
$Q^{1,1,1}$ and $M^{1,1,1}$. They have isometry $SU(2)^3\times U(1)$ and
$SU(3)\times SU(2)\times U(1)$, respectively. The isometry of these
manifolds corresponds to the global symmetry of the dual superconformal
theories, including the $U(1)$ R-symmetry of ${\cal N} = 2$
supersymmetry. The complete spectrum of 11-dimensional supergravity
compactified on $M^{1,1,1}$ has been recently computed
\cite{M111spectrum,superfieldsN2D3}.
The analogous spectrum for $Q^{1,1,1}$ has not been computed yet
\footnote{This spectrum is presently under construction \cite{merlatti}},
but several partial results exist in the literature \cite{popelast},
which will be enough for our purpose. The KK spectrum should match
the spectrum of the gauge theory operators of finite dimension in
the large $N$ limit. As a difference with the maximally supersymmetric case,
the KK spectrum contains both {\it short} and {\it long} operators;
this is a characteristic feature of ${\cal N}=2$ supersymmetry and
was already found  in $AdS_5\times T^{1,1}$ \cite{gubser,sergiotorino}.
\par
We will show that the spectra on $Q^{1,1,1}$ and $M^{1,1,1}$ share
several common features with their cousin $T^{1,1}$.
First of all, the KK spectrum is in perfect agreement with the spectrum of
operators of a superconformal theory with a set of fundamental {\it
supersingleton fields} inherited from the geometry of the manifold.
In the abelian case this is by no means a surprise because of the well
known relations among harmonic analysis, representation theory and
holomorphic line bundles over algebraic homogeneous spaces. The non-abelian 
case is more involved.
There is no straightforward method to identify the gauge theory living
on M2-branes placed at the singularity of
${\cal C}(M^7)$ when the space is not an orbifold. Hence we shall use intuition
from toric geometry to write candidate gauge theories that have the
right global symmetries and a spectrum of short operators which matches
the KK spectrum. Some points that still need to be clarified are pointed out.
\par
A second remarkable
property of these spaces is the existence of non-trivial cycles
and non-perturbative states, obtained by wrapping branes,
which are identified with baryons in the gauge theory \cite{bariowit}. 
The corresponding
baryonic $U(1)$ symmetry is associated with the so-called Betti multiplets
\cite{univer,spec321}. The conformal dimension of a baryon can be computed
in supergravity, following \cite{gubserkleb}, and unambiguously 
predicts the dimension of the
fundamental conformal fields of the theory in the IR. 
The result
from the baryon analysis is remarkably in agreement with
the expectations from the KK spectrum. This can be considered
as  a highly non-trivial
check of the AdS/CFT correspondence. Moreover we will also notice that,
as it happens on $T^{1,1}$ \cite{gubser,sergiotorino}, there exists a class of
long multiplets which, against expectations,  have a protected dimension
which is rational and agrees with a naive computation. There seems to be a
common pattern for the appearance of these operators in all the
various models.
\section{Conifolds and three-dimensional theories}
\subsection{The geometry of the conifolds}
Our purpose is to study a collection of M2-branes sitting at the singular point
of the conifold ${\cal C}(M^7)$, where $M^7=Q^{1,1,1}$ or $M^7=M^{1,1,1}$.
While for branes sitting at orbifold singularities there is a straightforward
method for identifying the gauge theory living on the world-volume
\cite{douglasmoore}, for
conifold singularities much less is known \cite{uranga,morpless}. The strategy
of describing the conifold as a deformation of an orbifold singularity
used in \cite{witkleb,morpless} and identifying the superconformal theory
as the IR limit of the deformed orbifold theory, seems more difficult to be applied in
three dimensions \footnote{See however \cite{tatar} where a similar approach
for $Q^{1,1,1}$ was attempted without, however, providing a match with Kaluza
Klein spectra. Another partial attempt in this direction was also given in
\cite{dallagat}.}. We will then use the intuition
from geometry in order to identify the fundamental degrees of freedom of the
superconformal theory and to compare them with the results of the KK expansion.
\par
We expect to find the superconformal fixed points dual to
$AdS$-compactifications as the IR limits of three-dimensional gauge theories.
In the maximally supersymmetric case $AdS_4\times S^7$, for example,
the superconformal theory is the IR limit of the ${\cal N}=8$
supersymmetric gauge theory \cite{maldapasto}. In three dimensions, the
gauge coupling constant is dimensionful and a gauge theory is certainly not
conformal. However, the theory becomes conformal in the IR, where the
coupling constant blows up. In this simple case, the identification of
the superconformal theory living on the world-volume of the M2-branes
follows from considering M-theory on a circle. The M2-branes
become D2-branes in type IIA, whose world-volume supports the  ${\cal N}=8$
gauge theory with a dimensionful coupling constant related to the radius
of the circle. The near horizon geometry of D2-branes is not anymore
AdS \cite{maldasonn}, since the theory is not conformal. The AdS
background and conformal invariance is recovered by sending the radius to
infinity; this corresponds to sending the gauge theory coupling to infinity
and probing the IR of the gauge theory.
\par
We expect a similar behaviour for other three dimensional gauge theories.
As a difference with four--dimensional CFT's corresponding to $AdS_5$
backgrounds, which always have exact marginal directions labeled
by the coupling constants (the type IIB dilaton is a free parameter of
the supergravity solution), these three dimensional fixed points
may also be isolated. The only universal parameter in  M-theory
compactifications is $\ell_P$, which is related to the number of colors
$N$, that is also the number of M2-branes. The
$1/N$ expansion in the gauge theory corresponds to the $R_{AdS}/\ell_P$
expansion of M-theory through the relation $R_{AdS}/\ell_P\sim N^{1/6}$
\cite{maldapasto}.
For large $N$, the M-theory solution is weakly coupled and supergravity
can be used for studying the gauge theory.
\par
The relevant degrees of freedom at the superconformal fixed points
are in general different from the elementary fields of the
supersymmetric gauge theory. For example, vector multiplets are not
conformal in three dimensions and they should be replaced by
some other multiplets of the superconformal group by dualizing
the vector field to a scalar. Let us again consider the simple example
of ${\cal N}=8$. The degrees of freedom at the superconformal point
(the singletons, in the language of representation theory of the
superconformal group) are contained in a supermultiplet with eight real
scalars and eight fermions, transforming in representations of the
global R-symmetry $SO(8)$. This is the same content of the ${\cal N}=8$
vector multiplet, when the vector field is dualized into a scalar.
The change of variable from a vector to a scalar, which is well-defined
in an abelian theory, is obviously a non-trivial and not even well-defined
operation in a non-abelian theory. The scalars in the supersingleton
parametrize the flat space transverse to the M2-branes. In this case,
the moduli space of vacua of the abelian ${\cal N}=8$ gauge theory,
corresponding to a single M2-brane, is isomorphic to the transverse space.
The case with $N$ M2-branes is obtained by promoting the theory to a
non-abelian one. We want to follow a similar procedure for the conifold cases.
\par
For branes at the conifold singularity of ${\cal C}(M^7)$ there is no obvious way
of reducing the system to a simple configuration of D2-branes in type IIA
and read the field content by using standard brane techniques \footnote{
This possibility exists for orbifold singularities and was exploited
in \cite{pz,fkpz,gomis} for ${\cal N}=4$ and in \cite{ahnC3} for
${\cal N}=2$.}. We can nevertheless use the intuition from geometry
for identifying the relevant degrees of freedom at the superconformal
point. We need an abelian gauge theory whose moduli space of vacua
is isomorphic to ${\cal C}(M^7)$. The moduli space of vacua of
${\cal N}=2$ theories have two different branches touching at a point, the
Coulomb branch parametrized by the {\it vev} of the scalars in the vector
multiplet and the Higgs branch parametrized by the {\it vev} of the
scalars in the chiral multiplets.
The Higgs branch is the one we are interested in.
Each of the two branches excludes the other, so we
can consistently set the scalars in the vector multiplets to zero
(see Appendix \ref{scalapot} for a discussion of the scalar potential
in general ${\cal N}=2$, $D=3$ theories).
We can find what we need in toric geometry.
Indeed, this latter describes certain complex manifolds as K\"ahler
quotients associated to symplectic actions of a product
of $U(1)$'s on some ${\mathbb C }^p$.
This is completely equivalent to imposing the
D-term equations for an abelian ${\cal N}=2,D=3$ gauge theory and dividing
by the gauge group or, in other words, to finding the moduli space
of vacua of the theory. Fortunately, both the cone over $Q^{1,1,1}$ and
that over $M^{1,1,1}$ have a toric geometry description. This description
 was already 
used for studying these spaces in \cite{tatar,dallagat}. In this paper,
we will consider
a different point of view. 
We can then easily find abelian gauge theories whose moduli space of vacua
(the Higgs branch component) is isomorphic to these two particular conifolds.
In the following subsections, we briefly discuss the geometry
of the two manifolds and the abelian gauge theory associated with the
toric description. More complete information about the geometry and the
homology of the manifolds are contained in Part \ref{geometry}.
Here we briefly recall
the basic information needed to discuss the matching of the KK spectrum
with the expectations from the conformal theory. 
\subsubsection{The case of $Q^{1,1,1}$}
\label{Qabel}
$Q^{1,1,1}$, originally introduced as a $D=11$ compactifying solution with
${\cal N}=2$ susy in \cite{dafrepvn}, is a specific instance in the family of 
the $Q^{p,q,r}$ manifolds, that are all of the form:
\begin{equation}
{SU(2)\times SU(2)\times SU(2)\over U(1)\times U(1)}.
\label{Q111}
\end{equation}
The cone over $Q^{1,1,1}$ is a toric manifold obtained as the K\"ahler
quotient of ${\mathbb C}^6$ by the symplectic action of two $U(1)$'s.
Explicitly, it is described as the solution of the following two
D--term equations
(momentum map equations in mathematical language)
\begin{equation}
\begin{array}{l}
|A_1|^2+|A_2|^2=|B_1|^2+|B_2|^2\\
|B_1|^2+|B_2|^2=|C_1|^2+|C_2|^2
\end{array}
\label{toricQ}
\end{equation}
modded by the action of the corresponding two $U(1)$'s,
the first acting only on $A_i$ with charge $+1$ and on
$B_i$ with charge $-1$, the second acting only on $B_i$ with charge $+1$ and on
$C_i$ with charge $-1$.
\par
The manifold $Q^{1,1,1}$ can be obtained by setting each term in ~(\ref{toricQ})
equal to 1, i.e. as $S^3\times S^3\times S^3/U(1)\times U(1)$.
This corresponds to taking a section of the cone at a fixed value of the
radial coordinate (an horizon in Morrison and Plesser's language
\cite{morpless}). Indeed, in full generality, this radial coordinate is
identified with
the fourth coordinate of $AdS_4$,
while the section is identified with
the internal manifold \footnote{In the solvable
Lie algebra parametrization
of $AdS_4$ \cite{g/hm2,noim2} the radial coordinate is algebraically
characterized as being associated
with the Cartan semisimple generator, while the remaining three are associated with
the three nilpotent generators spanning the brane world volume.
So we have a natural splitting of $AdS_4$ into $3+1$ which mirrors the natural splitting
of the eight dimensional conifold into $1+7$. The radial coordinate is shared by the two
spaces. This phenomenon, that is the algebraic basis for the existence of smooth M2 brane
solutions with horizon geometry $AdS_4 \times M^7$, was named dimensional transmigration in
\cite{g/hm2}.} $M^7$ \cite{kehagias,witkleb}.
\par
Given the toric description, the identification of an abelian ${\cal N}=2$
gauge theory whose Higgs branch reproduces the conifold is straightforward.
Equations ~(\ref{toricQ}) are the D-terms for the abelian theory $U(1)^3$
with doublets of chiral fields $A_i$ with charges $(1,-1,0)$, $B_i$
with charge $(0,1,-1)$ and $C_i$ with charges $(-1,0,1)$ and without
superpotential. The theory has an obvious global symmetry $SU(2)^3$
matching the isometry of $Q^{1,1,1}$. We introduced three $U(1)$ factors
(one more than those appearing in the toric data, as the attentive
reader certainly noticed) for symmetry reasons. One of the three
$U(1)$'s is decoupled and has no role in our discussion. Since
we do not expect a decoupled $U(1)$ in the world-volume theory of M2-branes
living at the conifold singularity, we should better consider
the theory $U(1)^3/U(1)_{{\rm DIAGONAL}}$.
\par
The fields appearing in the toric description should represent the
fundamental degrees of freedom of the superconformal theory, since they
appear as chiral fields in the gauge theory. They have definite
transformation properties under the gauge group. Out of
them we can also build some gauge invariant combinations, which should
represent the composite operators of the conformal theory and which should
be matched with the KK spectrum. Geometrically, this
corresponds to describing the cone as an affine subvariety of some
${\mathbb C}^p$.
This is a standard procedure, which converts the definition of
a toric manifold in terms of D-terms to an equivalent one in terms of
binomial equations in ${\mathbb C}^p$.
In this case, we have an embedding in ${\mathbb C}^8$. We first construct
all the $U(1)$ invariants (in this case there are $8=2\times 2 \times 2$ of them)
\begin{equation}
X^{ijk}=A^i B^j C^k, \qquad\qquad i,j,k=1,2.
\label{embedding}
\end{equation}
They satisfy a set of binomial equations which cut out the image of
our conifold ${\cal C}(Q^{1,1,1})$ in ${\mathbb C}^8$.
These equations are actually the $9$ quadrics explicitely written in
eq.s (\ref{sigXX}) of Part \ref{geometry}.
Indeed, there is a general method to obtain the embedding equations of the
cones over algebraic homogeneous varieties based on representation theory.
\footnote{The $9$ equations were
already mentioned in \cite{tatar} although their representation
theory interpretation was not given there.}  
If we want to summarize this general method in few words, we can say the
following. Through eq. (\ref{embedding}) we see that the coordinates
$X^{ijk}$ of ${\mathbb C}^8$ are assigned to a certain representation
$\cal R$ of the isometry group $SU(2)^3$. In our case such a
representation is ${\cal R}=(J_1=\ft{1}{2},J_2=\ft{1}{2},J_3=\ft{1}{2})$.
The products $X^{i_1j_1k_1}X^{i_2j_2k_2}$ belong to the symmetric product
$Sym^2({\cal R})$, which in general branches into various
representations, one of highest weight plus several subleading ones.
On the cone, however, only the highest weight representation survives
while all the subleading ones vanish.
Imposing that such subleading representations
are zero corresponds to writing the embedding equations.
This has far reaching consequences in the conformal field theory, since
provides the definition of the chiral ring. In principle all
the representations appearing in the $k$-th symmetric tensor power of ${\cal R}$
could correspond to primary conformal operators. Yet the attention should be restricted
to those that do not vanish modulo the equations of the cone, namely
modulo the ideal generated by the representations of subleading weights.
In other words, only the highest weight representation contained in
the $Sym^k({\cal R})$ gives a true chiral operator. 
This is what matches the Kaluza Klein spectra found through
harmonic analysis.  Two points should be stressed. In general the number of
embedding equations is larger than the codimension of the algebraic
locus. For instance $8-4 < 9$, i.e. the cone is not a complete 
intersection.
The $9$ equations (\ref{sigXX}) define the
ideal $I$ of ${\mathbb C}[X]:={\mathbb C}[X^{111},\ldots,X^{222}]$
cutting the cone ${\cal C}(Q^{1,1,1})$.
The second point to stress is the double interpretation of the embedding
equations.
The fact that $Q^{1,1,1}$ leads to ${\cal N}=2$
supersymmetry means that it is Sasakian, i.e. it is a circle bundle
over a suitable complex three--fold.
If considered in ${\mathbb C}^8$ the ideal $I$ cuts out the conifold
${\cal C}(Q^{1,1,1})$. Being homogeneous, it can also be
regarded as cutting out an algebraic variety in ${\mathbb P}^7$.
This is ${\mathbb P}^1 \times {\mathbb P}^1 \times {\mathbb P}^1$,
namely the base of the $U(1)$ fibre-bundle $Q^{1,1,1}$.
\par
It follows from this discussion that the invariant operators
$X^{ijk}$ of eq. (\ref{embedding}) can be naturally associated with the building blocks
of the gauge invariant composite operators of our CFT.
Holomorphic combinations of the $X^{ijk}$ should span the set of
chiral operators of the theory.
As we stated above, the set
of embedding equations (\ref{sigXX}) imposes restrictions on the allowed
representations of $SU(2)^3$ and hence on the existing operators.
If we put the definition of $X^{ijk}$ in terms of the fundamental
fields $A,B,C$ into the equations (\ref{sigXX}), we see that they are
automatically satisfied when the theory is abelian. Since we want
eventually to promote $A,B,C$ to non-abelian fields, these equations
become non-trivial because the fields do not commute
anymore. They essentially assert that the chiral operators
we may construct out of the $X^{ijk}$ are totally symmetric in the
exchange of the various $A,B,C$, that is they belong to
the highest weight representations we mentioned above.
\par
It is clear that the two different geometric descriptions of the conifold,
the first in terms of the variables $A,B,C$ and the second in terms of
the $X$, correspond to the two possible parametrization of the moduli
space of vacua of an ${\cal N}=2$ theory, one in terms of {\it vevs} of
the fundamental fields and the second in terms of gauge invariant
chiral operators.
\par
We notice that this discussion closely parallels the analogous one
in \cite{witkleb,nekr}. $Q^{1,1,1}$ is indeed a close relative of $T^{1,1}$.
\subsubsection{The case of $M^{1,1,1}$}
\label{Mabel}
$M^{1,1,1}$ is a specific instance in the family of 
the $M^{p,q,r}$ manifolds, that are all of the form (see \cite{noi321}):
\begin{equation}
{SU(3)\times SU(2)\times U(1)\over SU(2)\times U(1) \times U(1)}.
\label{M111}
\end{equation}
The details of the embedding are given in Section \ref{MP29}.
\par
The cone over $M^{1,1,1}$ is a toric manifold obtained as a K\"ahler
quotient of ${\mathbb C}^5$, described as the solution of the
 D-term equation
\begin{equation}
2\left( |U_1|^2+|U_2|^2+|U_3|^2\right) =3\left( |V_1|^2+|V_2|^2\right)
\label{toricM}
\end{equation}
modded by the action of a $U(1)$, acting on $U_i$ with charge $+2$ and on
$V_i$ with charge $-3$.
The manifold $M^{1,1,1}$ can be obtained by setting both terms of equation
~(\ref{toricM}) equal to 1.

Given the toric description, we can  identify the corresponding  abelian
${\cal N}=2$ gauge theory.
Equation ~(\ref{toricM}) is the D-term for the abelian theory $U(1)^2$
with a triplet of chiral fields $U_i$ with charges $(2,-2)$, a doublet $V_i$
with charge $(-3,3)$ and without
superpotential. The theory has an obvious global symmetry
$SU(3)\times SU(2)\times U(1)$ matching the isometry of $M^{1,1,1}$.
Again, we introduced two $U(1)$ factors for symmetry reasons. One of them
is decoupled and  we should better consider the theory
$U(1)^2/U(1)_{{\rm DIAGONAL}}$.
\par
The fields $U,V$ should represent the
fundamental degrees of freedom of the superconformal theory, since they
appear as chiral fields in the gauge theory. As before, we can find a second
representation of our manifold in terms of an embedding in some ${\mathbb C}^p$
with coordinates representing the chiral composite operators of our CFT.
In this case, we have an embedding in ${\mathbb C}^{30}$. We again construct
all the $U(1)$ invariants (in this case there are 30 of them) and we
find that they are assigned to the $({\bf 10},{\bf 3})$ of
$SU(3)\times SU(2)$. The embedding equations of the conifold into ${\mathbb C}^{30}$
correspond to the statement that in the Clebsch--Gordon expansion of
the symmetric product $({\bf 10},{\bf 3})\otimes_s ({\bf 10},{\bf3})$
all representations different from the highest
weight one should vanish. This yields $325$ equations grouped into
$5$ irreducible representations (see Section \ref{MP29} for details).
\par
As in the $Q^{1,1,1}$ case, the $X^{ij\ell\vert AB}$ can be associated
with the building blocks of the gauge invariant composite operators of our CFT
and the ideal generated by the embedding equations (\ref{embeddi})
(see Section \ref{MP29}) imposes many restrictions
on the existing conformal operators. Actually, as we try to make clear
in the explicit comparison with Kaluza Klein data (see Section
\ref{comparo}), the entire spectrum is fully determined by the
structure of the ideal above. Indeed,
as it should be clear from the previous group theoretical description of
the embedding equations, the result of the constraints is to select
chiral operators which are totally symmetrized in the $SU(3)$ and $SU(2)$
indices.
\section{The non-abelian theory and the comparison with KK spectrum}
In the previous Section, we explicitly constructed an abelian theory
whose moduli space of vacua
reproduces the cone over the two manifolds $Q^{1,1,1}$ and $M^{1,1,1}$.
These can be easily promoted to non-abelian ones.
Once this is done, we can compare the expected spectrum
of short operators in the CFT with the KK spectrum. In this Section
we compare only the chiral operators. The comparison of the full
spectrum, which is known only for $M^{1,1,1}$, will be done in Part
\ref{comparo}.
\subsection{The case of $Q^{1,1,1}$}
\begin{figure}[ht]
\begin{center}
\epsfxsize = 8cm
\epsffile{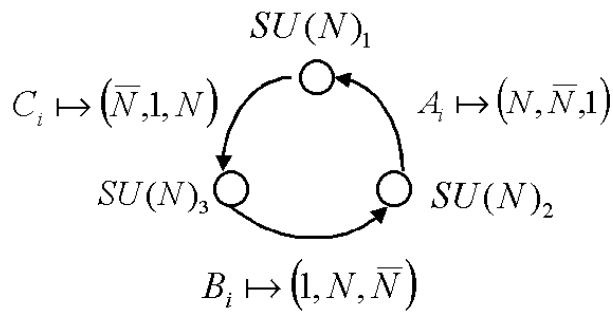}
\vskip  0.2cm
\hskip 2cm
\unitlength=1.1mm
\end{center}
\caption{Gauge group $SU(N)_1 \times SU(N)_2 \times SU(N)_3$ and color
representation assignments of the supersingleton  fields
$A_i, \, B_j,\, C_\ell$ in the $Q^{1,1,1}$ world volume gauge theory.
}
\label{ABCcolor}
\end{figure}

The theory for $Q^{1,1,1}$ becomes $SU(N)\times SU(N)\times SU(N)$ with
three series of chiral fields in the following representations of
the gauge group
\begin{equation}
A_i:\qquad ({\bf N},\bar {\bf N},{\bf 1}),\qquad\qquad
B_l:\qquad ({\bf 1,N},\bar {\bf N}),\qquad\qquad
C_p:\qquad (\bar {\bf N},{\bf 1,N})\, .\label{repps}
\end{equation}
The field content can be conveniently encoded in a quiver diagram,
where nodes represent the gauge groups and links matter fields in the
bi-fundamental representation of the groups they are connecting.
The quiver diagram for $Q^{1,1,1}$ is pictured in figure \ref{ABCcolor}.
The global symmetry of the gauge theory is $SU(2)^3$,
where each of the doublets of chiral fields transforms in the fundamental
representation of one of the $SU(2)$'s.

Notice that we are considering $SU(N)$ gauge group and not the naively expected
$U(N)$. The reason is that there is compelling evidence
\cite{witten,wittenaharon,bariowit} that the $U(1)$ factors
are washed out in the near horizon limit. Since in three dimensions
$U(1)$ theories may give rise to CFT's in the IR, it is an important point
to check whether $U(1)$ factors are described by the $AdS$-solution or not.
A first piece of evidence that the supergravity solutions are dual to
$SU(N)$ theories, and not $U(N)$, comes from the absence in the KK spectrum
(even in the maximal supersymmetric case) of KK modes corresponding
to color trace of single fundamental fields of the CFT, which are non
zero only for $U(N)$ gauge groups. A second evidence is the existence
of states dual to baryonic operators in the non-perturbative spectrum of
these Type II or M-theory compactifications; baryons
exist only for $SU(N)$ groups. We will find baryons in the spectrum of both
$Q^{1,1,1}$ and $M^{1,1,1}$: this implies that, for the compactifications
discussed in this paper, the  gauge group of the CFT is $SU(N)$.

In the non-abelian case, we expect that the
generic point of the moduli space corresponds to N separated branes.
Therefore,  the space of vacua of the theory should
reduce to the symmetrization of N copies
of $Q^{1,1,1}$. To get rid of unwanted light non-abelian degrees of freedom,
we would like to introduce, following \cite{witkleb}, a superpotential
for our theory. Unfortunately, the obvious candidate for this job
\begin{equation}
\epsilon^{ij}\epsilon^{mn}\epsilon^{pq}{\rm Tr}(A_iB_mC_pA_jB_nC_q)
\label{supoQ}
\end{equation}
is identically zero. Here the close analogy with $T^{1,1}$ and
reference \cite{witkleb} ends.

We consider now the spectrum of KK excitations of $Q^{1,1,1}$.
The full spectrum
of $Q^{1,1,1}$ is not known; however, the eigenvalues of the laplacian
were computed in \cite{popelast}. As shown in \cite{M111spectrum}, the
knowledge of the laplacian eigenvalues allows to compute the entire spectrum
of hypermultiplets of the theory, corresponding to the chiral
operators of the CFT. The result is that there is a
chiral multiplet in the $(k/2,k/2,k/2)$ representation of $SU(2)^3$
for each integer value of k, with dimension $E_0=k$. We naturally associate
these multiplets with the series of composite operators
\begin{equation}
{\rm Tr}(ABC)^k,
\label{chiralQ}
\end{equation}
where the $SU(2)$'s indices are totally symmetrized. A first
important result, following from the existence of these hypermultiplets
in the KK spectrum, is  that the dimension of the
combination $ABC$ at the superconformal point must be 1.

We see that the prediction from the KK spectrum are in perfect
agreement with the geometric discussion in the previous Section.
Operators which are not totally symmetric in the flavor indices do not
appear in the spectrum. The agreement with the proposed CFT, however,
is only partial. The chiral operators predicted by supergravity certainly
exist in the gauge theory. However, we can construct many more chiral
operators which are not symmetric in flavor indices.
They do not have any counterpart in the KK spectrum.
The superpotential in the case of $T^{1,1}$ \cite{witkleb} had the double
purpose of getting rid of the unwanted non-abelian degrees of freedom
and of imposing, via the equations of motion, the total symmetrization
for chiral and short operators which is predicted both by geometry and by
supergravity. Here, we are not so lucky, since there is no superpotential.
We can not consider superpotentials
of dimension bigger than that considered before
(for example, cubic or quartic in $ABC$) because
the superpotential ~(\ref{supoQ}) is
the only one which has dimension compatible with the supergravity
predictions. \footnote{For a three dimensional theory to be conformal
the dimension of the superpotential must be 2.}
We need to suppose that all the non symmetric operators
are not conformal primary. Since the relation between
R-charge and dimension is only valid for conformal chiral operators,
such operators are not protected  and
therefore may have enormous anomalous dimension, disappearing
from the spectrum.
Simple examples of chiral but not conformal operators are those
obtained by derivatives of the superpotential.
Since we do not have a superpotential here, we have to suppose that
both the elimination of the unwanted colored massless states as well
as the disappearing of the non-symmetric chiral operators
emerges as a non-perturbative IR effect.

\subsection{The case of $M^{1,1,1}$}
Let us now consider $M^{1,1,1}$.
The non-abelian theory is now $SU(N)\times SU(N)$ with
chiral matter in the following representations of
the gauge group
\begin{equation}
U^i\in Sym^2({\mathbb C}^N)\otimes Sym^2({\mathbb C}^{N*}),\qquad\qquad
V^A\in Sym^3({\mathbb C}^{N*})\otimes Sym^3({\mathbb C}^N).
\label{reppsM}
\end{equation}
The representations of the fundamental fields have been chosen in such a
way that they reduce to the abelian theory discussed in the previous
Section, match with the KK spectrum and imply the existence of baryons
predicted by supergravity.
Comparison with supergravity, which will be made soon,
justifies, in particular,
the choice of color symmetric representations.

\begin{figure}[ht]
\begin{center}
\epsfxsize = 8cm
\epsffile{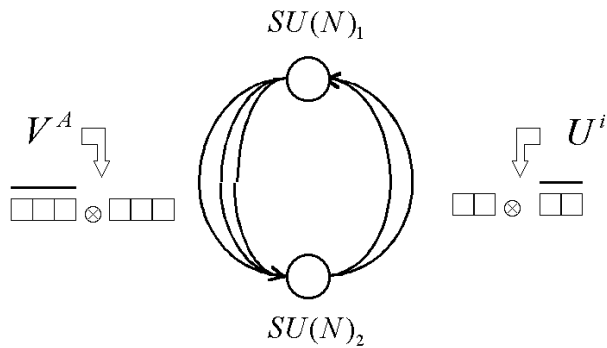}
\vskip  0.2cm
\hskip 2cm
\unitlength=1.1mm
\end{center}
\caption{Gauge group $U(N)_1 \times U(N)_2 $ and color
representation assignments of the supersingleton  fields
$V^A$ and $U^i$ in the $M^{1,1,1}$ world volume gauge theory.
}
\label{UVcolor}
\end{figure}

The field content can be conveniently encoded in the  quiver diagram
 in figure \ref{UVcolor}.

The global symmetry of the gauge theory is $SU(3)\times SU(2)$,
with the chiral fields $U$ and $V$ transforming in the fundamental
representation of $SU(3)$ and $SU(2)$, respectively.
\par
We next compare the expectations from gauge theory
with the KK spectrum \cite{M111spectrum}.
Let us start with the hypermultiplet spectrum (the full spectrum of KK
modes will be discussed in Part \ref{comparo}).
There is exactly one hypermultiplet in the  symmetric representation of
$SU(3)$ with $3k$ indices and the symmetric representation of $SU(2)$ with
$2k$ indices, for each integer $k\ge 1$.
The dimension of the operator is $E_0=2k$.
We naturally identify these states with the totally symmetrized
chiral operators
\begin{equation}
{\rm Tr} (U^3V^2)^k.
\label{chiralM}
\end{equation}
One immediate consequence of the supergravity analysis is that
the combination $U^3V^2$ has dimension 2 at the superconformal fixed point.
\par
Once again, we are not able to write any superpotential of dimension 2.
The natural candidate is the dimension two flavor singlet
\begin{equation}
\epsilon_{ijk}\epsilon_{AB}\left( U^i U^j U^k V^A V^B\right) _{\mbox{
color singlet}}
\label{supoM}
\end{equation}
which however vanishes identically.
There is no superpotential that might help in the elimination
of unwanted light colored degrees of freedom and that might
eliminate all the non symmetric chiral operators that we can construct out
of the fundamental fields. Once again, we have to suppose that, at the
superconformal fixed point in the IR, all the non totally symmetric
operators are not conformal primaries.
\subsection{The baryonic symmetries and the Betti multiplets}
\label{bariobetti}
There is one important property that $M^{1,1,1}$, $Q^{1,1,1}$ and
$T^{1,1}$ share. These manifolds have non-zero Betti numbers ($b_2=b_5=2$
for $Q^{1,1,1}$, $b_2=b_5=1$ for $M^{1,1,1}$ and $b_2=b_3=1$ for $T^{1,1}$).
This implies the existence of non-perturbative states in the supergravity
spectrum associated with branes wrapped on non-trivial cycles. They
can be interpreted as baryons in the CFT \cite{bariowit,gubserkleb}.

The existence of non-zero Betti numbers implies the existence of new
global $U(1)$ symmetries which do not come from the geometrical symmetries
of the coset manifold, as was pointed out long time ago.
The massless vector multiplets associated with these symmetries
were discovered and named Betti multiplets in \cite{univer,spec321}.
They  have the property that the entire KK spectrum is neutral
and only non-perturbative states can be charged.
The massless vectors, dual to the conserved currents, arise from the
reduction of the 11-dimensional 3-form
along the non-trivial 2-cycles. This definition implies that
non-perturbative objects made with M2 and M5 branes are charged
under these $U(1)$ symmetries.

We can identify the Betti multiplets
with baryonic symmetries. This was first pointed out in \cite{witkleb2,sergiotorino}
for the case of $T^{1,1}$ and discussed for orbifold models in
\cite{morpless}. The existence of baryons in the proposed CFT's
is due to the choice of $SU(N)$ (as opposed to $U(N)$) as gauge group.
In the $SU(N)$ case, we can form the
gauge invariant operators $\mbox{det}\, (A)$, $\mbox{det}\, (B)$ and $\mbox{det}\, (C)$ for
$Q^{1,1,1}$ and $\mbox{det}\, (U)$ and $\mbox{det}\, (V)$ for $M^{1,1,1}$.
The baryon symmetries act on fields in the same way as the $U(1)$
factors that we used for defining our abelian
theories in Sections \ref{Qabel} and \ref{Mabel}.
They disappeared in the non-abelian
theory associated to the conifolds, but the very same fact that they
can be consistently incorporated in the theory means that they must exist as
global symmetries. It is easy to check that no
operator corresponding to KK states is charged under these $U(1)$'s.
The reason is that the KK spectrum is made out with the combinations
$X=ABC$ or $X=U^3V^2$ defined in Sections \ref{Qabel} and
\ref{Mabel} which, by definition, are $U(1)$ invariant variables.
The only objects that are charged under the $U(1)$ symmetries
are the baryons.

Baryons have dimensions which diverge with $N$ and
can not appear in the KK spectrum. They are indeed non-perturbative
objects associated with wrapped branes \cite{bariowit,gubserkleb}.  We see that
the baryonic symmetries have the right properties to be associated
with the Betti multiplets: the only charged objects
are non-perturbative states. This identification can be strengthened
by noticing that the only non-perturbative branes in M-theory have
an electric or magnetic coupling to the eleven dimensional
three-form.
Since for our manifolds, both $b_2$ and $b_5$ are greater than 0,
we have the choice of wrapping both M2 and M5-branes.
M2 branes wrapped around a non-trivial two-cycle are certainly charged
under the massless vector in the Betti multiplet which is obtained by
reducing the three-form on the same cycle. Since a non-trivial 5-cycle
is dual to a 2-cycle, a similar remark applies also for M5-branes.
We identify M5-branes as baryons because they have a mass
(and therefore a conformal dimension) which goes like $N$, as discussed
in Section \ref{barioquila}.
\par
What follows from the previous discussion and is probably quite general,
is that there is a close relation between the $U(1)$'s entering
the brane construction of the gauge theory, the baryonic symmetries
and the Betti multiplets. The previous remarks apply as well to
CFT associated with orbifolds of $AdS_4\times S^7$. In the case of
$T^{1,1}$,$Q^{1,1,1}$ and $M^{1,1,1}$, the baryonic symmetries are also
directly related to the $U(1)$'s entering the toric description
of the manifold.
\subsection{Non trivial results from supergravity: a discussion}
\label{discussa}
In the previous Sections, we proposed non-abelian theories as dual candidates
for the M theory compactification on $Q^{1,1,1}$ and $M^{1,1,1}$.
We also pointed out the difficulties related to the existence of more
candidate conformal chiral operators than those expected from the KK
spectrum analysis. We have no good arguments for claiming that these
non flavor symmetric operators disappear
in the IR limit. If they survive, this certainly signals the need for
modifying our guess for the dual CFT's. In the latter case, new fields may be needed. The theories
we wrote down are based on the minimal assumption that there is no
superpotential in the abelian case\footnote{If there is a superpotential
the toric description may contain extra $U(1)$'s 
related to the F-terms of the theories, as it happens for orbifold models
\cite{douglastoric}.}; if we relax this assumption,
more complicated candidate dual
gauge theories may exist. In the case of $T^{1,1}$, the CFT was identified
in two different ways, by using the previous section arguments and also
by describing the conifold as a deformation of an orbifold singularity.
Since orbifold CFT can be often identified using standard techniques
\cite{douglasmoore}, this approach has the advantage of unambiguously
identifying the conifold CFT. It would be interesting to find
an analogous procedure for the case of $AdS_4$. It would provide
a CFT which flows in the IR to the conifold theory after a 
deformation \cite{witkleb,morpless} and it would help in checking
whether new fields are necessary or not 
for a correct description of the CFT's. 
Attempts to find associated orbifold models in the case of $Q^{1,1,1}$
have been made in \cite{tatar,dallagat}; the precise relation with
our approach is still to be clarified \footnote{A different
CFT was proposed for the case of $Q^{1,1,1}$ in \cite{tatar}; this
different proposal does not seem to solve the discrepancies with the KK
expectations.}.
\par
In any event, whatever is the microscopic description of the
gauge theory flowing to the superconformal points in the IR, it is reasonable 
to think all the relevant degrees of freedom at the superconformal
fixed point corresponding to the M theory on  $Q^{1,1,1}$ and $M^{1,1,1}$
has been identified in the previous geometrical analysis.  
We will make, from now on, the 
assumption that the fundamental singletons
of the CFT for $Q^{1,1,1}$ are the fields $A,B,C$ and for $M^{1,1,1}$
the fields $U,V$ with the previously discussed assignment of
color and flavor indices and that they always appear in totally symmetrized
flavor combinations. 
Given this simple assumption, inherited from the geometry of the conifolds,
we can make several non-trivial comparisons between the expectation
of a CFT (in which the singletons are totally symmetrized in flavor)
and the supergravity prediction. We leave for future work the clarification
of the dynamical mechanism (or possible modification of the 
three-dimensional gauge theories)  for suppressing the non-symmetric operators
as well as the search for a RG flow from an orbifold model.
\par
We already discussed the chiral operators of the two CFT's. We obtained
two  main
results from this analysis.
The first one states that all chiral operators are symmetrized in flavor indices.
The second one, more quantitative, predicts the conformal dimension of
some composite objects. When appearing in gauge invariant chiral operators,
the symmetrized combinations $ABC$  and $U^3V^2$ have dimensions 1 and 2,
respectively.

Having this information, there are two types of important and non-trivial
checks that we can make:
\begin{itemize}
\item The full spectrum of KK excitations should match with
composite operators in the CFT. Specifically, besides the hypermultiplets,
there are many other
short multiplets  in the spectrum. All these
multiplets should match with CFT operators with protected
dimension. This will be verified in Sections \ref{protcorto}, \ref{protlungo}.
\item We can
determine the dimension of a baryon operator by computing the volume of
the cycle the M5-brane is wrapping, following \cite{gubserkleb}.
From this, we can determine the dimension
of the fundamental fields of the CFT. This can be compared with the
expectations from the KK spectrum. The agreement of the two methods can be
considered as a non-trivial check of the AdS/CFT correspondence. This
will be discussed in Section \ref{barioformul}.
\end{itemize}

Leaving the actual computation and detailed comparison of spectra
for the second Part of this paper, here  we summarize the results
of our analysis.

The spectrum of $M^{1,1,1}$ is completely known \cite{M111spectrum}.
This allows a detailed comparison of all the states in supergravity
with CFT operators. Besides the hypermultiplets, which fit the
quantum field theory expectations in a straightforward manner, there
are various series of multiplets which are short and therefore
protected.
An highly non-trivial result is that we will be able
to identify all the KK short multiplets with candidate CFT operators
of requested quantum numbers and conformal dimension. Most of them
can be obtained by tensoring conserved currents with chiral operators.
The same analysis
was done for $T^{1,1}$ in \cite{sergiotorino}.
In ${\cal N}=2$ supersymmetric compactifications, the KK spectrum contains
both short and long multiplets.
We will notice that there is a common pattern in $Q^{1,1,1}$, $M^{1,1,1}$
as well as in $T^{1,1}$, of long multiplets which
have rational and protected dimension. In particular, 
following \cite{sergiotorino}, we can
identify in all these models rational long gravitons with products
of the stress energy tensor, conserved currents and chiral operators.
We suspect the existence of some field theoretical reason for
the unexpected protected dimension of these operators.

The dimension of the fundamental fields $A,B,C$ and $U,V$ at the superconformal
point can be computed and compared with the KK spectrum prediction.
In the KK spectrum, these fields always appear in
particular combinations. For example, we already know that
$ABC$ has dimension 1 and $U^3V^2$ has dimension 2. $A,B,C$ have clearly
the same dimension $1/3$ since there is a permutation symmetry.
But, what's about $U$ or $V$? From the CFT point of view,
we expect the existence of several baryon operators:
$\mbox{det} \, A$, $\mbox{det} \, B$, $\mbox{det} \, C$ for $Q^{1,1,1}$ and
$\mbox{det}\, U$, $\mbox{det}\, V$ for
$M^{1,1,1}$.
All of them should correspond to M5-branes wrapped on supersymmetric
five-cycles of $M^7$.
We can determine the dimension of the single fields $A$ or $U$
by computing the mass of a wrapped M5-brane \cite{gubserkleb}. 
This amounts to identifying
a supersymmetric 5-cycle and computing its volume.
The details of the identification of the cycles, the actual computation of
normalizations and volumes will be discussed in Sections \ref{bryn}, \ref{stab5M},
\ref{supsym5M}, \ref{brynQ}.
Here we give the results.
\par
In the case of $Q^{1,1,1}$, since the manifold is a $U(1)$ fibration
over $S^2\times S^2\times S^2$,
we can identify three distinct supersymmetric
5-cycles by considering the 5-manifolds obtained by selecting
a particular point in one of the three $S^2$.
The computation of volumes predicts a common
dimension $N/3$ for the three candidate baryons $\mbox{det} \, A$, $\mbox{det} \, B$ and
$\mbox{det} \, C$. We conclude that the three fundamental
fields $A,B,C$ have dimension $1/3$. Both the dimension
and the flavor representation of these baryons, which will
be determined in Section \ref{barioformul},
are in agreement with the KK expectations.
\par
In the case of $M^{1,1,1}$, there are two supersymmetric cycles.
$M^{1,1,1}$ is a $U(1)$ fibration over ${\mathbb P}^1\times {\mathbb P}^{2*}$.
A first non-trivial supersymmetric 5-cycle is obtained by selecting
a point in ${\mathbb P}^1$; the associated baryon carries flavor indices
of $SU(2)$. A second 5-cycle is obtained by selecting a ${\mathbb P}^1$
inside ${\mathbb P}^{2*}$ (see Section \ref{stab5M}); the associated baryon
only carries indices of $SU(3)$.
We can determine the dimensions of the
baryons $\mbox{det}\, U$ and $\mbox{det}\, V$, by computing the volume of these 5-cycles,
 and we find $4N/9$ and $N/3$,
predicting dimension $4/9$ and $1/3$ for $U$ and $V$. This strange numbers
are nevertheless in perfect agreement with the KK expectation:
the dimension of $U^3V^2$ is
\begin{equation}
3\times{4\over 9}+2\times{1\over 3}=2,
\label{check}
\end{equation}
as expected from the KK analysis.
We find that this is quite a non-trivial and remarkable check of the AdS/CFT
correspondence.

Let us finish this brief discussion, by considering the issue of possible
marginal deformations of our CFT's.  A natural question is whether the
proposed CFT's
belong to a line of fixed points or not. We already noticed
that in three dimensions there is no analogous of the $AdS_5$ dilaton
and therefore we may expect that, in general, the CFT's related to $AdS_4$ are
isolated fixed points, if we pretend to maintain the 
global symmetry and the number of supersymmetries
of our CFT's. If there is some marginal deformation 
we should be able
to see it in the KK spectrum as an operator of dimension 3.
We can certainly exclude the existence of marginal deformations
that preserves the global symmetries of the fixed point,
at least for $M^{1,1,1}$ where the KK spectrum is completely known:
there is no flavor singlet scalar of dimension 3 in the
supergravity spectrum. Other possible sources for exact marginal
deformations preserving the global symmetries
come from non-trivial cycles. In $T^{1,1}$, for example,
the second complex marginal deformation arises from the zero-mode
value of the B field on the non-trivial two-cycles of the manifold.
In our case, however, an analogous phenomenon requires
reducing the three form on a non-trivial 3-cycle,  which
 does not
exist. It is likely that marginal deformations
which break the flavor symmetry but maintain the same number of supersymmetries
exist in all these models, since non flavor
singlet multiplets with highest component of  dimension three can be found in 
the KK spectrum; whether these deformation are truly marginal or not
needs to be investigated in more details.
\par
The rest of this paper will be devoted
to an exhaustive comparison between quantum field theory and
supergravity and to a detailed description of the geometry involved in such a comparison.
\par
\part{Comparison between KK spectra and the gauge theory}
\label{comparo}
\section{Dimension of the supersingletons and the baryon operators}
\label{barioformul}
As we have anticipated in the introduction, the first basic check on
our conjectured conformal gauge theories comes from a direct
computation of the conformal weight of the singleton superfields
\begin{equation}
  \mbox{singleton superfields}=\cases{\begin{array}{cccl}
    U^i & V^A & \null &\mbox{in the $M^{1,1,1}$ theory}  \\
    \null & \null & \null \\
    A_i & B_j & C_\ell &\mbox{in the $Q^{1,1,1}$ theory}  \
  \end{array}}
\label{supsingleMQ}
\end{equation}
whose color index structure and $\theta$-expansion are explicitly
given in the later formulae (\ref{supsingM}), (\ref{supsingQ}). If the
non--abelian gauge theory has the $SU(N) \times \dots \times SU(N)$
gauge groups illustrated by the quiver diagrams of fig.s \ref{ABCcolor}
and \ref{UVcolor}, then we can consider the following chiral
operators:
\begin{eqnarray}
 \mbox{det}U&\equiv & U_{i_1\vert \Lambda^1_1\Sigma^1_1}^{\Lambda^2_1\Sigma^2_1} \,
 \dots \,
  U_{i_N\vert  \Lambda^1_N\Sigma^1_N}^{\Lambda^2_N\Sigma^2_N }\,
  \epsilon^{\Lambda^1_1 \dots \Lambda^1_N} \,\epsilon^{\Sigma^1_1 \dots \Sigma^1_N}
  \epsilon_{\Lambda^2_1 \dots \Lambda^2_N} \,\epsilon_{\Sigma^2_1 \dots \Sigma^2_N}
\label{operUUU}\\
  \mbox{det}V&\equiv & V_{A_1\vert \Lambda^1_1\Sigma^1_1\Gamma^1_1}^{\Lambda^2_1
  \Sigma^2_1\Gamma^2_1} \,
 \dots \,
  V_{A_N\vert  \Lambda^1_N\Sigma^1_N\Gamma^1_N}^{\Lambda^2_N\Sigma^2_N \Gamma^2_N}\,
  \epsilon^{\Lambda^1_1 \dots \Lambda^1_N} \,\epsilon^{\Sigma^1_1 \dots \Sigma^1_N}
  \epsilon^{\Gamma^1_1 \dots \Gamma^1_N}
  \epsilon_{\Lambda^2_1 \dots \Lambda^2_N} \,\epsilon_{\Sigma^2_1 \dots \Sigma^2_N}
  \epsilon_{\Gamma^2_1 \dots \Gamma^2_N}
\label{operVV} \\
\mbox{det}A&\equiv & A_{i_1\vert \Lambda^1_1}^{\Lambda^2_1} \, \dots \,
  A_{i_N\vert \Lambda^1_N}^{\Lambda^2_N }\, \epsilon^{\Lambda^1_1 \dots \Lambda^1_N} \,
  \epsilon_{\Lambda^2_1\dots \Lambda^2_N}
\label{q111baryopA}\\
\mbox{det}B&\equiv & B_{i_1\vert \Lambda^2_1}^{\Lambda^3_1} \, \dots \,
  B_{i_N\vert \Lambda^2_N}^{\Lambda^3_N }\, \epsilon^{\Lambda^2_1 \dots \Lambda^2_N} \,
  \epsilon_{\Lambda^3_1\dots \Lambda^3_N}
\label{q111baryopB}\\
  \mbox{det}C&\equiv & C_{i_1\vert \Lambda^3_1}^{\Lambda^1_1} \, \dots \,
  C_{i_N\vert \Lambda^3_N}^{\Lambda^1_N }\, \epsilon^{\Lambda^3_1 \dots \Lambda^3_N} \,
  \epsilon_{\Lambda^1_1\dots \Lambda^1_N}.
\label{q111baryopC}
\end{eqnarray}
If these operators are truly chiral primary fields, then their
conformal dimensions are obviously given by
\begin{equation}
\begin{array}{ccccccccccc}
h[\mbox{det} \, U] & = & h[U] \, \times \, N  & ; &
h[\mbox{det} \, V] & = & h[V] \, \times \, N  & \null & \null & \null & \null \\
h[\mbox{det} \, A] & = & h[A] \, \times \, N & ; &
h[\mbox{det} \, B] & = & h[B] \, \times \, N & ; &
h[\mbox{det} \, C] & = & h[C] \, \times \, N \\
\end{array}
\label{trulychir}
\end{equation}
and their flavor representations are:
\begin{eqnarray}
\mbox{det} \, U & \Rightarrow & (M_1=N, M_2=0, J=0) \label{repdetU},\\
\mbox{det} \, V & \Rightarrow & (M_1=0, M_2=0, J=N/2)  \label{repdetV},\\
\mbox{det} \, A & \Rightarrow & (J_1=N/2, J_2=0, J_3=0) \label{repdetA},\\
\mbox{det} \, B & \Rightarrow & (J_1=0, J_2=N/2, J_3=0)  \label{repdetB},\\
\mbox{det} \, C & \Rightarrow & (J_1=0, J_2=0, J_3=N/2)  \label{repdetC},
\end{eqnarray}
where the conventions for the flavor representation labeling are
those explained later in eq.s (\ref{su3young}), (\ref{su2young}).
\par
The interesting fact is that the conformal operators
(\ref{operUUU},...,\ref{q111baryopC}) can be reinterpreted as
solitonic supergravity states obtained by wrapping a $5$--brane on a
non--trivial supersymmetric $5$--cycle. This gives the possibility of calculating
directly the mass of such states and, as a byproduct, the conformal
dimension of the individual supersingletons. All what is involved is
a geometrical information, namely the ratio of the volume of the
$5$--cycles to the volume of the entire compact $7$--manifold. In
addition, studying the stability subgroup of the supersymmetric
$5$--cycles, we can also verify that the gauge--theory predictions
(\ref{repdetU},...,\ref{repdetC}) for the flavor representations are
the same one obtains in supergravity looking at the state as a
wrapped solitonic $5$--brane.
\par
To establish these results we need to  derive a general
mass--formula for baryonic states corresponding to wrapped
$5$--branes. This formula is obtained by considering various
relative normalizations.
\subsection{The M2 brane solution and normalizations of the seven
manifold metric and volume}
Using the conventions and  normalizations of
\cite{ricpie11,castdauriafre} for $D=11$ supergravity and  for its
Kaluza Klein expansions, a Freund Rubin solution on $AdS_4 \times M^7$
is described by the following three equations:
\begin{equation}
\begin{array}{ccccccc}
  R^{ab} & = &  -16 e^2 \, E^a \wedge E^b & \Rightarrow & R^{ a  b}_{ c b} & = &
  - 24 \, e^2 \, \delta^a_c \\
  R^{\hat a  \hat b} & = & R^{\hat a \hat b}_{\hat c \hat d} \, B^{\hat
c} \, B^{\hat d} & \mbox{with} & R^{\hat a \hat b}_{\hat c \hat
b} & = & 12 \, e^2 \, \delta^{\hat a}_{\hat c} \\
  F^{[4]} & = &  e \, \varepsilon_{abcd} \, E^a \, \wedge\, E^{b}\, \wedge\,
E^c \, \wedge\, E^d\,, & \null & \null & \null  & \null
\end{array}
\label{freurub}
\end{equation}
where $E^a$ ($a=0,1,2,3$) is the vielbein of anti de Sitter space
$AdS_4$, $R^{ab}$ is the corresponding curvature $2$--form, $B^{\hat
a}$ ($\hat a = 4,\dots, 10$) is the vielbein of $M^7$ and $R^{\hat a \hat b}$ is the
corresponding curvature. The parameter $e$, expressing the {\it vev}
of the $4$--form field strength, is called the Freund Rubin parameter.
In these normalizations, both the internal and space--time vielbeins
do not have their physical dimension of a length $[E^a]_{phys}=[B^{\hat a}]_{phys}=\ell$,
since one has reabsorbed the Planck length $\ell_P$ into their definition
by working in natural units where the $D=11$ gravitational constant $G_{11}$ has been
set equal to $1\over 8\pi$. Physical units are reinstalled through the following rescaling:
\begin{eqnarray}
  E^a &=& \frac{1}{\kappa^{2/9}}\, \hat{E}^a,\nonumber\\
  B^a &=& \frac{1}{\kappa^{2/9}}\, \hat{B}^a,\nonumber\\
  F^{[4]}_{abcd} &=& \kappa^{11/9}\, \hat{F}^{[4]}_{abcd},\nonumber\\
  \kappa^2 & = & 8 \pi G_{11} = \frac{(2\pi)^8}{2}\, \ell_P^9.
\label{riscal}
\end{eqnarray}
After such a rescaling, the relations between the Freund Rubin parameter and
the curvature scales for both $AdS_4$ and $M^7$ become
\begin{eqnarray}
\mbox{Ricci}_{\mu\nu}^{AdS} & = & - 2 \, \Lambda  \, g_{\mu\nu} \label{ricciout} \\
\mbox{Ricci}_{\hat \mu \hat \nu} \ &=& \Lambda  g_{\hat \mu \hat \nu}
\label{ricciin} \\
\Lambda & \stackrel{\mbox{\scriptsize def} }{=} & 24 \, \frac{e^2}{\kappa^{4/9}}.
\label{riccio}
\end{eqnarray}
Note that in eq. (\ref{riccio}) we have used the
normalization of the Ricci tensor which is standard in the general relativity
literature and is twice the normalization of the Ricci tensor $R^{ab}_{cb}$ appearing in
eq. (\ref{freurub}). Furthermore eq.s (\ref{freurub}) were written in
flat indices while eq.s (\ref{ricciout}, \ref{ricciin}) are written in
curved indices.
\par
For our further reasoning, it is convenient to write the anti de
Sitter metric in the solvable coordinates \cite{osp48,maldapasto}:
\begin{eqnarray}
  ds^2_{AdS_4} &=& R_{AdS}^2 \left[ \rho^2 \left( -dt^2 + dx_1^2 +
  dx_2^2 \right) + \frac{d\rho^2}{\rho^2} \right], \nonumber\\
  \mbox{Ricci}_{\mu\nu}^{AdS}&=& - \frac{3}{R^2_{AdS}}\, g_{\mu\nu},\nonumber\\
\label{adsolv}
\end{eqnarray}
which yields the relation
\begin{equation}
  R_{AdS}= \frac{\kappa^{2/9}}{4 \, e}=\frac{1}{2}\,
  \sqrt{\frac{6}{\Lambda}}.
\label{adsrad}
\end{equation}
\par
Next, following \cite{g/hm2} we can consider the exact M2--brane
solution of $D=11$ supergravity that has the cone ${\cal C}(M^7)$ over $M^7$
as transverse space. The $D=11$ bosonic action can be written as
\begin{equation}
I_{11}=\int d^{11}x \sqrt{-g} ~({R\over\kappa^2}-3\, \hat{F}^2_{[4]} ) + {288} \sigma\int
\hat{F}_{[4]} \wedge
\hat{F}_{[4]} \wedge \hat{A}_{[3]}
\label{elevenaction}
\end{equation}
(where the coupling constant for the last term is $\sigma=\kappa$) 
and the exact $M2$--brane solution is as follows:
\begin{eqnarray}
ds^2_{M2} & = & \left( 1+\frac{k}{r^6}\right) ^{-2/3}(-dt^2 + dx_1^2 +dx_2^2)+
\left( 1+\frac{k}{r^6}\right) ^{1/3} \, ds^2_{cone},  \nonumber\\
ds^2_{cone} & = & dr^2 +r^2\frac{\Lambda}{6} ds^2_{M^7}, \nonumber\\
A^{[3]} & = & dt\wedge dx_1 \wedge dx_2 \,\left( 1+\frac{k}{r^6}\right) ^{-1},
\label{ghm2bra}
\end{eqnarray}
where $ds^2_{M^7}$ is the Einstein metric on $M^7$, with Ricci tensor
as in eq. (\ref{riccio}), and $ds^2_{cone}$ is the corresponding Ricci
flat metric on the associated cone. When we go near the horizon, $r\to
0$, the metric (\ref{ghm2bra}) is approximated by
\begin{equation}
  ds^2_{M2} \approx r ^{4}(-dt^2 + dx_1^2 +dx_2^2) \, k^{-2/3} +
  k^{1/3} \, \frac{dr^2}{r^2} + k^{1/3} \, \frac{\Lambda}{6} \,
  ds_{M^7}^2.
\label{approxi}
\end{equation}
The Freund Rubin solution $AdS_4 \times M^7$ is obtained by setting
\begin{equation}
  \rho= {2\over\sqrt{k}} \, r^2
\label{rhotor}
\end{equation}
and by identifying
\begin{equation}
  R_{AdS}= \frac{k^{1/6}}{2} \quad \Leftrightarrow \quad \Lambda = 6 k^{-1/3}.
\label{agnisco}
\end{equation}
\subsection{The dimension of the baryon operators}
\label{barioquila}
Having fixed the normalizations, we can now compute the mass of a M5-brane
wrapped around a non-trivial supersymmetric cycle of $M^7$ and the conformal
 dimension of the associated baryon operator.
\par
The parameter $k$ appearing in the M2-solution is obviously proportional to
the number $N$ of membranes generating the $AdS$-background and, by dimensional
analysis, to $l_P^6$. The exact relation for the maximally supersymmetric case
$AdS_4\times S^7$ can be found in \cite{maldapasto} and
reads
\begin{equation}
  R_{AdS}= {l_P\over 2}\left (2^5\pi^2 N\right )^{1/6} \quad \Leftrightarrow \quad k=2^5 \pi^2 N l_P^6.
\label{rel}
\end{equation}

We can easily adapt this formula to the case of $AdS_4\times M^7$ by noticing
that, by definition,  the number of M2-branes $N$ is determined by the
flux of the RR
three-form through $M^7$, $\int_{M^7} *F^{[4]}$. As a consequence, $N$ and
the volume of $M^7$ will appear in all the relevant formulae in the
combination $N/{\rm Vol}(M^7)$. We therefore obtain the general formula
\begin{equation}
\sqrt{{\Lambda\over 6}}=\left ( {{\rm Vol}(M^7)\over {\rm Vol}(S^7)}\right )^{1/6}
{1\over l_P(2^5\pi^2 N)^{1/6}}.
\label{general}
\end{equation}
\par
We can now consider the solitonic particles in $AdS_4$ obtained by
wrapping M2- and
M5-branes on the
non-trivial 2- and 5-cycles of $M^7$, respectively.
They are associated with boundary operators with conformal
dimensions that diverge in the
large $N$ limit. The exact dependence on $N$ can be easily
estimated.
Without loss of generality, we can put
$\Lambda=1$ using a conformal transformation; its only role in the game
is to fix a reference scale and it will eventually cancel in the
final formulae. The mass of a p-brane wrapped on a p-cycle is given by
$T_p\times {\rm Vol(p-cycle)}\sim l_P^{-\left(p+1\right)}\Lambda^{-{p\over 2}}\sim 
l_P^{-\left(p+1\right)}$. Once the
mass of the non-perturbative states is known, the dimension $E_0$ of the
associated
boundary operator is given by the relation $m^2=(2\Lambda/3)(E_0-1)(E_0-2)
\sim 2E_0^2/3$. From equation ~(\ref{general}) we learn that
$l_P\sim N^{-1/6}$.
We see that M2-branes correspond to operators with dimension $\sqrt{N}$
while M5-branes to operators with dimension of order $N$. The natural
candidates for the baryonic operators we are looking for are therefore
the wrapped five-branes.
\par
We can easily write a more precise formula for the dimension of the baryonic
operator associated with a wrapped  M5-brane, following the analogous 
computation in \cite{gubserkleb}. For this, we need the exact
expression for the M5 tension which can be found, for example, in
\cite{dealwis}. We find
\begin{equation}
m={1\over (2\pi )^5l_P^6}{\rm Vol(5-cycle)}.
\label{int}
\end{equation}

Using equation~(\ref{general}) and the above discussed relation
between mass of the bulk particle and
conformal dimension of the associated boundary operators, we obtain the
formula for the dimension of a baryon,
\begin{equation}
E_0={\pi N\over \Lambda}{{\rm Vol(5-cycle)}\over {\rm Vol}(M^7)},
\label{baryondim}
\end{equation}
where the volume is evaluated with the internal metric normalized so that
(\ref{ricciin}) is true.
\par
As a check, we can compute the dimension of a Pfaffian operator in the
${\cal N} = 8$ theory with gauge group $SO(2N)$. The theory contains
adjoint scalars which can be represented as antisymmetric matrices
$\phi_{ij}$ and we can form the gauge invariant baryonic operator
$\epsilon_{i_1,...,i_{2N}}\phi_{i_1i_2}....\phi_{i_{2N-1}i_{2N}}$
with dimension $N/2$. The internal manifold
is ${\mathbb {RP}}^7$ \cite{bariowit,ahn}, a supersymmetric preserving ${\mathbb Z}_2$
projection of original $AdS_4\times S^7$ case, corresponding to the $SU(N)$
gauge group. We obtain the Pfaffian by wrapping an M5-brane on a ${\mathbb {RP}}^5$
submanifold. Equation ~(\ref{baryondim}) gives
\begin{equation}
E_0= {\pi N\over \Lambda}{{\rm Vol}{\mathbb {RP}}^5\over {\rm Vol}{\mathbb {RP}}^7}=
{\pi N\over \Lambda}{{\rm Vol}S^5\over {\rm Vol}S^7}= N/2,
\end{equation}
as expected.
\par
Let us now apply the above formula to the case of the theories
$M^{1,1,1}$ and $Q^{1,1,1}$.
In Section \ref{bryn} we show that the
Sasakian manifold $M^{1,1,1}$ has two homology $5$ cycles ${\cal
C}^1$ and ${\cal C}^2$ (see their definition in eq.s (\ref{cycca1}, \ref{cycca2}))
belonging to the unique homology class, but distinguished
by their stability subgroups
$H({\cal C}^{1,2}) \subset SU(3)\times SU(2) \times U(1)$, respectively
given in eq.s (\ref{hcyc1}) and (\ref{hcyc2}). Furthermore in Section
\ref{supsym5M} we show that, after being pulled back to these cycles, the
$\kappa$--supersymmetry projector of the $5$--brane is non vanishing
on the killing spinors of the supersymmetries preserved by
$M^{1,1,1}$. This proves that the $5$--brane wrapped on these cycles
is a $BPS$--state with mass equal to its own $6$--form charge or,
briefly stated, that the $5$--cycles are supersymmetric.
Wrapping the $5$--brane on these cycles, we obtain good candidates for
the supergravity representation of the baryonic operators
(\ref{operUUU}) and (\ref{operVV}). To understand which is which, we
have to decide the flavor representation. This is selected by the
stability subgroup $H({\cal C}^i)$. Following an argument
introduced by Witten \cite{bariowit}, the collective degrees of
freedom $c$ of the wrapped $5$--brane soliton live on the coset manifold
$G/H({\cal C}^i)$, where $G$ is the isometry group of $M^7$. The
wave--function $\Psi(c)$ of the soliton must be expanded in harmonics on
$G/H({\cal C}^i)$ characterized by having  charge $N$ under the
baryon number $U(1)_B \subset H({\cal C}^i)$. Minimizing the energy
operator (the laplacian) on such harmonics one obtains the
corresponding $G$ representation and hence the flavor assignment of
the baryon. In Section \ref{stab5M},
applying such a discussion to the pair of $5$--cycles under
consideration, we find that they are respectively associated with the flavor
representations
\begin{eqnarray}
{\cal C}^1 & \Leftrightarrow & (M_1=0,M_2=0,J=N/2) \label{repre1M}\\
{\cal C}^2 & \Leftrightarrow & (M_1=N,M_2=0,J=0) \label{repre2M}
\end{eqnarray}
(see eq.s (\ref{JisN}), and (\ref{m1m2N})). Comparing
eq.s (\ref{repre1M}, \ref{repre2M}) with eq.s (\ref{repdetU},
\ref{repdetV}), we  see that the first cycle
is a candidate to represent the operator $\mbox{det}\, U$, while the
second cycle is a candidate to represent the operator
$ \mbox{det}\, V$. The final check comes from the evaluation of the
cycle volumes. This is done in eq.s (\ref{volcyc1}) and
(\ref{volcyc2}). Inserting these results and the formula
(\ref{volm111}) for the $M^{1,1,1}$ volume into the general
formula (\ref{baryondim}), we obtain
\begin{eqnarray}
  E_0\left( \mbox{det} \, U \right) & =& \ft{4}{9}\, \times \, N
  \quad \Rightarrow h[U] = \ft{4}{9}, \\
  E_0\left( \mbox{det} \, U \right) & =& \ft{1}{3}\, \times \, N
  \quad \Rightarrow h[V] = \ft{1}{3}.
\label{sbingus}
\end{eqnarray}
As we have already stressed, it is absolutely remarkable that these
non--perturbatively determined conformal weights are in perfect
agreement with the Kaluza Klein spectra as we show in Section
\ref{supertesso}.
\par
In Section \ref{brynQ} we show that the manifold $Q^{1,1,1}$ has
three homology cycles ${\cal C}^{A,B,C}$ permuted by the $\Sigma_3$ symmetry
that characterizes this manifold. Their volume is calculated in eq.
(\ref{volcycq}) and their stability subgroups in
eq. (\ref{hcycq111}). Applying the same argument as above, we show in
Section \ref{brynQ} that the flavor representations associated with
these three cycles are indeed those of eq.s
(\ref{repdetA},...,\ref{repdetC}), so that these three cycles are
candidates as supergravity representations of the conformal operators
${\rm det} \, A$, ${\rm det} \, B$,  and ${\rm det} \, C$. Inserting
the volume (\ref{volcycq}) of the cycles and the volume
(\ref{volq111}) of $Q^{1,1,1}$ into the baryon formula
(\ref{baryondim}), we find that the conformal dimension of the $A,B,C$
supersingletons is
\begin{equation}
  h[A_i]=h[B_j]=h[C_\ell]=\frac{1}{3}
\label{ABCunterz}
\end{equation}
as stated in eq. (\ref{hABfield}).
\section{Conformal superfields of the $M^{1,1,1}$ and $Q^{1,1,1}$ theories}
\label{supertesso}
Starting from the choice of the supersingleton fields and of the
chiral ring (inherited from the geometry of the
compact Sasakian manifold), we can build all sort of candidate conformal superfields
for both theories $M^{1,1,1}$ and $Q^{1,1,1}$. In the first case,
where the full spectrum of $Osp(2|4)\times SU(3)\times SU(2)$
supermultiplets has already been determined through harmonic analysis
\cite{M111spectrum}, relying on the conversion vocabulary between $AdS_4$
bulk supermultiplets and boundary superfields established in
\cite{superfieldsN2D3}, we can make a detailed comparison
of the Kaluza Klein predictions with the candidate conformal
superfields available in the gauge theory. In particular we find the
gauge theory interpretation of the entire spectrum of short
multiplets. The corresponding short superfields are in the right
$SU(3)\times SU(2)$ representations and have the right conformal
dimensions. Applying the same scheme to the case of $Q^{1,1,1}$, we
can use the gauge theory to make predictions about the spectrum of
short multiplets one should find in Kaluza Klein harmonic expansions.
The partial results already known from harmonic analysis on
$Q^{1,1,1}$ are in agreement with these predictions.
\par
In addition, looking at the results of \cite{M111spectrum}, one finds
that there is a rich collection of long multiplets whose conformal
dimensions are rational and seem to be protected from acquiring
quantum corrections. This is in full analogy with results
obtained in the four--dimensional theory associated with the
$T^{1,1}$ manifold \cite{gubser,sergiotorino}. 
Actually, we find an even larger class of
such {\sl rational} long multiplets. For a subclass of them the gauge
theory interpretation is clear while for others it is not immediate.
Their presence, which seems universal in all coset models, indicates
some general protection mechanism that has still to be clarified.
\par
Using  the notations of \cite{superfieldsN2D3},
the singleton  superfields of the $M^{1,1,1}$ theory are the
following ones:
\begin{eqnarray}
U^{i\vert\Lambda\Sigma}_{\phantom{i\vert\Lambda\Sigma} \underline{\Gamma\Delta}}(x,\theta)
&=&u^{i\vert\Lambda\Sigma}_{\phantom{i\vert\Lambda\Sigma} \underline{\Gamma\Delta}}(x)+
\left(\lambda_u^{\alpha}\right)^{i\vert\Lambda\Sigma}_{\phantom{i\vert\Lambda\Sigma}
\underline{\Gamma\Delta}}(x)~\theta^+_{\alpha}\nonumber,\\
V^{A\vert\underline{\Gamma\Delta\Theta}}_{\phantom{A\vert\underline{\Gamma\Delta\Theta}}
\Lambda\Sigma\Pi}(x,\theta)
&=&v^{A\vert\underline{\Gamma\Delta\Theta}}_{\phantom{A\vert\underline{\Gamma\Delta\Theta}}
\Lambda\Sigma\Pi}(x)+
\left(\lambda_v^{\alpha}\right)^{A\vert
\underline{\Gamma\Delta\Theta}}_{\phantom{A\vert\underline{\Gamma\Delta\Theta}}
\Lambda\Sigma\Pi}(x)~\theta^+_{\alpha},
\label{supsingM}
\end{eqnarray}
where $(i,A)$ are $SU(3)\times SU(2)$ {\sl flavor} indices, $(\Lambda,
\underline{\Lambda})$ are
$SU(N)\times SU(N)$ {\sl color} indices while  $\alpha$ is a world volume
spinorial index of $SO(1,2)$.
The {\sl supersingletons} are chiral superfields, so they satisfy $E_0=|y_0|$.
\par
$U^i$ is in the fundamental representation ${\bf 3}$  of
$SU(3)_{\rm flavor}$ and in the $\left( \Box\!\Box,{\Box\!\Box}^\star\right) $ of
$(SU(N)\times SU(N))_{\rm color}$.
$V^A$ is in the fundamental representation ${\bf 2}$  of
$SU(2)_{\rm flavor}$ and in the $\left(
{\Box\!\Box\!\Box}^\star,
\Box\!\Box\!\Box\right)$
of $\left(SU(N)\times SU(N)\right)_{\rm color}$. In eq.s (\ref{supsingM})
we have followed the conventions that lower $SU(N)$ indices transform
in the fundamental representation, while upper $SU(N)$ indices
transform in the complex conjugate of the fundamental representation.
\par
Studying the non perturbative baryon state, obtained by wrapping the
$5$--brane on the supersymmetric cycles of $M^{1,1,1}$, we have
unambiguously established the conformal weights of the
supersingletons (or, more precisely, the conformal weights of the
Clifford vacua $u,v$) that are:
\begin{equation}
E_0(u)=y_0(u)={4\over 9},~~~E_0(v)=y_0(v)={1\over 3}.
\label{hsupsingM}
\end{equation}
For the $Q^{1,1,1}$ theory the singleton superfields are instead the
following ones:
\begin{eqnarray}
A^{\phantom{\Lambda_1}\Gamma_2}_{i_1\vert\Lambda_1}(x,\theta)
&=&a^{\phantom{\Lambda_1}\Gamma_2}_{i_1\vert\Lambda_1}(x)+
\left(\lambda_a^{\alpha}\right)^{\phantom{\Lambda_1}\Gamma_2}_{i_1\vert\Lambda_1}
(x)~\theta^+_{\alpha}\nonumber,\\
B^{\phantom{\Lambda_2}\Gamma_3}_{i_2\vert\Lambda_2}(x,\theta)
&=&b^{\phantom{\Lambda_2}\Gamma_3}_{i_2\vert\Lambda_2}(x)+
\left(\lambda_b^{\alpha}\right)^{\phantom{\Lambda_2}\Gamma_3}_{i_2\vert\Lambda_2}
(x)~\theta^+_{\alpha}\nonumber,\\
C^{\phantom{\Lambda_3}\Gamma_1}_{i_3\vert\Lambda_3}(x,\theta)
&=&c^{\phantom{\Lambda_3}\Gamma_1}_{i_3\vert\Lambda_3}(x)+
\left(\lambda_c^{\alpha}\right)^{\phantom{\Lambda_3}\Gamma_1}_{i_3\vert\Lambda_3}
(x)~\theta^+_{\alpha},
\label{supsingQ}
\end{eqnarray}
where $i_\ell$ $(\ell=1,2,3)$ are flavor indices of $SU(2)_1 \times
SU(2)_2 \times SU(2)_3$, while $\Lambda_\ell$  $(\ell=1,2,3)$ are
color indices of $SU(N)_1 \times SU(N)_2 \times SU(N)_3$. Also in
this case we know the conformal dimension of the supersingleton
fields through the calculation of the conformal dimension of the
baryon operators. We have:
\begin{equation}
E_0(a)=E_0(b)=E_0(c)=y_0(a) =y_0(b)=y_0(c) ={1\over 3}.
\label{hsupsingQ}
\end{equation}
We now discuss short and long multiplets and the corresponding operators.
Our analysis closely parallels the one in \cite{sergiotorino}.
\subsection{Chiral operators}
\par
When the gauge group is $U(1)^N$, there is a simple interpretation for
the ring of the chiral superfields: they  describe the oscillations
of the $M2-$branes in the $7$ compact transverse directions, so
they should have the form of a parametric description of the manifold.
As we explain in Section \ref{MP29},
$M^{1,1,1}$ embedded in ${\mathbb P}^{29}$, can be parametrized
by
\begin{equation}
X^{ijl\vert AB}=U^iU^jU^kV^AV^B.
\end{equation}
Furthermore, the embedding equations can be reformulated in the following
way. In a product
\begin{equation}
X^{i_1j_1l_1\vert A_1B_1}\, X^{i_2j_2l_2\vert A_2B_2}\dots X^{i_kj_kl_k\vert A_kB_k}
\label{prodchir}
\end{equation}
only the highest weight representation of $SU(3)\times SU(2)$,
that is the completely symmetric in the $SU(3)$ indices
and completely symmetric in the $SU(2)$ indices, survives.
So, as advocated in eq. (\ref{3k2k}), the ring
of the chiral superfields should be composed by superfields
of the form
\begin{equation}
\label{hyper}
\Phi^{\left(i_1j_1l_1\dots i_kj_kl_k\right)\left(A_1B_1\dots A_kB_k\right)}=
\underbrace{U^{i_1}U^{j_1}U^{l_1}V^{A_1}V^{B_1}
\dots U^{i_k}U^{j_k}U^{l_k}V^{A_k}V^{B_k}}_{k}.
\end{equation}
First of all, we note that a product of supersingletons is always
a chiral superfield, that is, a field satisfying the equation
(see \cite{superfieldsN2D3})
\begin{equation}
\label{eqchiral}
{\cal D}^+_{\a}\Phi=0,
\end{equation}
whose general solution has the form
\begin{equation}
\Phi(x,\theta)=S(x)+\lambda^{\a}(x)\theta^+_{\a}+\pi(x)\theta^{+\a}\theta^+_{\a}.
\label{phisupfi}
\end{equation}
Following the notations of \cite{M111spectrum},
we identify the flavor representations with three nonnegative integers $M_1,~M_2,~2J$,
where $M_1$, $M_2$ count the boxes of an $SU(3)$ Young
diagram according to
\begin{eqnarray}
\begin{array}{l}
\begin{array}{|c|c|c|c|c|c|}
\hline
             \hskip .3 cm & \cdots & \hskip .3 cm &
             \hskip .3 cm & \cdots & \hskip .3 cm \\
\hline
\end{array}\\
\begin{array}{|c|c|c|}
             \hskip .3 cm & \cdots & \hskip .3 cm \\
\hline
\end{array}
\end{array}\ \\
\underbrace{\hskip 2.2 cm}_{M_2}
\underbrace{\hskip 2.2 cm}_{M_1}\,,
\label{su3young}
\end{eqnarray}
while $J$ is the usual isospin quantum number and counts the boxes of an
$SU(2)$ Young diagram as follows
\begin{eqnarray}
\begin{array}{|c|c|c|}
\hline
             \hskip .3 cm & \cdots & \hskip .3 cm \\
\hline
\end{array}\ \ \\
\underbrace{\hskip 2.2 cm}_{2J}.
\label{su2young}
\end{eqnarray}
The superfields (\ref{hyper}) are in the same  $Osp(2|4)\times SU(3)\times SU(2)$
representations as the bulk hypermultiplets that were determined in \cite{M111spectrum}
through harmonic analysis:
\begin{equation}
\label{reprhyper}
\cases{
M_1=3k\cr
M_2=0\cr
J=k\cr
E_0=y_0=2k\cr}
~~~k>0\,.
\end{equation}
In particular, it is worth noticing that every block $UUUVV$ is in the
$(\Box\!\Box\!\Box,\Box\!\Box)_{\rm flavor}$ and has conformal weight
\begin{equation}
3\cdot\left(4\over 9\right)+2\cdot\left(1\over 3\right)=2,
\end{equation}
as in the Kaluza Klein spectrum.
As a matter of fact, the conformal weight of a product of chiral fields
equals the sum of the weights of the single components, as in a free
field theory.
This is due to the relation $E_0=|y_0|$ satisfied by the chiral superfields and
to the additivity of the hypercharge.
\par
When the gauge group is promoted to $SU(N)\times SU(N)$, the coordinates
become tensors (see (\ref{supsingM})).
Our conclusion about the composite operators is that the only primary chiral
superfields are those which preserve the structure (\ref{hyper}).
So, for example, the lowest lying operator is:
\begin{equation}
U^{\Lambda\Sigma}_{\phantom{\Lambda\Sigma}i\vert(\underline{\Lambda\Sigma}}
U^{\Gamma\Delta}_{\phantom{\Gamma\Delta}j\vert\underline{\Gamma\Delta}}
U^{\Theta\Xi}_{\phantom{\Theta\Xi}\ell\vert\underline{\Theta\Xi})}
V^{\underline{\Lambda\Sigma\Gamma}}_{\phantom{\underline{\Lambda\Sigma\Gamma}}A\vert
(\Lambda\Sigma\Gamma}
V^{\underline{\Delta\Theta\Xi}}_{\phantom{\underline{\Delta\Theta\Xi}}B\vert
\Delta\Theta\Xi)},
\end{equation}
where the color indices of every $SU(N)$ are symmetrized.
The generic primary chiral superfield has the form (\ref{hyper}),
with all the color indices symmetrized before being contracted.
The choice of symmetrizing the color indices is not arbitrary:
if we impose symmetrization on the
flavor indices, it necessarily follows that also the color
indices are symmetrized (see Appendix \ref{AppC} for a proof
of this fact). Clearly, the $Osp(2|4)
\times SU(3)\times SU(2)$ representations (\ref{reprhyper}) of these
fields are the same as in the abelian case, namely those
predicted by the $AdS/CFT$ correspondence.
\par
It should be noted that in the $4$--dimensional analogue of these
theories, namely in the $T^{1,1}$ case \cite{witkleb,sergiotorino},
the restriction of the primary conformal fields to the
geometrical chiral ring occurs through the derivatives of the quartic
superpotential. As we already noted, in the $D=3$ theories there is no
superpotential of dimension $2$ which can be introduced and,
accordingly, the embedding equations defining the vanishing
ideal cannot be given as derivatives of a single holomorphic
"function". It follows that there is some other non perturbative
and so far unclarified mechanism that suppresses the chiral
superfields not belonging to the highest weight representations.
\par
Let us know consider the case of the $Q^{1,1,1}$ theory. Here, as
already pointed out, the complete Kaluza Klein spectrum is still under
construction \cite{merlatti}. Yet the information available in the
literature is sufficient to make a comparison between the
Kaluza Klein predictions and the gauge theory at the level of the
chiral multiplets (and also of the graviton multiplets as we show below).
Looking at table 7 of \cite{M111spectrum}, we learn that, in a generic
$AdS_4 \times M^7$ compactification, each hypermultiplet contains a
scalar state $S$ of energy label $E_0=|y_0|$, which is actually the
Clifford vacuum of the representation and corresponds to the world volume
field $S$ of eq.(\ref{phisupfi}). From the general bosonic mass--formulae of
\cite{bosmass,univer}, we know that $S$ is related to traceless
deformations of the internal metric and its mass is determined by the
spectrum of the scalar laplacian on $M^7$. In the notations of \cite{univer},
we normalize the scalar harmonics as
\begin{equation}
  \begin{array}{|c|}
\hline
             \hskip .2 cm   \\
\hline
\end{array}_{\,(0)^3} \, Y \, = \, H_0 \, Y
\label{scallapla}
\end{equation}
and we have the mass--formula (see \cite{univer} or eq.(B.3) of \cite{M111spectrum})
\begin{equation}
  m^2_S=H_0 +176 -24 \, \sqrt{H_0+36}
\label{massadiS}
\end{equation}
which, combined with the general $AdS_4$ relation between scalar
masses and energy labels $16(E_0 -2)(E_0-1)=m^2$, yields the formula
\begin{equation}
  E_0=\ft{3}{2} + \ft{1}{4}\sqrt{180+H_0 -24\sqrt{36+H_0}}
\label{eoformul}
\end{equation}
for the conformal weight of candidate hypermultiplets in terms of the
scalar laplacian eigenvalues. These are already known for $Q^{1,1,1}$
since they were calculated by Pope in \cite{popelast}. In our
normalizations, Pope's result reads as follows:
\begin{equation}
  H_0 =32 \, \left( J_1 (J_1+1) + J_2(J_2+1) + J_3(J_3+1) -\ft{1}{4}
  y^2 \right ),
\label{H0q111}
\end{equation}
where $(J_1,J_2,J_3)$ denotes the $SU(2)^3$ flavor representation and
$y$ the $R$--symmetry $U(1)$ charge. From our knowledge of the geometrical chiral ring
of $Q^{1,1,1}$ (see Section \ref{q111chirri}) and from our
calculation of the conformal weights of the supersingletons, on the
gauge theory side we expect the following chiral operators:
\begin{equation}
  \Phi_{i_1 j_1 \ell_1,\dots\, i_k j_k \ell_k} = \mbox{Tr}\, \left(
  A_{i_1} B_{j_1} C_{\ell_1} \, \dots \, A_{i_k} B_{j_k} C_{\ell_k} \right)
\label{phiq111}
\end{equation}
in the following $Osp(2\vert 4) \times SU(2) \times SU(2) \times
SU(2)$ representation:
\begin{eqnarray}
 Osp(2\vert 4) &:& \mbox{hypermultiplet with} \cases{ \matrix{E_0 & = & k\cr
 y_0 &=& k \cr}}\label{osprep}\\
 SU(2)\times SU(2) \times SU(2) &:& J_1=J_2=J_3 =\ft{1}{2} \, k\\
 &&k\ge 1.\nonumber
\label{hypq111}
\end{eqnarray}
Inserting the representation (\ref{hypq111}) into eq. (\ref{H0q111})
we obtain $H_0=16 k^2 + 48 k$ and, using this value in
eq. (\ref{eoformul}), we retrieve the conformal field theory prediction
$E_0 = k$. This shows that the hypermultiplet spectrum found in
Kaluza Klein harmonic expansions on $Q^{1,1,1}$ agrees with the chiral superfields
predicted by the conformal gauge theory.
\subsection{Conserved currents of the world volume gauge theory}
The supergravity mass--spectrum on $AdS_4 \times M^7$, where $M^7$ is
Sasakian, contains a number of {\it ultrashort} or {\it massless}
$Osp(2\vert N)$ multiplets that correspond to the unbroken local gauge
symmetries of the vacuum. These are:
\begin{enumerate}
  \item The massless ${\cal N}=2$ graviton multiplet $\left( 2, 2(\ft{3}{2}),1\right) $
  \item The massless ${\cal N}=2$ vector multiplets of the flavor group
  $G_{\rm flavor}$
  \item The massless ${\cal N}=2$ vector multiplets associated with
  the non--trivial harmonic 2--forms of $M^7$ (the Betti multiplets).
\end{enumerate}
Each of these massless multiplets must have a suitable gauge theory
interpretation. Indeed, also on the gauge theory side, the ultra--short
multiplets are associated with the symmetries of the theory (global
in this case) and are given by the corresponding conserved Noether
currents.
\par
We begin with the  stress--energy  superfield
$T_{\alpha\beta}$ which has a pair of symmetric $SO(1,2)$ spinor
indices and satisfies the conservation equation
\begin{equation}
\label{eqemtenssor}
{\cal D}^+_{\a}T^{\alpha\beta}={\cal D}^-_{\a}T^{\alpha\beta}=0.
\end{equation}
In components, the $\theta$--expansion of this superfield yields  the
stress energy tensor $T_{\mu\nu}(x)$, the ${\cal N}=2$ supercurrents
$j_\mu^{A\alpha}(x)$ ($A=1,2$) and the $U(1)$ R--symmetry current
$J_\mu^R(x)$. Obviously $T^{\alpha\beta}$ is a singlet with respect to the flavor
group $G_{\rm flavor}$ and it has
\begin{equation}
E_0=2,~~y_0=0,~~s_0=1.
\label{stressE0}
\end{equation}
This corresponds to the massless graviton multiplet of the bulk and explains
the first entry in the above enumeration.
\par
To each generator of the flavor symmetry group there corresponds, via Noether
theorem, a conserved vector supercurrent. This is a scalar superfield
$J^I(x,\theta)$ transforming in the adjoint representation of $G_{\rm flavor}$
and satisfying the conservation equations
\begin{equation}
\label{eqcurrents}
{\cal D}^{+\alpha}{\cal D}^+_{\alpha}J^I={\cal D}^{-\alpha}{\cal D}^-_{\alpha}J^I=0.
\end{equation}
These superfields have
\begin{equation}
E_0=1,~~y_0=0,~~s_0=0
\end{equation}
and correspond to the ${\cal N}=2$ massless vector multiplets of
$G_{\rm flavor}$ that propagate in the bulk. This explains the second
item of the above enumeration.
\par
In the specific theories under consideration, we can easily construct
the flavor currents  in terms of the supersingletons:
\begin{equation}
\begin{array}{rl}
  M^{1,1,1} & \cases{ \begin{array}{ccc}
     J^{\phantom{SU(3)\vert j}i}_{SU(3)\vert j} &
    = & U^{i\vert\Lambda\Sigma}_{\phantom{i\vert\Lambda\Sigma}
    \underline{\Lambda\Sigma}} \, \bar{U}_{j\vert\Lambda\Sigma}^{\phantom{i\vert\Lambda\Sigma}
    \underline{\Lambda\Sigma}}\, - \,
    {1\over 3}\delta^i_j \, U^{\ell\vert\Lambda\Sigma}_{\phantom{\ell\vert\Lambda\Sigma}
    \underline{\Lambda\Sigma}} \,
    \bar{U}_{\ell\vert\Lambda\Sigma}^{\phantom{\ell\vert\Lambda\Sigma}
    \underline{\Lambda\Sigma}} \\
    \null & \null & \\
   J^{\phantom{SU(2)\vert B}A}_{SU(2)\vert B} &
    = & V^{A\vert\underline{\Lambda\Sigma\Gamma}}_{
    \phantom{A\vert\underline{\Lambda\Sigma\Gamma}}
     \Lambda\Sigma\Gamma} \, \bar{V}_{B\vert\underline{\Lambda\Sigma\Gamma}}^{
    \phantom{B\vert\underline{\Lambda\Sigma\Gamma}}
     \Lambda\Sigma\Gamma} \, - \,
    {1\over 2}\delta^A_B \, V^{C\vert\underline{\Lambda\Sigma\Gamma}}_{
    \phantom{A\vert\underline{\Lambda\Sigma\Gamma}}
     \Lambda\Sigma\Gamma} \, \bar{V}_{C\vert\underline{\Lambda\Sigma\Gamma}}^{
    \phantom{C\vert\underline{\Lambda\Sigma\Gamma}}
     \Lambda\Sigma\Gamma} \, \\
  \end{array}} \\
  \null & \null \\
  Q^{1,1,1} & \cases{\begin{array}{ccc}
    J^{\phantom{SU(2)_1\vert j_1}i_1}_{SU(2)_1\vert j} & = &
    A^{i_1\vert\Gamma_1}_{
    \phantom{i_1\vert\Gamma_1}\Lambda_2}\, \bar{A}_{j_1\vert\Gamma_1}^{
    \phantom{j_1\vert\Gamma_1}\Lambda_2} \, - \, \ft{1}{2} \,
    \delta^{i_1}_{j_1} \, A^{\ell_1\vert\Gamma_1}_{
    \phantom{i_1\vert\Gamma_1}\Lambda_2}\, \bar{A}_{\ell_1\vert\Gamma_1}^{
    \phantom{j_1\vert\Gamma_1}\Lambda_2}
      \\
      \null & \null & \null \\
   J^{\phantom{SU(2)_2\vert j_2}i_2}_{SU(2)_1\vert j_2} & = &
    B^{i_2\vert\Gamma_2}_{
    \phantom{i_2\vert\Gamma_2}\Lambda_3}\, \bar{B}_{j_1\vert\Gamma_2}^{
    \phantom{j_1\vert\Gamma_2}\Lambda_3} \, - \, \ft{1}{2} \,
    \delta^{i_2}_{j_2} \, B^{\ell_2\vert\Gamma_2}_{
    \phantom{i_1\vert\Gamma_2}\Lambda_3}\, \bar{B}_{\ell_2\vert\Gamma_2}^{
    \phantom{j_1\vert\Gamma_2}\Lambda_3}
      \\
     \null & \null & \null \\
   J^{\phantom{SU(2)_3\vert j_3}i_3}_{SU(2)_3\vert j_3} & = &
    C^{i_1\vert\Gamma_3}_{
    \phantom{i_3\vert\Gamma_2}\Lambda_1}\, \bar{C}_{j_3\vert\Gamma_3}^{
    \phantom{j_1\vert\Gamma_3}\Lambda_1} \, - \, \ft{1}{2} \,
    \delta^{i_3}_{j_3} \, C^{\ell_3\vert\Gamma_3}_{
    \phantom{i_1\vert\Gamma_3}\Lambda_1}\, \bar{C}_{\ell_3\vert\Gamma_3}^{
    \phantom{j_1\vert\Gamma_3}\Lambda_1}~.
      \\
  \end{array}
  }
\end{array}
\label{flaviocurro}
\end{equation}
These currents satisfy eq.(\ref{eqcurrents}) and are in the right
representations of $SU(3)\times SU(2)$. Their hypercharge is $y_0=0$.
The conformal weight is not the one obtained by a naive sum,
being the theory interacting.
As shown in \cite{superfieldsN2D3}, the conserved currents satisfy
$E_0=|y_0|+1$, hence $E_0=1$.
\par
Let us finally identify the gauge theory superfields associated with
the Betti multiplets. As we stressed in the introduction, the non
abelian gauge theory has $SU(N)^p$ rather than $U(N)^p$ as gauge
group. The abelian gauge symmetries that were used to obtain the
toric description of the manifold $M^{1,1,1}$ and $Q^{1,1,1}$ in the one--brane case
$N=1$ are not promoted to gauge symmetries in the many brane regime
$N\to \infty$. Yet, they survive as exact global symmetries of the
gauge theory. The associated conserved currents provide the
superfields corresponding to the massless Betti multiplets
found in the Kaluza Klein spectrum of the bulk. As the reader can
notice, the $b_2$ Betti number of each manifold always agrees with the
number of independent $U(1)$ groups needed to give a toric
description of the same manifold. It is therefore fairly easy to
identify the Betti currents of our gauge theories. For instance
for the $M^{1,1,1}$ case the Betti current is
\begin{equation}
 J_{\rm Betti} \, = \, 2\,  U^{\ell\vert\Lambda\Sigma}_{\phantom{\ell\vert\Lambda\Sigma}
    \underline{\Lambda\Sigma}} \,
    \bar{U}_{\ell\vert\Lambda\Sigma}^{\phantom{\ell\vert\Lambda\Sigma}
    \underline{\Lambda\Sigma}} \, - \, 3 \, V^{C\vert\underline{\Lambda\Sigma\Gamma}}_{
    \phantom{A\vert\underline{\Lambda\Sigma\Gamma}}
     \Lambda\Sigma\Gamma} \, \bar{V}_{C\vert\underline{\Lambda\Sigma\Gamma}}^{
    \phantom{C\vert\underline{\Lambda\Sigma\Gamma}}
     \Lambda\Sigma\Gamma} \,.
\label{betcurro}
\end{equation}
The two Betti currents of $Q^{1,1,1}$ are similarly
written down from the toric description. Since the Betti
currents are conserved, according to what shown in
\cite{superfieldsN2D3},  they satisfy
$E_0=|y_0|+1$. Since the hypercharge is zero, we have $E_0=1$ and the
Betti currents provide the gauge theory interpretation of the
massless Betti multiplets.
\subsection{Gauge theory interpretation of the short multiplets}
\label{protcorto}
Using the massless currents reviewed in the previous Section and the
chiral superfields, one has all the building blocks necessary to
construct the constrained superfields that correspond to all the
short multiplets found in the Kaluza Klein spectrum.
\par
As originally discussed in \cite{multanna} and applied to the
explicitly worked out spectra in
\cite{M111spectrum,superfieldsN2D3}, short $Osp(2\vert 4)$ multiplets
correspond to the saturation of the unitarity bound that relates the
energy (or conformal dimension) $E_0$ and hypercharge $y_0$ of the
Clifford vacuum to the highest spin $s_{max}$ contained in the multiplet.
Hence short multiplets occur when:
\begin{equation}
E_0=|y_0|+s_{max},~~~\cases{s_{max} = 2 \quad \mbox{short graviton}\cr
s_{max} = \ft{3}{2} \quad \mbox{short gravitino}\cr
s_{max} = 1 \quad \mbox{short vector}\cr}\,.
\label{shortening}
\end{equation}
In abstract representation theory condition (\ref{shortening}) implies that
a subset of states of the Hilbert space have zero norm and decouple from
the others. Hence the representation is shortened. In  superfield
language, the $\theta$--expansion of the superfield is shortened by
imposing a suitable differential constraint, invariant with respect
to Poincar\'e supersymmetry \cite{superfieldsN2D3}. Then
eq. (\ref{shortening}) is the necessary condition for
such a constraint to be invariant also under superconformal
transformations. Using chiral superfields and conserved currents as
building blocks, we can construct candidate short superfields that
satisfy the appropriate differential constraint and
eq. (\ref{shortening}). Then we can compare their flavor representations with
those of the short multiplets obtained in Kaluza Klein expansions.
In the case of
the $M^{1,1,1}$ theory, where the Kaluza Klein spectrum is known, we
find complete agreement and hence we explicitly verify the $AdS/CFT$ correspondence.
For the
$Q^{1,1,1}$ manifold we make instead a prediction in the reverse
direction: the gauge theory realization predicts the outcome of
harmonic analysis. While we wait for the construction of the complete
spectrum \cite{merlatti}, we can partially verify the correspondence
using the information available at the moment, namely the spectrum of
the scalar laplacian \cite{popelast}.
\subsubsection{Superfields corresponding to the short graviton multiplets}
The gauge theory interpretation of these multiplets is quite simple.
Consider the superfield
\begin{equation}
  \Phi_{\a\b}(x,\theta)=T_{\alpha\beta}(x,\theta) \, \Phi_{\rm chiral}(x,\theta),
\label{Phicorto}
\end{equation}
where $T_{\alpha\beta}$ is the stress energy tensor
(\ref{eqemtenssor}) and $\Phi_{\rm chiral}(x,\theta)$ is a chiral
superfield. By construction, the superfield (\ref{Phicorto}), at least in 
the abelian case, satisfies the equation
\begin{equation}
\label{eqshortgraviton}
{\cal D}^+_{\a}\Phi^{\a\b}=0
\end{equation}
and then, as shown in \cite{superfieldsN2D3}, it corresponds to a short graviton
multiplet of the bulk. It is natural to extend this identification to the
non-abelian case.
\par
Given the chiral multiplet spectrum (\ref{reprhyper})
and the dimension of the stress energy current (\ref{reprhyper}), we
immediately get the spectrum of superfields (\ref{Phicorto}) for the case
$M^{1,1,1}$:
\begin{equation}
\label{reprshortgraviton}
\cases{
M_1=3k\cr
M_2=0\cr
J=k\cr
E_0=2k+2,~~y_0=2k\cr}
~~~k>0\,.
\end{equation}
This exactly coincides with the spectrum of short graviton multiplets
found in Kaluza Klein theory through harmonic analysis
\cite{M111spectrum}.
\par
For the $Q^{1,1,1}$ case  the same analysis gives the
following prediction for the short graviton multiplets:
\begin{equation}
\label{reprshortgravq111}
\cases{
J_1=J_2=J_3=\ft{1}{2}k\cr
E_0=k+2,~~y_0=k\cr}
~~~k>0\,.
\end{equation}
We can make a consistency check on this prediction just relying on the
spectrum of the laplacian (\ref{H0q111}). Indeed, looking at table 4
of \cite{M111spectrum}, we see that in a short graviton multiplet the
mass of the spin two particle is
\begin{equation}
  m_h^2 = 16 y_0 (y_0+3).
\label{tab4predi}
\end{equation}
Looking instead at eq. (B.3) of the same paper, we see that such a mass
is equal to the eigenvalue of the scalar laplacian $m^2_h= H_0$.
Therefore, for consistency of the prediction (\ref{reprshortgravq111}),
we should have  $H_0=16k(k+3)$ for the representation $J_1=J_2=J_3=k/2;
Y=k$. This is indeed the value provided by eq. (\ref{H0q111}).
\par
It should be noted that when we write the operator (\ref{Phicorto}),
it is understood that {\it all color indices are symmetrized before taking
the contraction.}
\subsubsection{Superfields corresponding to the short vector multiplets}
\par
Consider next the superfields of the following type:
\begin{equation}
\label{shortvectora}
\Phi (x,\theta)=J(x,\theta) \, \Phi_{\rm chiral}(x,\theta),
\end{equation}
where $J$ is a conserved vector current of the type analyzed in
eq. (\ref{flaviocurro}) and $\Phi_{\rm chiral}$ is a chiral
superfield. By construction, the superfield (\ref{shortvectora}), at least in
the abelian case,
satisfies the constraint
\begin{equation}
\label{eqshortvector}
{\cal D}^{+\a}{\cal D}^+_{\a}\Phi =0
\end{equation}
and then, according to the analysis of \cite{superfieldsN2D3}, it can describe
a  short vector multiplet propagating into the  bulk.
\par
In principle, the flavor irreducible representations occurring in the superfield
(\ref{shortvectora}) are those originating from the tensor product decomposition
\begin{equation}
  ad  \otimes  {\cal R}_{\rho_k} = {\cal R}_{\chi_{max}} \oplus
  \sum_{\chi < \chi_{max}} {\cal R}_{\chi},
\label{chicchimax}
\end{equation}
where $ad$ is the adjoint representation, $\rho_k$ is the flavor weight of the
chiral field at level $k$, $\chi_{max}$ is the highest weight occurring in
the product $ad\otimes {\cal R}_{\rho_k}$ and $\chi < \chi_{max}$ are the lower
weights occurring in the same decomposition.
\par
Let us assume that the quantum mechanism that  suppresses
all the candidate chiral superfields of subleading weight does the
same suppression also on the short vector superfields
(\ref{shortvectora}). Then in the sum appearing on the l.h.s of eq. (\ref{chicchimax})
we keep only the first term and, as we show in a moment, we reproduce
the Kaluza Klein spectrum of short vector multiplets.
As we see, there is just a universal rule that presides at the
selection of the flavor representations in all sectors of the
spectrum. It is the restriction to the maximal weight. This is the
group theoretical implementation of the ideal that defines
the conifold as an algebraic locus in ${\mathbb C}^p$. We already
pointed out that, differently from the $D=4$ analogue of these conformal gauge
theories, the ideal cannot be implemented through a superpotential.
An equivalent way of imposing the result is to assume that the color indices have
to be completely symmetrized: such a symmetrization automatically
selects the highest weight flavor representations.
\par
Let us now explicitly verify the matching with Kaluza Klein spectra.
We begin with the $M^{1,1,1}$ case. Here the highest weight representations
occurring in the tensor product of the adjoint $(M_1=M_2=1,J=0)\oplus
(M_1=M_2=0,J=1)$ with the chiral spectrum (\ref{reprhyper}) are
$M_1=3k+1,M_2=1,J=k$ and $M_1=k,M_2=0,J=k+1$. Hence the spectrum of
vector fields (\ref{shortvectora}) limited to highest weights is
given by the following list of $Osp(2|4)\times SU(2)
\times SU(3)$ {\it irreps}:
\begin{equation}
\label{reprshortvectorsu3}
\cases{
M_1=3k+1\cr
M_2=1\cr
J=k\cr
E_0 =2k+1,~~y_0=2k \, \parallel\, \mbox{short vector multiplet}\cr}
~~~k>0
\end{equation}
and
\begin{equation}
\label{reprshortvectorsu2}
\cases{
M_1=3k\cr
M_2=0\cr
J=k+1\cr
E_0=2k+1,~~y_0=2k\, \parallel \,\mbox{short vector multiplet}\cr}
~~~k>0\,.
\end{equation}
This is precisely the result found in \cite{M111spectrum}.
\par
For the $Q^{1,1,1}$ case our gauge theory realization predicts the following short
vector multiplets:
\begin{equation}
\label{reprshortvecq111}
\cases{
J_1=\ft{1}{2}k+1\cr
J_2=\ft{1}{2}k\cr
J_3=\ft{1}{2}k\cr
E_0=k+1,~~y_0=k\, \cr}
~~~k>0
\end{equation}
and all the other are obtained from
(\ref{reprshortvecq111}) by permuting the role of the three $SU(2)$
groups. Looking at table 6 of \cite{M111spectrum}, we see that in
every ${\cal N}=2$ short multiplet emerging from M--theory compactification on
$AdS_4 \times M^7$ the lowest energy state is a scalar $S$ with
squared mass
\begin{equation}
  m_S^2=16 y_0 (y_0 -1).
\label{ceccovect}
\end{equation}
Hence, recalling eq. (\ref{massadiS}) and combining it with
(\ref{ceccovect}), we see that for consistency of our predictions we
must have
\begin{equation}
  H_0 +176 -24 \sqrt{H_0+36} = 16 k (k-1)
\label{consishort}
\end{equation}
for the representations (\ref{reprshortvecq111}). The quadratic
equation (\ref{consishort}) implies $H_0=16 k^2 + 80 k +64$ which is
precisely the result  obtained by inserting the values (\ref{reprshortgravq111})
into Pope's formula (\ref{H0q111}) for the laplacian eigenvalues.
Hence, also the short vector multiplets follow a general pattern
identical in all Sasakian compactifications.
\par
We can finally wonder why there are no short vector multiplets
obtained by multiplying the Betti currents with chiral superfields.
The answer might be the following. From the flavor view point these
would not be highest weight representations occurring in the tensor
product of  the  constituent supersingletons. Hence they are suppressed from
the spectrum.
\subsubsection{Superfields corresponding to the short gravitino multiplets}
\par
The spectrum of $M^{1,1,1}$ derived in \cite{M111spectrum} contains
various series of short gravitino multiplets. We can provide their
gauge theory interpretation through the following superfields.
Consider:
\begin{eqnarray}
\label{shortgravitino1}
&
{\Phi'}_{jB}^{\left(ii_1j_1\ell_1\dots i_kj_k\ell_k\right)
\left(AC_1D_1\dots C_kD_k\right)}=
&\nonumber\\
&=\left(U\bar{U}\left({\cal D}^+_{\a}V\bar{V}\right)+
V\bar{V}\left({\cal D}^+_{\a}U\bar{U}\right)\right)^{i~A}_{~j~~B}
\underbrace{U^{i_1}U^{j_1}U^{\ell_1}V^{C_1}V^{D_1}
\dots U^{i_k}U^{j_k}U^{\ell_k}V^{C_k}V^{D_k}}_{k}&\nonumber\\
\end{eqnarray}
and
\begin{eqnarray}
\label{shortgravitino2}
&{\Phi''}^{\left(ij\ell i_1j_1\ell_1\dots i_kj_k\ell_k\right)
\left(C_1D_1\dots C_kD_k\right)}=&\nonumber\\
&
=\left(U^iU^jU^\ell V^A{\cal D}^-_{\a}V^B\epsilon_{AB}\right)
\underbrace{U^{i_1}U^{j_1}U^{\ell_1}V^{C_1}V^{D_1}
\dots U^{i_k}U^{j_k}U^{\ell_k}V^{C_k}V^{D_k}}_{k}\,,&\nonumber\\
\end{eqnarray}
where all the color indices are symmetrized before being contracted.
By construction the superfields (\ref{shortgravitino1},\ref{shortgravitino2}),
at least in the abelian case,
satisfy  the equation
\begin{equation}
\label{eqshortgravitino}
{\cal D}^+_{\a}\Phi^{\a}=0
\end{equation}
and then, as explained in  \cite{superfieldsN2D3}, they correspond to short gravitino
multiplets propagating in the bulk.
We can immediately check that their highest weight flavor
representations yield the spectrum of $Osp(2|4)\times SU(2)
\times SU(3)$ short gravitino multiplets found by means of harmonic analysis
in \cite{M111spectrum}. Indeed for (\ref{shortgravitino1}),(\ref{shortgravitino2})
we respectively have:
\begin{equation}
\label{reprshortgravitino1}
\cases{
M_1=3k+1\cr
M_2=1\cr
J=k+1\cr
E_0=2k+{5\over 2},~~y_0=2k+1\cr}
~~~k\geq 0\,,
\end{equation}
and
\begin{equation}
\label{reprshortvectorgravitino2}
\cases{
M_1=3k+3\cr
M_2=0\cr
J=k\cr
E_0=2k+{5\over 2},~~y_0=2k+1\cr}
~~~k\geq 0\,.
\end{equation}
We postpone the analysis of short gravitino multiplets on $Q^{1,1,1}$
to \cite{merlatti} since this requires a more extended knowledge of
the spectrum.
\subsection{Long multiplets with rational protected dimensions}
\label{protlungo}
Let us now observe that, in complete analogy to what happens for the
$T^{1,1}$ conformal spectrum one dimension above \cite{gubser,sergiotorino},
also in the case of $M^{1,1,1}$ there is a large class of long
multiplets with rational conformal dimensions. Actually this seems to
be a general phenomenon in all Kaluza Klein compactifications on
homogeneous spaces $G/H$. Indeed, although the $Q^{1,1,1}$ spectrum is
not yet completed \cite{merlatti}, we can already see from its laplacian spectrum
(\ref{H0q111}) that a similar phenomenon occurs also there. More
precisely, while the short multiplets saturate the unitarity bound
and have a conformal weight related to the hypercharge and maximal
spin by eq. (\ref{shortening}), the {\it rational long multiplets}
satisfy a quantization condition of the conformal dimension of the
following form
\begin{equation}
E_0=|y_0|+s_{max}+\lambda,~~~\lambda \, \in \, {\mathbb N}.
\label{quantcondo}
\end{equation}
\par
Inspecting the  $M^{1,1,1}$ spectrum determined in
\cite{M111spectrum}, we find the following long rational multiplets:
\begin{itemize}
\item {\it Long rational graviton multiplets}
\par
In the series
\begin{equation}
\cases{
M_1=0,~M_2=3k,~J=k+1\cr
M_1=1,~M_2=3k+1,~J=k\cr}
\end{equation}
and conjugate ones we have
\begin{equation}
y_0=2k,~~E_0=2k+3=|y_0|+3
\end{equation}
corresponding to
\begin{equation}
\lambda=1.
\end{equation}
\item {\it Long rational gravitino multiplets}
\par
In the series of representations
\begin{equation}
M_1=1,~M_2=3k+1,~J=k+1
\end{equation}
(and conjugate ones) for the gravitino  multiplets of type $\chi^-$
 we have
\begin{equation}
y_0=2k+1,~~E_0=2k+{9\over 2}=|y_0|+{7\over 2},
\end{equation}
while in the series
\begin{equation}
M_1=0,~M_2=3k,~J=k-1
\end{equation}
(and conjugate ones)
for the same type of gravitinos we get
\begin{equation}
y_0=2k-1,~~E_0=2k+{5\over 2}=|y_0|+{7\over 2}.
\end{equation}
Both series fit into the quantization rule (\ref{quantcondo}) with:
\begin{equation}
\lambda=2.
\end{equation}
\item {\it Long rational vector multiplets}
\par
In the series
\begin{equation}
M_1=0,~M_2=3k,~J=k
\end{equation}
(and conjugate ones) for the  vector multiplets of type $W$ we have
\begin{equation}
y_0=2k,~~E_0=2k+4=|y_0|+4,
\end{equation}
that fulfills the quantization condition (\ref{quantcondo}) with
\begin{equation}
\lambda=3.
\end{equation}
For the same vector multiplets of type $W$, in the series
\begin{equation}
\cases{
M_1=0,~M_2=3k,~J=k+1\cr
M_1=1,~M_2=3k+1,~J=k\cr}
\end{equation}
(and conjugate ones) we have
\begin{equation}
y_0=2k,~~E_0=2k+10=|y_0|+10,
\end{equation}
that satisfies the quantization condition (\ref{quantcondo}) with
\begin{equation}
\lambda=9.
\end{equation}
\end{itemize}
The generalized presence of these rational long multiplets hints at
various still unexplored quantum mechanisms that, in the conformal
field theory, protect certain operators from acquiring anomalous
dimensions. At least for the long graviton multiplets, characterized by
$\lambda=1$, the corresponding protected superfields can be
guessed, in analogy with 
\cite{sergiotorino}. If we take the superfield of a short vector multiplet
$J(x,\theta) \, \Phi_{\rm chiral}(x,\theta)$ and we multiply it by a
stress--energy superfield $T_{\alpha\beta}(x,\theta)$, namely if we consider a
superfield of the form
\begin{equation}
  \Phi \sim \mbox{conserved vector current} \, \times \, \mbox{stress energy
  tensor} \, \times \, \mbox{chiral operator},
\label{guessa}
\end{equation}
we reproduce the right $Osp(2\vert 4)\times SU(3) \times SU(2)$
representations of the long rational graviton multiplets of
$M^{1,1,1}$. The soundness of such an interpretation can be checked
by looking at the graviton multiplet spectrum on $Q^{1,1,1}$. This is
already available since it is once again determined by the laplacian
spectrum. Applying formula eq. (\ref{guessa}) to the $Q^{1,1,1}$ gauge theory
leads to predict the following spectrum of long rational multiplets:
\begin{equation}
\label{ratlongq111}
\cases{
J_1=\ft{1}{2}k+1\cr
J_2=\ft{1}{2}k\cr
J_3=\ft{1}{2}k\cr
E_0=k+1,~~y_0=k\, \cr}
~~~k>0
\end{equation}
and all the other are obtained from
(\ref{ratlongq111}) by permuting the role of the three $SU(2)$
groups. Looking at table 1 of \cite{M111spectrum}, we see that in a
graviton multiplet the spin two particle has mass
\begin{equation}
  m_h^2=16(E_0+1)(E_0-2),
\label{emmemgrave}
\end{equation}
which for the candidate multiplets (\ref{emmemgrave}) yields
\begin{equation}
  m_h^2=16(k+4)(k+1).
\label{emmecandid}
\end{equation}
On the other hand, looking at eq. (B.3) of \cite{M111spectrum} we see
that the squared mass of the graviton is just the eigenvalue of the scalar
laplacian $m_h^2=H_0$. Applying Pope's formula (\ref{H0q111}) to the
representations of (\ref{ratlongq111}) we indeed find
\begin{equation}
  H_0 = 16 k^2 +80 k +64 = 16(k+4)(k+1).
\label{bingo!}
\end{equation}
It appears, therefore, that the generation of rational long multiplets
is based on the universal mechanism codified by the ansatz
(\ref{guessa}), proposed in \cite{sergiotorino} and
 applicable to all compactifications. Why these
superfields have protected conformal dimensions is still to be
clarified within the framework of the superconformal gauge theory.
The superfields  leading to rational long multiplets with much
higher values of $\lambda$, like the cases $\lambda=3$ and
$\lambda=9$ that we have found, are more difficult to guess. Yet
their appearance seems to be a general phenomenon and this, as we
have already stressed, hints at general protection mechanisms that
have still to be investigated.
\part{Detailed Geometrical Analysis}
\label{geometry}
In the third Part of this paper we give a more careful
discussion of the geometry of the homogeneous Sasakian manifolds on
which we compactify M--theory in order to obtain the conformal gauge
theories we have been discussing. In the general classification
\cite{castromwar} of seven dimensional coset manifolds that can be used as internal
manifolds for Freund Rubin solutions $AdS_4 \times M^7$, all the
supersymmetric cases have been determined and found to be in finite
number. There is one ${\cal N}=8$ case corresponding to the seven sphere
$S^7$, three ${\cal N}=1$ cases and three ${\cal N}=2$ cases. The
reason why these $G/H$ manifolds admit ${\cal N}=2$ is that they are
Sasakian, namely the metric cone constructed over them is a
Calabi--Yau conifold.  We give a unified geometric description of
these manifolds emphasizing all the features of their algebraic, topological and
differential structure which are relevant in deriving the properties
of the associated superconformal field theories.
The cases, less relevant for the present paper,
of $N^{0,1,0}$, $V_{5,2}$ and $S^7$
are discussed in Appendix \ref{AppB}.
\section{Algebraic geometry, topology and metric structure
 of the homogeneous Sasakian $7$-manifolds}
We want to describe all the Sasakian $7$-manifolds entering the game as
fibrations $ \pi :  M^7 = G/H \to G/\tilde H = M_a$ with fibre $\tilde H /H
= U(1)$, where $G$ is a
semisimple compact Lie group and $\tilde H \subset G$ is a compact
subgroup containing a maximal torus $T$ of $G$.
As such, the base $M_a$ is a compact real
$6$-dimensional manifold.
\par
Since  $\tilde H/H\simeq U(1)$, $H$ is the kernel of the non-trivial
homomorphism $\chi :\tilde H \rightarrow U(1)$ given by the natural
projection. The bundle $G \times_\chi
U(1) \to G/\tilde H$  associated to this character is the space of
orbits of $\tilde H$ acting
on $G\times U(1)$ as $(g,u)\tilde h=(g\tilde h, \chi (\tilde h)^{-1}u)$.
Since the character is non-trivial, the total space of this bundle
is homogeneous for $G$ and the stabilizer of the base point is precisely
the kernel of $\chi$. Accordingly, we have an isomorphism
\begin{equation}
  G/H\simeq G \times_\chi U(1) \to G/\tilde H.
\label{isomchi}
\end{equation}
\par
The rationale for this description is that
there is a holomorphic version; one
first complexifies $G$ to $G_{\mathbb C}$ in the standard way, next one
chooses an orientation of the roots of $Lie G_{\mathbb C}$ in such a way that
the character $\chi$ is the exponential of an antidominant weight and
finally one completes the complexification $\tilde H_{\mathbb C}$ by
exponentiating
the missing positive roots. This gives a parabolic subgroup $P\subset
G_{\mathbb C}$ and $G_{\mathbb C}/P\simeq G/\tilde H$. Giving to  $M_a$ the complex
structure of $G_{\mathbb C}/P$ we get a compact complex $3$-fold.
\par
The character of $\tilde H$  determined above extends to a
(holomorphic) character of the parabolic subgroup $P$
and this induces a holomorphic line bundle $L$ over $M_a$,
which is homogeneous for $G_{\mathbb C}$ and has plenty holomorphic sections
spanning the {\it irrep} with highest weight $-\log(\chi )$.
Restricting to the compact
form $G$, $L$ acquires a fibre metric and $M^7$ is simply the unit circle
bundle inside $L$.
It turns out that $L$ produces a
Kodaira embedding of $M_a$ in ${\mathbb P}(V^*)$, the linear space $V=H^0(M_a,L)$
being precisely the space of holomorphic sections of $L$.
\vskip 0.3cm
\leftline{{\underline {\it Embedding Quadrics and
Representation Theory}}}
\vskip 0.3cm
One can also write down the equations for the image of $M_a$ in
${\mathbb P}(V^*)$ by means of representation theory. Being
$M_a$ a homogeneous variety, it is cut out by homogeneous equations
of degree at most two. To find them one proceeds as follows.
The space of quadrics in ${\mathbb P}(V^*)$ is the symmetric tensor
product $Sym^2(V)$. As a representation of
$G_{\mathbb C}$ this is actually reducible (for a generic dominant
character $\chi^{-1}$) and decomposes as
\begin{equation}
Sym^2(V)=W_{\chi^{-2}}\oplus _\rho W_\rho,
\label{sym2V}
\end{equation}
$W_\rho$ being the {\it irrep} induced
by the character $\rho$ of $P$. It turns out that the weight
vectors spanning the  addenda $W_\rho,\; \rho \ne \chi^{-2}$, considered as
quadratic relations among the homogeneous coordinates of
${\mathbb P}(V^*)$, generate the ideal $I$ of $M_a$.
Generically the image of
the embedding is not a complete intersection.
\vskip 0.3cm
\leftline{ {\underline {\it Coordinate Ring versus Chiral Ring}}}
\vskip 0.3cm
Finally, the homogeneous
coordinate ring of $M_a\subset {\mathbb P}(V^*)$ is
\begin{equation}
  {\mathbb C}[W_{\chi^{-1}}]/I\simeq \oplus_{k\geq 0} W_{\chi^{-k}}.
\label{coordring}
\end{equation}
The physical interpretation of this coordinate ring in the context of
$3D$ conformal field theories emerging from an M2 brane compactification on
a Sasakian $M^7_{S}$ is completely analogous to the
interpretation of the coordinate ring in the context of $2D$
conformal field theories emerging from string compactification on an
algebraic Calabi Yau threefold $M^6_{CY}$. In the second case let $X$ be
the projective coordinates of ambient ${\mathbb P}^4$ space and $W(X)=0$ the algebraic equation
cutting out the Calabi--Yau locus. Then the ring
\begin{equation}
  \frac{{\mathbb C}[X]}{\partial W_{CY}(X)}
\label{cyringo}
\end{equation}
is isomorphic to the  ring of  {\it primary conformal chiral
operators} of the $(2,2)$ CFT with $c=9$ realized on the world sheet.
These latter are characterized by being invariant under one of the
two world--sheet supercurrents (say $G^-(z)$) and by having their
conformal weight $h=|y|/2$ fixed in terms of their $U(1)$ charge.
Geometrically, this is also the ring of Hodge structure deformations.
In a completely analogous way the coordinate ring (\ref{coordring})
is isomorphic to the ring of conformal hypermultiplets of the ${\cal N}=2$,
$D=3$ superconformal theory. The hypermultiplets are short
representations \cite{M111spectrum,superfieldsN2D3}
of the conformal group $Osp(2\vert 4)$ and are
characterized by $E_0=|y_0|$, where $E_0$ is the conformal weight
while $y_0$ is the R--symmetry charge.
\vskip 0.3cm
\leftline{{\underline {\it Homology}}}
\vskip 0.3cm
For applications to brane geometry, it is also important to know the
homology (equivalently, cohomology) of $M^7$. Being $M^7$ a circle bundle, we
can use the Gysin sequence in cohomology \cite{BT}. Since the base $M_a$ is
acyclic in odd dimensions, the Gysin sequence splits into subsequences of
the form
\begin{equation}
0 \to H^{2k-1}(M^7, \zz) \to
H^{2k-2}(M_a, \zz) \buildrel c_1 \over \longrightarrow
H^{2k}(M_a, \zz) \to
H^{2k}(M^7, \zz) \to 0,
\label{gysingen}
\end{equation}
where the map $c_1$ is the product by the Euler class of the fibration $M^7$,
which equals the first Chern class of $L$.
\par
We now apply these standard results \cite{FH,BoHi,LT12,BT} to the various
manifolds which enter our physical problem.
\subsection{The manifold $M^{1,1,1}$}
The first case
is given by
\begin{equation}
M^{1,1,1} = SU(3) \times SU(2)\times U(1) / SU(2) \times U(1) \times
U(1) =  G'/ H'.
\label{m111cesar}
\end{equation}
\subsubsection{Generalities}
Let us call
$h_i,\,i=1,2$, $h$ and $Y$ the generators of the Lie algebras of the
standard maximal tori of $SU(3)$, $SU(2)$ and $U(1)$ respectively,
all normalized with periods $2\pi$. Then $SU(2)$ is embedded in
$SU(3)$ as the stabilizer of the last basis vector of
${\mathbb C}^3$, and the $U(1)$'s are generated by
$Z'=(h_1 + 2 h_2) - h - 4 Y $, and $Z'' = (h_1 +2 h_2) + 3 h $.
\par
To reconstruct the general structure described at the beginning of this
Section, notice that the image of $H'$ under the projection of $G'$ onto
$SU(3)$ is isomorphic to $S(U(2) \times U(1))$.
This projection gives an exact sequence $0 \to K \to H' \to H \to 0$
and, since $K$ is normal in $H'$, we have an isomorphism $G/H = (G'/K)/
(H'/K) \simeq G' / H'$. The elements of $H'$ have the form
\begin{equation}
\left(
\left( \begin{array}{cc}
g & 0 \\
0 & 1
\end{array} \right)
\exp(i (\theta + \phi ) (h_1 +2 h_2) ) ,
\exp(i (3 \phi - \theta ) h) ,
\exp(- i 4 \theta Y)
\right),
\label{hprimele}
\end{equation}
where $g\in SU(2)$ and $\theta,\;\phi\in [0,2\pi]$.
So $K$ is given by $g= (-1)^l$, $ \theta + \phi =  \pi l $ and its generic
element is $( 1 , \pm \exp (-4 i \theta h) , \exp (-4 i \theta Y))$.
Taking the quotient by K, the third factor of
$G'$ is factored out and there is an extra
$ {\mathbb Z}_2$ acting on the maximal torus of $SU(2)$. Consequently,
we find that $G/K = SU(3)\times SO(3)$ and the image of $H$ in $G/K$
has a component on the maximal torus of $SO(3)$ (generated by
$\exp {2\pi i t}$) corresponding
to the infinitesimal character $\lambda(h_1)=0$, $\lambda(h_2)=-3t$.
\par
Summing up, we have $G= SU(3) \times SO(3)$, $H= S(U(2)\times U(1))$ and
$\tilde H = S(U(2)\times U(1))\times U(1)$. Accordingly
\begin{equation}
M_a=Gr(2,3)\times {\mathbb P}^1\simeq {\mathbb P}^{2*}\times {\mathbb
P}^1,
\end{equation}
${\mathbb P}^n$ being the complex $n-$dimensional space and $Gr(2,3)$
being the Grassmannian of $2-$planes in ${\mathbb C}^3$.
\footnote{The isomorphism
between $Gr(2,3)$ and the dual projective space ${\mathbb P}^{2*}$ comes
because giving a $2-$plane in ${\mathbb C}^3$ is the same as giving the
homothety class of linear functionals vanishing on it, i.e. a line in
the dual space ${\mathbb C}^{3*}$.}

Now we have to recognize the character. This comes by projecting
$\tilde H$ onto $\tilde H/H$. We get
\begin{eqnarray}
&&\nonumber \exp(i\theta h_1)\cdot eH = eH,\\
&&\nonumber\exp(i\theta h_2)\cdot eH = \exp(-3i\theta t)H,\\
&&\nonumber \exp(i\theta t)\cdot eH = \exp(i\theta t)H,
\end{eqnarray}
where $t$
has been identified with the generator of $\tilde H/H$. The
character restricted to $SO(3)$ is the fundamental one, which
corresponds
to the adjoint representation of $SU(2)$ and therefore is the square
of the fundamental character of $SU(2)$.
We see then that
$M^{1,1,1}$ is the circle bundle
\footnote{If $L_i\rightarrow X_i,\; (i=1,2)$ are two vector bundles,
one denotes for short by $L_1\boxtimes L_2$ the vector bundle on the
product $X_1\times X_2$ given by $p_1^*L_1\otimes p_2^*L_2$, where
$p_i:X_1\times X_2\rightarrow X_i$ is the projection on the $i$-th
factor.}
inside
$ L={\cal O}(3)\boxtimes{\cal O}(2)$ over ${\mathbb P}^{2*}\times{\mathbb P}^1$.
\par
The fundamental group of the circle bundle associated to the infinitesimal
character $\chi_*(h_1)=0$, $\chi_*(h_2)=m$, $\chi_*(h)=n$ is ${\mathbb
Z}_{\gcd(m,n)}$
Applying the same analysis to $M^{p,q,r}$, the character corresponds to
$$m=-{{3p}\over{2r}}\lcm\left({{r}\over{\gcd(r,q)}}
{{2r}\over{\gcd(2r,3p)}}\right)$$
$$n=-{{q}\over{r}}\lcm\left({{r}\over{\gcd(r,q)}},
{{2r}\over{\gcd(2r,3p)}}\right)$$
for $r\ne 0$ and
$$m=-{{3p}\over{\gcd(3p,2q)}}$$
$$n=-{{2q}\over{\gcd(3p,2q)}}$$
for $r=0$.
Notice that $M^{1,1,1}=M^{1,1,r}$ is simply connected.
Although in \cite{castromwar} it is stated that $M^{1,1,1}=M^{1,1,0}/{\mathbb Z}_4$,
a closer analysis shows that the ${\mathbb Z}_4$ action is trivial.
In particular $H_1(M^{1,1,1},{\mathbb Z})$ is torsionless.
\subsubsection{The algebraic embedding equations and the chiral ring of $M^{1,1,1}$}
\label{MP29}
As for the algebraic embedding of $M^{1,1,1}$, since
$\dim W_{\chi{-1}}= 30$, $L$ embeds
\begin{equation}
  M_a \simeq {\mathbb P}^{2*} \times {\mathbb P}^{1} \, \hookrightarrow \, {\mathbb P}^{29}
\label{embeddi}
\end{equation}
by
\begin{equation}
  X^{ijk\vert AB} = U^i \, U^j \, U^k \, V^A \, V^B \qquad \left( \quad  i,j,k=0,2,3
  \quad ; \quad A,B =1,2 \quad \right)\,,
\label{coorde}
\end{equation}
namely by writing the $30$ homogeneous coordinates $ X^{ijk\vert AB}$
of ${\mathbb P}^{29}$ as polynomials
in the homogeneous coordinates $U^i$, $i=0,1,2$ of ${\mathbb P}^{2*}$
and $V^A$ ($A=0,1$) of  ${\mathbb P}^{1}$.
 The image of $M_a$ is cut out by $\dim
Sym^2(W_{\chi{-1}})- \dim W_{\chi{-2}}= 465-140=325$ equations.
  Alternatively the same $325$ equations can be seen as the embedding
of the cone ${\cal C}(M^{1,1,1})$ over the Sasakian $U(1)$ bundle  into ${\mathbb C}^{30}$.
\par
For further clarification we describe the explicit form of these
embedding equations in the language of Young tableaux.
From eq. (\ref{embeddi}) it follows that the $30$ homogeneous coordinates of
${\mathbb P}^{29}$ are assigned to the representation $({\bf 10},{\bf 3})$ of
$SU(3) \times SU(2)$:
\begin{equation}
  X^{ijk\vert AB} \,  \mapsto \, ({\bf 10},{\bf 3}) \, \equiv \,
  \begin{array}{|c|c|c|}
\hline
             \hskip .3 cm & \hskip .3 cm & \hskip .3 cm  \\
\hline
\end{array} \, \otimes \, \begin{array}{|c|c|}
\hline
             \times & \times \\
\hline
\end{array}\,.
\label{assegno}
\end{equation}
This means that the quadric monomials $X^2$
 span the following symmetric tensor product:
\begin{eqnarray}
 X^2 &=&  \Bigl (\quad \left[\quad \begin{array}{|c|c|c|}
\hline
             \hskip .3 cm & \hskip .3 cm & \hskip .3 cm  \\
\hline
\end{array} \, \otimes \, \begin{array}{|c|c|}
\hline
             \times & \times \\
\hline
\end{array}\quad\right] \,
\otimes \,\left[\quad\begin{array}{|c|c|c|}
\hline
             \hskip .3 cm & \hskip .3 cm & \hskip .3 cm  \\
\hline
\end{array} \, \otimes \, \begin{array}{|c|c|}
\hline
             \times & \times \\
\hline
\end{array} \quad\right] \quad \Bigl) _{sym}
\label{xsquare}
\end{eqnarray}
In general the number of independent components of $X^2$ is just
\begin{equation}
  \mbox{dim} \, X^2 = \frac{30 \times 31}{2} = 465,
\label{465}
\end{equation}
which corresponds to the sum of dimensions of all the irreducible
representations of $SU(3) \times SU(2)$ contained in the symmetric
product (\ref{xsquare}), but on the locus defined by the explicit
embedding (\ref{embeddi}) only $28 \times 5 = 140$ of these
components are independent. These components fill the representation
of highest weight
\begin{equation}
  ({\bf 28},{\bf 5}) \,\equiv \, \begin{array}{|c|c|c|c|c|c|}
\hline
             \hskip .3 cm & \hskip .3 cm & \hskip .3 cm &
              \hskip .3 cm & \hskip .3 cm & \hskip .3 cm \\
\hline
\end{array} \, \times \,\begin{array}{|c|c|c|c|}
\hline
             \times & \times & \times &
             \times \\
\hline
\end{array}\,.
\label{higwei}
\end{equation}
The remaining $325$ components are the quadric equations of the locus.
They are nothing else but the statement that
all the representations of $SU(3) \times SU(2)$ contained in the
symmetric product (\ref{xsquare}) should vanish with the exception of
the representation (\ref{higwei}).
\par
Let us work out the representations that should vanish. To this effect we begin
by writing the complete decomposition into irreducible
representations of $SU(3)$ of the tensor product ${\bf 10} \times
{\bf 10}$:
\begin{eqnarray}
\underbrace{\begin{array}{|c|c|c|}
\hline
             \hskip .3 cm & \hskip .3 cm & \hskip .3 cm  \\
\hline
\end{array} }_{10} \, \otimes \, \underbrace{\begin{array}{|c|c|c|}
\hline
             \hskip .3 cm & \hskip .3 cm & \hskip .3 cm  \\
\hline
\end{array} }_{10} \,  & = & \underbrace{\begin{array}{|c|c|c|c|c|c|}
\hline
             \hskip .3 cm & \hskip .3 cm & \hskip .3 cm &
              \hskip .3 cm & \hskip .3 cm & \hskip .3 cm \\
\hline
\end{array} }_{28} \, \oplus \, \underbrace{\begin{array}{l}
\begin{array}{|c|c|c|c|c|}
\hline
             \hskip .3 cm & \hskip .3 cm & \hskip .3 cm &
             \hskip .3 cm & \hskip .3 cm \\
\hline
\end{array}\,\\
\begin{array}{|c|}
             \hskip .3 cm  \\
\hline
\end{array}
\end{array}}_{35} \nonumber\\
\null & \null & \oplus \underbrace{\begin{array}{l}
\begin{array}{|c|c|c|c|}
\hline
             \hskip .3 cm & \hskip .3 cm & \hskip .3 cm &
             \hskip .3 cm   \\
\hline
\end{array}\,\\
\begin{array}{|c|c|}
             \hskip .3 cm &  \hskip .3 cm \\
\hline
\end{array}
\end{array}}_{27} \,\oplus \, \underbrace{\begin{array}{l}
\begin{array}{|c|c|c|}
\hline
             \hskip .3 cm & \hskip .3 cm & \hskip .3 cm    \\
\hline
\end{array}\,\\
\begin{array}{|c|c|c|}
             \hskip .3 cm & \hskip .3 cm & \hskip .3 cm     \\
\hline
\end{array}
\end{array}}_{\overline {10}}
\label{10x10}
\end{eqnarray}
Next in eq. (\ref{10x10}) we separate the symmetric and antisymmetric
parts of the decomposition, obtaining
\begin{eqnarray}
\left( \underbrace{\begin{array}{|c|c|c|}
\hline
             \hskip .3 cm & \hskip .3 cm & \hskip .3 cm  \\
\hline
\end{array} }_{10} \, \otimes \, \underbrace{\begin{array}{|c|c|c|}
\hline
             \hskip .3 cm & \hskip .3 cm & \hskip .3 cm  \\
\hline
\end{array} }_{10} \,  \right) _{sym}& = & \underbrace{\begin{array}{|c|c|c|c|c|c|}
\hline
             \hskip .3 cm & \hskip .3 cm & \hskip .3 cm &
              \hskip .3 cm & \hskip .3 cm & \hskip .3 cm \\
\hline
\end{array} }_{28} \,  \oplus \underbrace{\begin{array}{l}
\begin{array}{|c|c|c|c|}
\hline
             \hskip .3 cm & \hskip .3 cm & \hskip .3 cm &
             \hskip .3 cm   \\
\hline
\end{array}\,\\
\begin{array}{|c|c|}
             \hskip .3 cm &  \hskip .3 cm \\
\hline
\end{array}
\end{array}}_{27} 
\label{10x10sym}
\end{eqnarray}
and
\begin{eqnarray}
\left( \underbrace{\begin{array}{|c|c|c|}
\hline
             \hskip .3 cm & \hskip .3 cm & \hskip .3 cm  \\
\hline
\end{array} }_{10} \, \otimes \, \underbrace{\begin{array}{|c|c|c|}
\hline
             \hskip .3 cm & \hskip .3 cm & \hskip .3 cm  \\
\hline
\end{array} }_{10} \, \right) _{antisym} & = &  \underbrace{\begin{array}{l}
\begin{array}{|c|c|c|c|c|}
\hline
             \hskip .3 cm & \hskip .3 cm & \hskip .3 cm &
             \hskip .3 cm & \hskip .3 cm \\
\hline
\end{array}\,\\
\begin{array}{|c|}
             \hskip .3 cm  \\
\hline
\end{array}
\end{array}}_{35}  \oplus \underbrace{\begin{array}{l}
\begin{array}{|c|c|c|}
\hline
             \hskip .3 cm & \hskip .3 cm & \hskip .3 cm    \\
\hline
\end{array}\,\\
\begin{array}{|c|c|c|}
             \hskip .3 cm & \hskip .3 cm & \hskip .3 cm     \\
\hline
\end{array}
\end{array}}_{\overline {10}} \,
\label{10x10asym}
\end{eqnarray}
As a next step we do the same decomposition for the tensor product
${\bf 3} \times {\bf 3}$ of $SU(2)$ representations. We have
\begin{eqnarray}
 \underbrace{\begin{array}{|c|c|}
\hline
             \times & \times \\
\hline
\end{array} }_{3} \, \otimes \,  \underbrace{\begin{array}{|c|c|}
\hline
             \times & \times \\
\hline
\end{array} }_{3}  & = &  \underbrace{
\begin{array}{|c|c|c|c|}
\hline
             \times & \times & \times &
             \times \\
\hline
\end{array}}_{5} \, \oplus \, \underbrace{\begin{array}{l}
\begin{array}{|c|c|c|}
\hline
            \times & \times & \times   \\
\hline
\end{array}\,\\
\begin{array}{|c|}
             \times     \\
\hline
\end{array}
\end{array}}_{3} \, \oplus \, \underbrace{\begin{array}{l}
\begin{array}{|c|c|}
\hline
            \times & \times    \\
\hline
\end{array}\,\\
\begin{array}{|c|c|}
\hline
            \times & \times    \\
\hline
\end{array}\,\\
\end{array}}_{1}
\label{3x3}
\end{eqnarray}
and separating the symmetric and antisymmetric parts we respectively
obtain
\begin{eqnarray}
 \left( \underbrace{\begin{array}{|c|c|}
\hline
             \times & \times \\
\hline
\end{array} }_{3} \, \otimes \,  \underbrace{\begin{array}{|c|c|}
\hline
             \times & \times \\
\hline
\end{array} }_{3} \right) _{sym} & = &  \underbrace{
\begin{array}{|c|c|c|c|}
\hline
             \times & \times & \times &
             \times \\
\hline
\end{array}}_{5} \,  \oplus \, \underbrace{\begin{array}{l}
\begin{array}{|c|c|}
\hline
            \times & \times    \\
\hline
\end{array}\,\\
\begin{array}{|c|c|}
\hline
            \times & \times    \\
\hline
\end{array}\,\\
\end{array}}_{1}
\label{3x3symm}
\end{eqnarray}
and
\begin{eqnarray}
\left(  \underbrace{\begin{array}{|c|c|}
\hline
             \times & \times \\
\hline
\end{array} }_{3} \, \otimes \,  \underbrace{\begin{array}{|c|c|}
\hline
             \times & \times \\
\hline
\end{array} }_{3} \right) _{antisym} & = & \underbrace{\begin{array}{l}
\begin{array}{|c|c|c|}
\hline
            \times & \times & \times   \\
\hline
\end{array}\,\\
\begin{array}{|c|}
             \times     \\
\hline
\end{array}
\end{array}}_{3}
\label{3x3asym}
\end{eqnarray}
The symmetric product we are interested in is  given by the sum
\begin{equation}
  \left( sym_{SU(3)} \, \times \, sym_{SU(2)} \right) \oplus
  \left(antisym_{SU(3)}\, \times \, antisym_{SU(2)} \right)
\label{symsym}
\end{equation}
so that we can write
\begin{eqnarray}
   465 & = & \underbrace{ ({\bf 28},{\bf 5})}_{140} \oplus\nonumber\\
       &   & \underbrace{({\bf 28},{\bf 1}) \oplus ({\bf 27},{\bf 5})
             \oplus ({\bf 27},{\bf 1})
             \oplus({\bf 35},{\bf 3})  \oplus({\overline {\bf 10}},{\bf3})}_{325}.\nonumber\\
\label{somma}
\end{eqnarray}
Hence the equations can be arranged into $5$ representations
corresponding to the list appearing in the second row of
eq. (\ref{somma}). Indeed, eq. (\ref{somma}) is the explicit form, in
the case $M^{1,1,1}$, of the general equation (\ref{sym2V}) and the
addenda in its second line are what we named $W_\rho,\; \rho \ne \chi^{-2}$,
in the general discussion.
\par
Coming now to the coordinate ring (\ref{coordring}) it is obvious
from the present discussion that, in the $M^{1,1,1}$ case, it takes the
following form:
\begin{equation}
 {\mathbb C}[W_{\chi^{-1}}]/I\simeq \oplus_{k\geq 0} W_{\chi^{-k}} =
 \sum_{k\ge 0} \,\left( \underbrace{
\begin{array}{|c|c|c|c|c|}
\hline
\phantom{\times} & \phantom{\times} &\phantom{\times} & \dots &\phantom{\times}   \\
\hline
\end{array}}_{3k}\, \otimes \,
\underbrace{\begin{array}{|c|c|c|c|}
\hline
            \times & \times& \dots &\times    \\
\hline
\end{array}}_{2k} \right)\,.
\label{3k2k}
\end{equation}
In eq. (\ref{3k2k}) we recognize the spectrum of $SU(3) \times SU(2)$
representations of the $Osp(2\vert 4)$ hypermultiplets as determined
by harmonic analysis on $M^{1,1,1}$. Indeed, recalling the results of
\cite{M111spectrum,superfieldsN2D3}, the hypermultiplet of conformal
weight (energy label) $E_0 = 2k$ and hypercharge $y_0= 2k$ is in the
representation
\begin{equation}
  M_1 = 3k \quad ; \quad M_2=0 \quad ; \quad J=k.
\label{figaretto}
\end{equation}

\subsubsection{Cohomology of $M^{1,1,1}$}
Let us now compute the cohomology
of $M^{1,1,1}$. The
first Chern class of $L$ is $c_1 = 2 \omega_1 + 3 \omega_2$, where
$\omega_1$ (resp. $\omega_2$) is the generator of the second cohomology
group of ${\mathbb P}^1$
(resp. ${\mathbb P}^{2*}$).
 In this case the Gysin sequence gives:
\begin{eqnarray}
&H^0(\m1) = H^7(\m1)= \zz,&\nonumber\\
& 0 \to H^1 (\m1) \to \zz \ra{c_1} \zz \oplus \zz
\to H^2(\m1) \to 0,
&\nonumber\\
& 0 \to H^3 (\m1) \to \zz \oplus \zz \ra{ c_1} \zz
\oplus \zz \to H^4(\m1) \to 0,
&\nonumber\\
& 0 \to H^5 (\m1) \to \zz\oplus \zz \ra{c_1} \zz
 \to H^6(\m1) \to 0.
&
\label{m111gys}
\end{eqnarray}
The first $c_1$ sends $1 \in H^0(M_a)$ to $c_1 \in H^2 (M_a)$. Its
kernel is zero, and its image is $\zz$. Accordingly, $H^2 (\m1) =
\zz \cdot \pi^*(\omega_1 + \omega_2) $.
The second $c_1$ sends $(\omega_1 ,\omega_2) \in \zz \oplus \zz =
H^2(M_a)$ to $(3 \omega_1 \omega_2, 2\omega_1 \omega_2 +3 \omega_2^2) \in
\zz \oplus \zz = H^4 (M_a)$. Its kernel vanishes and therefore $H^3(\m1)= 0$.
Its cokernel is $\zz_9 = H^4(\m1)$ generated by $\pi^*(\omega_1 \omega_2
+ \omega_2^2)$.
Finally, the last $c_1$ sends $\omega_1 \omega_2$ and $ \omega_2^2
\in H^4 (M_a)= \zz \oplus \zz $ respectively to $ 3 \omega_1 \omega_2^2 $
and $ 2 \omega_1 \omega_2^2 \in H^6(M_a)$. This map is surjective,
so $H^6(\m1)=0$ and its kernel is generated by $ \beta = -2 \omega_1
\omega_2 + 3\omega_2^2$. Hence $H^5(\m1) = \zz \cdot \alpha$, with $\pi_*
\alpha = \beta$.
\subsubsection{Explicit description of the Sasakian fibration for $M^{1,1,1}$}
We proceed next to an explicit description
of the fibration structure of  $M^{1,1,1}$  as
 a $U(1)$-bundle over ${\mathbb P}^{2*}\times{\mathbb P}^1$.
 We construct an atlas of local trivializations and we give the
 appropriate transition functions. This is important for our
  discussion of the supersymmetric cycles leading to the
 baryon states.
 \par
We take $\tau\in[0,4\pi)$ as a local coordinate on the fibre and
$(\tilde\theta,\tilde\phi)$ as local coordinates on ${\mathbb P}^1\simeq S^2$.
To describe ${\mathbb P}^{2*}$ we have to be a little bit careful.
${\mathbb P}^{2*}$ can be covered by the three patches $W_\alpha\simeq{\mathbb C}^2$ in
which one of the three homogeneous coordinates, $U_\alpha$, does not vanish.
The set not covered by one of these $W_\alpha$ is homeomorphic to $S^2$.
We choose to parametrize $W_3$ as in \cite{gibbonspope}:
\begin{equation}
\label{C2coord}
\left\{\begin{array}{c}
\zeta^1=U_1/U_3=\tan\mu\,\cos(\theta/2)\,e^{i(\psi+\phi)/2}\\
\zeta^2=U_2/U_3=\tan\mu\,\sin(\theta/2)\,e^{i(\psi-\phi)/2}
\end{array}\right.,
\end{equation}
where
\begin{equation}
\left\{\begin{array}{l}
\mu\in(0,\pi/2)\\
\theta\in(0,\pi)\\
0\leq(\psi+\phi)\leq 4\pi\\
0\leq(\psi-\phi)\leq 4\pi
\end{array}\right.\,.
\label{psiphi}
\end{equation}
These coordinates cover the whole $W_3\simeq{\mathbb C}^2$ except for the
trivial coordinate singularities $\mu=0$ and $\theta=0,\pi$.
Furthermore $\theta$ and $\phi$ can be extended to the complement of $W_3$.
Indeed, the ratio
\begin{equation}
z=\zeta^1/\zeta^2=\tan^{-1}(\theta/2)\,e^{i\phi}
\label{fubstud}
\end{equation}
is well defined in the limit $\mu\to\pi/2$ and it constitutes
the usual stereographic map of $S^2$ onto the complex plane
(see the next discussion of $Q^{1,1,1}$ and in particular figure
\ref{S2patches}).
\par
\begin{figure}[ht]
\begin{center}
\epsfxsize = 8cm
\epsffile{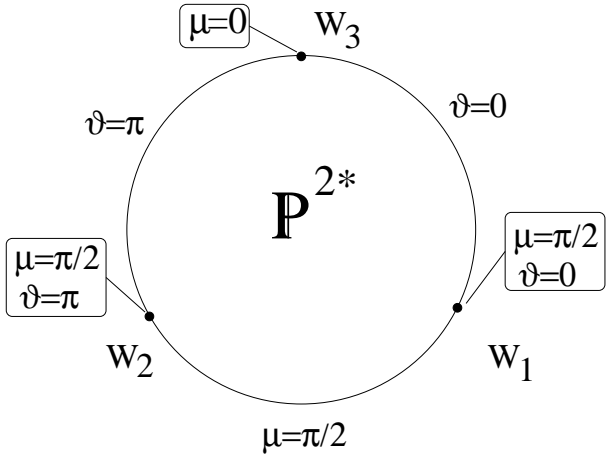}
\vskip  0.2cm
\hskip 2cm
\unitlength=1.1mm
\end{center}
\caption{Schematic representation of the atlas on ${\mathbb P}^{2*}$.
The three patches $W_\alpha$ cover the open ball and part of the
boundary circle, which constitutes the set of coordinate singularities.
This latter is made of three $S^2$'s: $\{\theta=0\}$,
$\{\theta=\pi\}$ and $\{\mu=\pi/2\}$, which touch each other
at the three points marked with a dot.
Each $W_\alpha$ covers the whole ${\mathbb P}^{2*}$ except for one of the spheres
(for example, $W_3$ does not cover $\{\mu=\pi/2\}$).
The three \emph{most singular} points are covered by only one
patch (for example, $\{\mu=0\}$ is covered by the only $W_3$).
}
\label{CP2patches}
\end{figure}

Just as  for the sphere, we must be careful in treating
some one-forms near the coordinate singularities.
In particular, $d\psi$ and $d\phi$ are not well defined on the
three $S^2$ which are not covered by one of the patches $W_\alpha$:
$\{\mu=\pi/2\}$, $\{\theta=0\}$ and $\{\theta=\pi/2\}$
(see figure  \ref{CP2patches}.)
Actually, except for the three points of these spheres that
are covered by only one patch ($\{\mu=0\}\in W_3$, $\{\mu=\pi/2,
\theta=0\}\in W_1$, $\{\mu=\pi/2,\theta=\pi\}\in W_2$), one particular
combination of $d\psi$ and $d\phi$ survives, as it is illustrated in
table (\ref{1forms}).
\begin{equation}
\label{1forms}
\begin{array}{|c|c|c|}
\hline
{\rm coordinate}&{\rm regular}&{\rm singular}\\
{\rm singularity}&{\rm one-form}&{\rm one-forms}\\
\hline\hline
\theta=0&d\psi+d\phi&\alpha d\psi+\beta d\phi~(\alpha\neq\beta)\\
\theta=\pi&d\psi-d\phi&\alpha d\psi-\beta d\phi~(\alpha\neq\beta)\\
\mu=\pi/2&d\phi&\alpha d\psi\\
\hline
\end{array}
\end{equation}
The singular one-forms become well defined if we multiply them
by a function having a double zero at the coordinate
singularities.
\par
We come now to the description of the fibre bundle $M^{1,1,1}$.
We cover the base ${\mathbb P}^{2*}\times{\mathbb P}^1$ with six open
charts ${\cal U}_{\alpha\pm}=W_\alpha\times H_\pm$ ($\alpha=1,2,3$) on which we
can define a local fibre coordinate $\tau_{\alpha\pm}\in[0,4\pi)$.
The transition functions are given by:
\begin{equation}\label{Mtrans}
\left\{\begin{array}{l}
\tau_{1\beta}=\tau_{3\gamma}-3(\psi+\phi)+2(\beta-\gamma)\tilde\phi\,,\qquad
(\beta,\gamma=\pm1)\\
\tau_{1\beta}=\tau_{2\gamma}-6\phi+2(\beta-\gamma)\tilde\phi\,.
\end{array}\right.
\end{equation}
On this principal fibre bundle we can easily introduce a $U(1)$ Lie
algebra valued connection which, on the various patches
of the base space, is described by the following one--forms:
\begin{eqnarray}
\left\{\begin{array}{l}
{\cal A}_{1\pm}=-\frac{3}{2}(\cos2\mu+1)(d\psi+d\phi)
-\frac{3}{2}(\cos2\mu-1)(\cos\theta-1)d\phi
+2(\pm 1-\cos\tilde\theta)d\tilde\phi\,,\\
{\cal A}_{2\pm}=-\frac{3}{2}(\cos2\mu+1)(d\psi-d\phi)
-\frac{3}{2}(\cos2\mu-1)(\cos\theta+1)d\phi
+2(\pm 1-\cos\tilde\theta)d\tilde\phi\,,\\
{\cal A}_{3\pm}=-\frac{3}{2}(\cos2\mu-1)(d\psi+\cos\theta d\phi)
+2(\pm 1-\cos\tilde\theta)d\tilde\phi\,.\\
\end{array}\right.
\label{m111connec}
\end{eqnarray}
Due to (\ref{Mtrans}), the one-form $(d\tau\!-\!{\cal A})$ is a global angular form
\cite{BT}.
It can then be taken as the $7$-th vielbein of the following
$SU(3)\times SU(2)\times U(1)$
invariant metric on $M^{1,1,1}$:
\begin{equation}
ds^2_{M^{1,1,1}}=c^2(d\tau\!-\!{\cal A})^2+ds^2_{\,{\mathbb P}^{2*}}+
ds^2_{\,{\mathbb P}^1}.
\label{fibrametra}
\end{equation}
The one-form ${\cal A}$ is the connection of the Hodge-K\"ahler bundle on 
${\mathbb P}^{2*}\times{\mathbb P}^1$.
\vskip 0.3cm
\leftline{ {\underline {\it Einstein Metric}}}
\vskip 0.3cm
The Einstein metric on the homogeneous space $M^{1,1,1}$ was originally
constructed by Castellani et al in \cite{noi321} using the intrinsic geometry of coset
manifolds. In such a language, which is that employed in \cite{M111spectrum}
to develop harmonic analysis and construct the Kaluza Klein spectrum we
have
\begin{eqnarray}
  ds^2_{M^{1,1,1}} &=&\frac{3}{8 \Lambda} \, (\sqrt{3} \Omega^8 + \Omega^3 +
  \Omega^\bullet) \otimes (\sqrt{3} \Omega^8 + \Omega^3 +
  \Omega^\bullet)\nonumber\\
  &&+ \frac{1}{8 \Lambda} \,  \sum_{A=4}^{7}\, \Omega^A \otimes \Omega^A
  \frac{3}{4 \Lambda} + \sum_{m=1}^{2} \, \Omega^{m} \otimes
  \Omega^{m},
\label{olda}
\end{eqnarray}
where $\Omega^A$ ($A=1,\dots,8$) are the left--invariant 1--forms in
the adjoint of $SU(3)$, $\Omega^m$ ($m=1,\dots,3$) are the
left--invariant 1--forms in the adjoint of $SU(2)$ and
$\Omega^\bullet$ is the left-invariant 1--form on $U(1)$. The
$\Lambda$ dependent rescalings appearing in (\ref{olda}) were obtained
by imposing that the Ricci tensor is proportional to the metric which
yields a cubic equation with just one real root \cite{noi321}. Such
a cubic equation was retrieved a little later also by Page and Pope
\cite{pagepopeM}. These authors wrote the Einstein metric in the
coordinate frame we have just utilized to describe the fibration
structure and which is convenient for our discussion of the
supersymmetric $5$--cycles. In this  frame eq. (\ref{olda}) becomes
\begin{eqnarray}
\label{metrica}
ds^2_{M^{1,1,1}}&=&\frac{3}{32 \Lambda}\Bigl [
d\tau-3 \sin^2 \mu \,\left(d\psi
+\cos{\theta}d\phi\right)+2\cos{\tilde{\theta}}
d\tilde{\phi} \Bigr ]^2 \nonumber\\
&+& \frac{9}{2 \, \Lambda}\left[d\mu^2+{1\over 4}\sin^2{\mu}\cos^2{\mu}^2\left(d\psi
+\cos{\theta}d\phi\right)^2\right.\nonumber\\
&+&\left.{1\over 4}\sin^2{\mu}\left(d\theta^2
+\sin^2{\theta}d\phi^2\right)\right]
+\frac{3}{4 \Lambda
}\left( d\tilde{\theta}^2+\sin^2{\tilde{\theta}}d\tilde{\phi}^2\right),
\end{eqnarray}
where the three addenda of (\ref{metrica}) are one by one identified
with the three addenda of (\ref{olda}). The second and the third
addenda are the ${\mathbb P}^{2*}$ and $S^2$ metric on the base manifold of the
$U(1)$ fibration, while the first term is the fibre metric. In other
words, one recognizes the structure of the metric anticipated in
(\ref{fibrametra}). The parameter $\Lambda$ appearing in the metric
(\ref{metrica}) is the internal cosmological constant defined by
eq. (\ref{ricciin}).
\subsubsection{The baryonic $5$--cycles of $M^{1,1,1}$ and their volume}
\label{bryn}
As we saw above, the relevant homology group of $M^{1,1,1}$ for the
calculation of the baryonic masses is
\begin{equation}
H_5(M^{1,1,1},{\mathbb R})={\mathbb R}\,.
\end{equation}
Let us consider the following two five-cycles, belonging to the same homology class:
\begin{eqnarray}
{\cal C}^1:\left\{\begin{array}{c}
\tilde\theta={\tilde \theta}_0 =const\\
\tilde\phi={\tilde \phi}_0=const
\end{array}\right.\,, \label{cycca1}\\
{\cal C}^2:\left\{\begin{array}{c}
\theta=\theta_0 =const\\
\phi=\phi_0 = const
\end{array}\right.\,.
\label{cycca2}
\end{eqnarray}
The two representatives
(\ref{cycca1}, \ref{cycca2}) are distinguished by their
different stability subgroups which we calculate in the next subsection.
\vskip 0.3cm
\leftline{ {\underline {\it Volume of the $5$--cycles}}}
\vskip 0.3cm
The volume of the cycles (\ref{cycca1}, \ref{cycca2}) is easily computed by pulling back
the metric (\ref{metrica}) on ${\cal C}^1$ and ${\cal C}^2$, that
have the topology of a $U(1)$-bundle over ${\mathbb P}^{2*}$ and
${\mathbb P}^1\times{\mathbb P}^1$ respectively:
\begin{eqnarray}
{\rm Vol}({\cal C}^1)=\oint_{{\cal C}^1}\sqrt{g_1}=
9\left(8\Lambda/3\right)^{-5/2}\int\sin^3\mu\cos\mu\sin\theta\,
d\tau d\mu d\psi d\theta d\phi
=\frac{9\pi^3}{2}\left(\frac{3}{2\Lambda}\right)^{5/2}
\label{volcyc1}
\end{eqnarray}
\begin{eqnarray}
{\rm Vol}({\cal C}^2)=\oint_{{\cal C}^2}\sqrt{g_2}=
6\left(8\Lambda/3\right)^{-5/2}\int\sin\mu\cos\mu\sin\tilde\theta\,
d\tau d\mu d\psi d\tilde\theta d\tilde\phi
=6\pi^3\left(\frac{3}{2\Lambda}\right)^{5/2}.
\label{volcyc2}
\end{eqnarray}
The volume of $M^{1,1,1}$ is instead given by
\begin{eqnarray}
{\rm Vol}(M^{1,1,1})=\oint_{M^{1,1,1}}\sqrt{g}=
18\left(8\Lambda/3\right)^{-7/2}\int\sin^3\mu\cos\mu\sin\theta
\sin\tilde\theta\, d\tau d\mu d\psi d\theta d\phi d\tilde\theta d\tilde\phi
\nonumber\\
=\frac{27\pi^4}{2\Lambda}\left(\frac{3}{2\Lambda}\right)^{5/2}.
\label{volm111}
\end{eqnarray}
The results (\ref{volcyc1}, \ref{volcyc2}, \ref{volm111}) can be
inserted into the general formula (\ref{baryondim}) to calculate the
conformal weights (or energy labels) of five-branes wrapped on the
cycles ${\cal C}^1$ and ${\cal C}^2$. We obtain:
\begin{equation}
\left\{\begin{array}{c}
E_0({\cal C}^1)=N/3\\
\\
E_0({\cal C}^2)=4N/9
\end{array}\right.\,.
\label{risultone}
\end{equation}
As stated above, the result (\ref{risultone}) is essential in proving
that the conformal weight of the elementary world--volume fields
$V^A$, $U^i$ are
\begin{equation}
  h\left[ V^4 \right]=1/3 \quad , \quad h\left[ U^i \right]=4/9
\label{hUV}
\end{equation}
respectively. To reach such a conclusion we need to
identify the states obtained by wrapping the five--brane on ${\cal
C}^1, {\cal C}^2$ with operators in the flavor representations
$M_1=0,M_2=0,J=N/2$ and $M_1=N,M_2,J=0$, respectively. This
conclusion, as anticipated in the introductory Sections is reached by
studying the stability subgroups of the supersymmetric $5$--cycles.
\subsubsection{Stability subgroups of the baryonic $5$-cycles of $M^{1,1,1}$
\label{stab5M}
and the flavor representations of the baryons}
Let us now consider the stability subgroups
\begin{equation}
  H({\cal C}_i) \subset G = SU(3) \times SU(2) \times U(1)
\label{stabalg}
\end{equation}
of the two cycles (\ref{cycca1}, \ref{cycca2}). Let us begin with the
first cycle defined by (\ref{cycca1}). As we have previously said, this is the
restriction of the $U(1)$-fibration to ${\mathbb P}^{2*}\times\{p\}$, $p$ being a
point of ${\mathbb P}^1$.
Hence, the stability subgroup of the cycle ${\cal C}^1$ is:
\begin{equation}
  H\left({\cal C}^1\right) = SU(3) \times U(1)_R \times U(1)_{B,1}
\label{hcyc1}
\end{equation}
where $U(1)_R$ is the R--symmetry $U(1)$ appearing as a factor in
$SU(3) \times SU(2) \times U(1)_R$ while $U(1)_{B,1} \subset SU(2)$  is
a maximal torus.

Turning to the case of the second cycle (\ref{cycca2}), which is the restriction of the
$U(1)$-bundle to the product of a hyperplane of ${\mathbb P}^{2*}$ and ${\mathbb P}^1$,
its stabilizer is
\begin{equation}
  H\left({\cal C}^2\right) = S(U(1)_{B,2}\times U(2)) \times SU(2)\times U(1)_R,
\label{hcyc2}
\end{equation}
where $SU(2) \times U(1)_R$ is the group appearing as a factor in
$SU(3) \times SU(2) \times U(1)_R$, $U(1)_{B,2} \subset SU(3)$ is the
subgroup generated by $h_1=\diag(1,-1,0)$ and $S(U(1)_{B,2}\times U(2))\subset SU(3)$ is the
stabilizer of the first basis vector of ${\mathbb C}^3$.

Following the procedure introduced by Witten in \cite{bariowit} we
should now quantize the {\it collective coordinates} of the
non--perturbative baryon state obtained by {\it wrapping} the
five--brane on the $5$--cycles we have been discussing. As explained
in Witten's paper this leads to  quantum mechanics on the
homogeneous manifold $G/H(\cal C)$. In our case the collective
coordinates of the baryon live on the following spaces:
\begin{equation}
  \mbox{space of collective coordinates} \,\quad \rightarrow \quad \frac{G}{H(\cal C)}=
   \cases{\begin{array}{cc}
    \frac{SU(2)}{U(1)_{B,1}} \simeq {\mathbb P}^1 & \mbox{for ${\cal C}^1$} \\
    \null & \null \\
    \frac{SU(3)}{S(U(1)_{B,2}\times U(2))} \simeq {\mathbb P}^2 & \mbox{for ${\cal C}^2$} \\
  \end{array}\cr}.
\label{collecti}
\end{equation}
The wave function $\Psi\left( \mbox{collec. coord.}\right) $ is in
Witten's phrasing a section of a line bundle of degree $N$. This
happens because the baryon has {\it baryon number} $N$, namely it has
charge $N$ under the additional massless vector multiplet that is associated with
a harmonic $2$--form and appears
in the Kaluza Klein spectrum since $\mbox{dim} H_2(M^{1,1,1}) = 1 \ne
0$. These are the Betti multiplets mentioned in Section \ref{bariobetti}.
Following Witten's reasoning there is a morphism
\begin{equation}
 \mu^i: \quad  U(1)_{Baryon} \hookrightarrow H({\cal C}^i) \quad
 i=1,2
\label{morfismo}
\end{equation}
of the non perturbative baryon number group  into the stability subgroup
of the $5$--cycle. Clearly the image of such a morphism must be a
$U(1)$--factor in $H({\cal C})$ that has a non trivial action on the
collective coordinates of the baryons. Clearly in the case of our two
baryons we have:
\begin{equation}
  \mbox{Im} \, \mu^i =\, U(1)_{B,i} \quad  i=1,2\,.
\label{immorf}
\end{equation}
The name given to these groups anticipated the conclusions of such an
argument.
\par
Translated into the language of harmonic analysis, Witten's statement
that the baryon wave function should be a section of a line bundle
with degree $N$ means that we are supposed to consider harmonics on
$G/H({\cal C})$ which, rather than being scalars of $H({\cal C})$, are
in the $1$--dimensional representation of $U(1)_B$ with charge $N$.
According to the general rules of harmonic analysis
(see \cite{spectfer,univer,M111spectrum}) we are supposed to collect
all the representations of $G$ whose reduction with respect to $H({\cal
C})$ contains the prescribed representation of $H({\cal C})$. In the
case of the first cycle, in view of eq. (\ref{hcyc1}) we want all
representations of $SU(2)$ that contain the state $2 J_3 =N$. Indeed
the generator of $U(1)_{B,1}$ can always be regarded as
the third component of angular momentum by means of a change of
basis. The representations with this property are those characterized
by:
\begin{equation}
  2 J = N+2k,\qquad k \ge 0.
\label{boundo}
\end{equation}
Since the laplacian on $G/H({\cal C})$ has eigenvalues proportional to
the Casimir
\begin{equation}
  \Box_{SU(2)/U(1)} = \mbox{const} \, \times \, J(J+1),
\label{laplacio}
\end{equation}
the harmonic satisfying the constraint (\ref{boundo}) and with
minimal energy is just that with
\begin{equation}
  2J=N.
\label{JisN}
\end{equation}
This shows that under the flavor group the baryon associated with the
first cycle is neutral with respect to $SU(3)$ and transforms in the
$N$--times symmetric representation of $SU(2)$. This perfectly
matches, on the superconformal field theory side, with our candidate
operator (\ref{operVV}).
\par
Equivalently the choice of the representation $2 J= N$ corresponds
with the identification of the baryon wave--function with a {\it
holomorphic section (=zero mode)} of the $U(1)$--bundle under
consideration, i.e. with a section of the corresponding line bundle.
Indeed such a line bundle is, by definition, constructed over
${\mathbb P}^1$ and declared to be of degree $N$, hence it is ${\cal
O}_{{\mathbb P}^1}(N)$. Representation-wise a section of ${\cal
O}_{{\mathbb P}^1}(N)$ is just an element of the $J=N/2$
representation, namely it is the $N$ times symmetric of $SU(2)$.
\par
Let us now consider the case of the second cycle. Here the same
reasoning instructs us to consider all representations of $SU(3)$
which, reduced with respect to $U(1)_{B,2}$, contain a state of charge
$N$. Moreover, directly aiming at zero mode,
we can assign the baryon wave--function
to a holomorphic sections of a line bundle on ${\mathbb P}^2$, which
must correspond to characters of the parabolic subgroup $S(U(1)_{B,2}\times U(2))$.
As before the degree $N$ of this line bundle uniquely characterizes it as
${\cal O}(N)$.
In the language of Young tableaux, the corresponding $SU(3)$ representation is
\begin{equation}
  M_1 = 0 \, ; \, M_2 =N,\label{m1m2N}
\end{equation}
i.e. the representation of this baryon state is the $N$--time
symmetric of the dual of $SU(3)$ and this perfectly matches with the
complex conjugate of the candidate
conformal operator \ref{operUUU}. In other words we have constructed
the antichiral baryon state. The chiral one obviously has the same
conformal dimension.
\subsubsection{These $5$--cycles are supersymmetric}
\label{supsym5M}
The $5$--cycles we have been considering in the above subsections
have to be supersymmetric in order for the conclusions we have been
drawing to be correct. Indeed all our arguments have been based on
the assumption that the $5$--brane wrapped on such cycles is a
$BPS$--state. This is true if the $5$--brane action localized on the
cycle is $\kappa$--supersymmetric.
\par
The $\kappa$-symmetry projection operator for a five-brane is
\begin{equation}\label{kprojector}
P_\pm=\frac{1}{2}\left(\unity\pm {\rm i}\,\frac{\rm 1}{5!\,\sqrt{g}}\epsilon^
{\alpha\beta\gamma\delta\varepsilon}
\partial_\alpha X^M\partial_\beta X^N\partial_\gamma X^P
\partial_\delta X^Q\partial_\varepsilon X^R\,\Gamma_{MNPQR}\right)\,,
\end{equation}
where the functions $X^M(\sigma^\alpha)$ define the embedding of
the five-brane into the eleven dimensional spacetime, and $\sqrt{g}$
is the square root of the determinant of the induced metric on the
brane.
The gamma matrices $\Gamma_{MNPQR}$, defining the spacetime
spinorial structure, are the pullback through the vielbeins
of the constant gamma matrices $\Gamma_{ABCDE}$ satisfying the
standard Clifford algebra:
\begin{equation}
\Gamma_{MNPQR}=e^A_{\,M}e^B_{\,N}e^C_{\,P}e^D_{\,Q}e^E_{\,R}
\Gamma_{ABCDE}\,.
\end{equation}
A possible choice of vielbeins for ${\cal C}(M^{1,1,1})\times{\msbm
M}^3$, namely
the product of  the
metric cone over $M^{1,1,1}$  times three dimensional Minkowski space is the following
one:
\begin{equation}\label{Mvielbein}
\left\{\begin{array}{ccl}
e^1&=&\frac{1}{2\sqrt2}\,r\,d\tilde\theta\\
e^2&=&\frac{1}{2\sqrt2}\,r\sin\tilde\theta d\tilde\phi\\
e^3&=&\frac{1}{8}\,r\left(d\tau+3\sin^2\mu(d\psi+\cos\theta d\phi)
+2\cos\tilde\theta d\tilde\phi\right)\\
e^4&=&\frac{\sqrt3}{2}\,r\,d\mu\\
e^5&=&\frac{\sqrt3}{4}\,r\sin\mu\cos\mu
\left(d\psi+\cos\theta d\phi\right)\\
e^6&=&\frac{\sqrt3}{4}\,r\sin\mu
\left(\sin\psi d\theta-\cos\psi\sin\theta d\phi\right)\\
e^7&=&\frac{\sqrt3}{4}\,r\sin\mu
\left(\cos\psi d\theta+\sin\psi\sin\theta d\phi\right)\\
e^{8}&=&dr\\
e^9&=&dx^1\\
e^{10}&=&dx^2\\
e^0&=&dt\\
\end{array}\right.\,.
\end{equation}
In these coordinates the embedding equations of the two cycles
(\ref{cycca1}), (\ref{cycca2}) are very simple, so we have
\begin{equation}
\frac{1}{5!}\epsilon^{\alpha\beta\gamma\delta\varepsilon}
\partial_\alpha X^M\partial_\beta X^N\partial_\gamma X^P
\partial_\delta X^Q\partial_\varepsilon X^R\,\Gamma_{MNPQR}=
\left\{\begin{array}{c}
\Gamma_{\tau\mu\theta\psi\phi}\\
\Gamma_{\tau\mu\tilde\theta\psi\tilde\phi}
\end{array}\right.\,,
\end{equation}
for ${\cal C}^1$ and ${\cal C}^2$ respectively.
By means of the vielbeins (\ref{Mvielbein}) these gamma matrices
are immediately computed:
\begin{equation}
\left\{\begin{array}{c}
\Gamma_{\tau\mu\theta\psi\phi}=\left(\frac{3}{32}\right)^2
r^5\sin^3\mu\cos\mu\sin\theta\,
\Gamma_{34567}\\
\Gamma_{\tau\mu\tilde\theta\psi\tilde\phi}=\frac{3}{512}r^5
\sin\mu\cos\mu\sin\tilde\theta\,\Gamma_{31245}
\end{array}\right.\,,
\end{equation}
while the square root of the determinant of the metric on the
two cycles is easily seen to be
\begin{equation}
\left\{\begin{array}{c}
\sqrt{g_1}=\left(\frac{3}{32}\right)^2r^5\sin^3\mu\cos\mu\sin\theta\\
\sqrt{g_1}=\frac{3}{512}r^5\sin\mu\cos\mu\sin\tilde\theta
\end{array}\right.\,.
\end{equation}
So, for both cycles, the $\kappa$-symmetry projector
(\ref{kprojector})
reduces to the projector of a five dimensional hyperplane embedded
in flat spacetime:
\begin{equation}
P_\pm=\left\{\begin{array}{c}
\frac{1}{2}\left(\unity\pm{\rm i}\,\Gamma_{34567}\right)\\
\frac{1}{2}\left(\unity\pm{\rm i}\,\Gamma_{31245}\right)
\end{array}\right.\,.
\label{project}
\end{equation}
The important thing to check is that the projectors (\ref{project})
are non--zero on the two Killing spinors of the space ${\cal C}(M^{1,1,1})\times{\msbm
M}^3$. Indeed, this latter has not $32$ preserved supersymmetries, rather it
has only $8$ of them. In order to avoid long and useless calculations
we just argue as follows. Using the gamma--matrix basis of \cite{noi321}, the
Killing spinors are already known. We have:
\begin{equation}
\begin{array}{rclcrcl}
  \Gamma_0&=&\gamma_0 \,\otimes \,{\bf 1}_{8 \times 8} & ; &
  \Gamma_8&=&\gamma_1 \,\otimes \,{\bf 1}_{8 \times 8} \\
  \Gamma_9&=&\gamma_2 \,\otimes \,{\bf 1}_{8 \times 8} & ; &
  \Gamma_{10}&=&\gamma_3 \,\otimes \,{\bf 1}_{8 \times 8}\\
  \Gamma_i&=&\gamma_5 \,\otimes \,\tau_i & (i=1,\dots,7) &\null&\null&\null  \\
\end{array}
\label{gammole}
\end{equation}
where $\gamma_{0,1,2,3}$ are the usual $4 \times 4$ gamma matrices in
four--dimensional space--time, while $\tau_i$ are the $8\times 8$
gamma--matrices satisfying the $SO(7)$ Clifford algebra in the form:
$\{ \tau_i \, , \, \tau_j \}= -\delta_{ij}$. For these matrices we
take the representation given in the Appendix of \cite{noi321}, which
is well adapted to the intrinsic description of the $M^{1,1,1}$
metric through Maurer--Cartan forms as in eq. (\ref{olda}). In this
basis the Killing spinors were calculated in \cite{noi321} and have
the following form:
\begin{eqnarray}
  \mbox{Killing spinors} &=& \epsilon (x)\,\otimes \,\eta \quad ;
  \quad \eta = \left(\begin{array}{c}
   {\bf 0} \\
\hline
    {\bf u} \\
    \hline
   {\bf 0} \\
    \hline
    \epsilon {\bf u}^\star \\
\end{array}\right)\,,
\label{kilspino}
\end{eqnarray}
where
\begin{eqnarray}
 {\bf u} &=& \left( \begin{array}{c}
   a+{\rm i} b \\
   0 \
 \end{array}\right) \, \quad ; \quad \,\epsilon {\bf u}^\star= \left( \begin{array}{cc}
   0 & 1 \\
   -1 & 0 \
 \end{array}\right) \,{\bf u}^\star=\left( \begin{array}{c}
   0 \\
   -a+{\rm i} b \
 \end{array}\right)
\label{uspino}
\end{eqnarray}
and where the $8$--component spinor was written in $2$--component blocks.
\par
In the same basis, using notations of \cite{noi321}, we have:
\begin{equation}
\begin{array}{rcccl}
\Gamma_{34567} & = & \gamma_5 \, \otimes \, U_8 \, U_4 \, U_5 \, U_6 \, U_7
\otimes \sigma_3 &=&
{\rm i} \, \gamma_5 \, \otimes \, \left( \begin{array}{c|c|c|c}
 - {\bf 1}_{2\times 2} & 0 & 0 & 0 \\
  \hline
  0 &  {\bf 1}_{2\times 2} & 0 & 0 \\
  \hline
  0 & 0 &  {\bf 1}_{2\times 2} & 0 \\
  \hline
  0 & 0 & 0 &  - {\bf 1}_{2\times 2}
\end{array}\right)\,,\\
\null & \null & \null & \null & \null \\
\Gamma_{31245} & = & \gamma_5 \, \otimes \,{\rm i} U_8 \, U_4 \, U_5 \,
\otimes \, {\bf 1} &=&{\rm i} \, \gamma_5 \, \otimes \, \left( \begin{array}{c|c|c|c}
 \sigma_3 & 0 & 0 & 0 \\
  \hline
  0 &  \sigma_3 & 0 & 0 \\
  \hline
  0 & 0 &  \sigma_3 & 0 \\
  \hline
  0 & 0 & 0 &  \sigma_3
\end{array}\right)\,.\\
\end{array}
\label{expligamma}
\end{equation}
As we see, by comparing eq. (\ref{project}) with eq. (\ref{kilspino})
and (\ref{expligamma}), the $\kappa$--supersymmetry projector reduces
for both cycles to a chirality projector on the $4$--component
space--time part $\epsilon(x)$. As such, the $\kappa$--supersymmetry
projector always admits non vanishing eigenstates implying that the
cycle is supersymmetric. The only flaw in the above argument is
that the Killing spinor (\ref{kilspino}) was determined in
\cite{noi321} using as vielbein basis the suitably rescaled Maurer--Cartan forms
$\Omega^3$, $\Omega^{m}$,  $(m=1,2)$ and $\Omega^A$, $(A=4,5,6,7)$.
Our choice (\ref{Mvielbein}) does not correspond to the same vielbein basis.
However, a little inspection shows that it differs only by some
$SO(4)$ rotation in the space of ${\mathbb P}^{2*}$ vielbein $4,5,6,7$.
Hence we can turn matters around and ask what happens to the Killing spinor
(\ref{kilspino}) if we apply an $SO(4)$ rotation in the directions
$4,5,6,7$. It suffices to check the form of the gamma--matrices
$[\tau_A \, , \, \tau_B]$ which are the generators of such rotations.
Using again the Appendix of \cite{noi321} we see that such $SO(4)$
generators are of the form
\begin{equation}
 {\rm i} \,  \left( \begin{array}{c|c|c|c}
 \sigma_i & 0 & 0 & 0 \\
  \hline
  0 &  \sigma_i & 0 & 0 \\
  \hline
  0 & 0 &  \sigma_i & 0 \\
  \hline
  0 & 0 & 0 &  \sigma_i
\end{array}\right) \quad  \mbox{or} \quad  {\rm i} \,  \left( \begin{array}{c|c|c|c}
 \sigma_i & 0 & 0 & 0 \\
  \hline
  0 &  -\sigma_i & 0 & 0 \\
  \hline
  0 & 0 &  \sigma_i & 0 \\
  \hline
  0 & 0 & 0 &  -\sigma_i
\end{array}\right),
\end{equation}
so that  the $SO(4)$ rotated Killing spinor is of the same form as in
eq.(\ref{kilspino}) with, however, ${\bf u}$ replaced by ${\bf u}^\prime = A {\bf u}$
where $A\in SU(2)$. It is obvious that such an $SU(2)$ transformation does
not alter our conclusions. We can always decompose ${\bf u}^\prime$
into $\sigma_3$ eigenstates and associate the $\sigma_3$--eigenvalue
with the chirality eigenvalue, so as to satisfy the
$\kappa$--supersymmetry projection. Hence, our $5$--cycles are indeed
supersymmetric.
\par
\subsection{The manifold $Q^{1,1,1}$}
This is defined by
$G= SU(2)\times SU(2)\times SU(2), H = U(1)\times U(1)$.
If we call $h_i$ the generators of the maximal tori of the $SU(2)$'s,
normalized with periods $2 \pi$, $H$ is generated by $h_1 - h_2$ and
$h_1 - h_3$, i.e. the complement of  $Z = h_1 +h_2 +h_3$, while
$\tilde H= U(1)\times U(1)\times U(1)$ is the product of the
three maximal tori. So the base is
\begin{equation}
M_a = {\mathbb P}^1\times {\mathbb P}^1\times {\mathbb P}^1.
\label{q111base}
\end{equation}
The generator of $\tilde H/H$ can be taken to be $h_1$.
Now
\begin{equation}
\exp(i\theta h_k) \cdot e H = \exp(i\theta h_1) H,
\label{q111char}
\end{equation}
showing that $Q^{1,1,1}$ is the circle bundle inside
${\cal O} (1) \boxtimes{\cal O} (1) \boxtimes {\cal O} (1)$ over
${\mathbb P}^1\times {\mathbb P}^1\times {\mathbb P}^1$.
\subsubsection{The algebraic embedding equations and the chiral ring}
\label{q111chirri}
Since $\dim H^0(M_a, L)=8$, $L$ embeds
\begin{equation}
  M_a \simeq {\mathbb P}^1 \times {\mathbb P}^1 \times {\mathbb P}^1
  \hookrightarrow {\mathbb P}^{7}
\label{q111embed}
\end{equation}
by setting the $8$ homogeneous coordinates  of ${\mathbb P}^{7}$ equal to
three--linear expressions in the homogeneous coordinates of the
three ${\mathbb P}^1$, namely $A_i$, $B_j$, $C_k$:
\begin{eqnarray}
 X^{ijk} &=& A^i \, B^j \, C^k
 \quad \left(  i,j,k=1,2 \right)\,.
 \label{XABC}
\end{eqnarray}
By the same argument as in the $M^{1,1,1}$ case, we find that
the image is cut out by $36-27=9$ equations.
Indeed, eq. (\ref{XABC}) states that the  ${\mathbb P}^7$ homogeneous
coordinates are assigned to the following {\it irrep}
 of $SU(2) \times SU(2) \times SU(2)$:
\begin{eqnarray}
  X^{ijk} \,  \mapsto \, ({\bf 2},{\bf 2},{\bf 2}) \, \equiv \,
  \begin{array}{|c|}
\hline
             \hskip .3 cm  \\
\hline
\end{array} \, \otimes \, \begin{array}{|c|}
\hline
             \times   \\
\hline
\end{array} \otimes \, \begin{array}{|c|}
\hline
             \,\cdot\,   \\
\hline
\end{array}\,.
\label{222q111}
\end{eqnarray}
In angular momentum notation we have
\begin{eqnarray}
X^{ijk} = \left( j_1=\ft12, j_2=\ft12, j_3=\ft12\right)
\label{3unmez}
\end{eqnarray}
and it is easy to find the structure of the embedding equations.
Here we have
\begin{eqnarray}
&\left [ \left( \ft 12 , \ft 12 , \ft 12 \right) \times \left( \ft 12 , \ft 12 , \ft 12
\right)\right ]_{sym}=\underbrace{\left( 1,1,1\right)}_{27}
\oplus \underbrace{\left( 1,0,0\right)+ \left( 0,1,0\right)+ \left(
0,0,1\right)}_{9}\,.&
\label{27+9}
\end{eqnarray}
The $9$ embedding equations $\Lambda_i$ are given by the vanishing of the
irreducible representations not of highest weight, namely:
\begin{eqnarray}
0&=& \left( \epsilon \sigma^A \right)_{ij} \, X^{i \ell p} \, X^{j m q} \,
\epsilon_{\ell m} \, \epsilon_{pq}\,, \nonumber\\
0&=& \left( \epsilon \sigma^A\right)_{\ell m} \, X^{i \ell p} \, X^{j m q} \,
\epsilon_{ij} \, \epsilon_{pq}\,, \nonumber\\
0&=& \left( \epsilon \sigma^A\right)_{pq} \, X^{i \ell p} \, X^{j m q} \,
\epsilon_{ij} \, \epsilon_{\ell m}\,.
\label{sigXX}
\end{eqnarray}
\par
\par
Coming now to the coordinate ring (\ref{coordring}), it follows
that in the $Q^{1,1,1}$ case  it takes the
following form:
\begin{equation}
 {\mathbb C}[W_{\chi^{-1}}]/I\simeq \oplus_{k\geq 0} W_{\chi^{-k}} =
 \sum_{k\ge 0} \,\left( \underbrace{
\begin{array}{|c|c|c|}
\hline
\phantom{\times} & \dots &\phantom{\times}   \\
\hline
\end{array}}_{k}\, \otimes \,
\underbrace{\begin{array}{|c|c|c|}
\hline
            \times &  \dots &\times    \\
\hline
\end{array}}_{k} \otimes \,
\underbrace{\begin{array}{|c|c|c|}
\hline
            \,\cdot\, &  \dots &\,\cdot\,    \\
\hline
\end{array}}_{k} \right)\,.
\label{kkk}
\end{equation}
In eq. (\ref{kkk}) we predict the spectrum of $SU(2) \times SU(2) \times SU(2)$
representations of the $Osp(2\vert 4)$ hypermultiplets as determined
by harmonic analysis on $Q^{1,1,1}$. We find that the hypermultiplet of conformal
weight (energy label) $E_0 = k$ and hypercharge $y_0= k$ should be in the
representation:
\begin{equation}
  J_1 = \frac{k}{2} \quad ; \quad J_2 = \frac{k}{2} \quad ; \quad \quad J_3 = \frac{k}{2}\,.
\label{donalfonso}
\end{equation}
\subsubsection{Cohomology of $Q^{1,1,1}$}
As for the cohomology, the
first Chern class of $L$ is $c_1 = \omega_1 + \omega_2 + \omega_3$, where
$\omega_i$ are the generators of the second cohomology
group of the ${\mathbb P}^1$'s.
Reasoning as for $\m1$, one gets
\begin{eqnarray}
& H^1(Q^{1,1,1}, \zz) =
H^3(Q^{1,1,1}, \zz) =
H^6(Q^{1,1,1}, \zz) = 0,&
\nonumber\\ & H^2(Q^{1,1,1}, \zz) = \zz \cdot \omega_1 \oplus \zz \cdot \omega_2,
&
\nonumber\\ &
H^4(Q^{1,1,1}, \zz) = \zz_2 \cdot (\omega_1 \omega_2 +\omega_1 \omega_3
+ \omega_2 \omega_3),
&
\nonumber\\ &
H^5(Q^{1,1,1}, \zz) = \zz \cdot \alpha \oplus \zz \cdot \beta,&
\label{q111gys}
\end{eqnarray}
where $\pi_* \alpha = \omega_1 \omega_2 - \omega_1 \omega_3$,
$\pi_* \beta = \omega_1 \omega_2 - \omega_2 \omega_3$ and the pullbacks
are left implicit.
\subsubsection{Explicit description of the Sasakian fibration for $Q^{1,1,1}$}
The coset space $Q^{1,1,1}$ is a $U(1)$-fibre bundle over
${\mathbb P}^1\times{\mathbb P}^1\times{\mathbb P}^1\simeq S^2\times S^2\times S^2$.
We can parametrize the base manifold with polar coordinates
$(\theta_i,\phi_i)$, $i=1,2,3$.
We cover the base with eight coordinate patches,
$H_{\alpha\beta\gamma}$ $(\alpha,\beta,\gamma=\pm 1)$ and choose local coordinates
for the fibre, $\psi_{\alpha\beta\gamma}\in[0,4\pi)$.
Every patch is the product of three open sets, $H^i_{\pm}$,
each one describing a coordinate patch for a single two-sphere,
as indicated in fig. \ref{S2patches}:
\begin{equation}
H_{\alpha\beta\gamma}=H^1_\alpha\times H^2_\beta\times H^3_\gamma.
\end{equation}
\iffigs
\begin{figure}[ht]
\begin{center}
\epsfxsize = 5cm
\epsffile{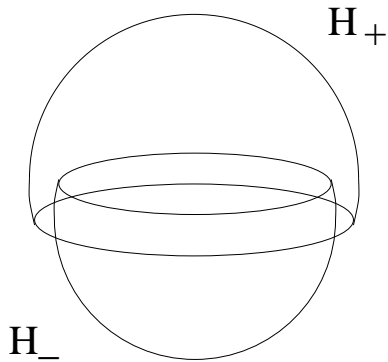}
\vskip  0.2cm
\hskip 2cm
\unitlength=1.1mm
\end{center}
\caption{Two coordinate patches for the sphere.
They constitute the base for a local trivialization of a fibre
bundle on $S^2$.
Each patch covers only one of the poles, where the coordinates
$(\theta,\phi)$ are singular.
}
\label{S2patches}
\end{figure}
\fi
To describe the total space we have to specify the transition maps
for $\psi$ on the intersections of the patches.
These maps for the generic $Q^{p,q,r}$ space are
\begin{equation}\label{maps}
\psi_{\alpha_1\beta_1\gamma_1}=\psi_{\alpha_2\beta_2\gamma_2}+p(\alpha_1-\alpha_2)\phi_1+q(\beta_1-\beta_2)\phi_2
+r(\gamma_1-\gamma_2)\phi_3\,.
\end{equation}
For example, in the case of interest, $Q^{1,1,1}$, we have
\begin{equation}
\psi_{+-+}=\psi_{++-}-2\phi_2+2\phi_3\,.
\end{equation}
We note that these maps are well defined, being all
the $\psi$'s and $\phi$'s defined modulo $4\pi$ and $2\pi$ respectively.

It is important to note that $\theta$ and $\phi$ are clearly not good
coordinates for the whole $S^2$.
The most important consequence of this fact is that the one-form
$d\phi$ is not extensible to the poles.
To extend it to one of the poles, $d\phi$ has to be multiplied
by a function which has a double zero on that pole, such as
$\sin^2\frac{\theta}{2}\,d\phi$.

We can define a $U(1)$-connection $\cal A$ on the base
$S^2\times S^2\times S^2$ by specifying it on each patch $H_{\alpha\beta\gamma}$
\footnote{It is worth noting that the connection $\cal A$ is
chosen to be well defined on the coordinate singularities of each patch,
i.e. on the product of the three $S^2$ poles covered by the patch.}:
\begin{eqnarray}
{\cal A}_{\alpha\beta\gamma}=(\alpha-\cos\theta_1)d\phi_1+(\beta-\cos\theta_2)d\phi_2+
(\gamma-\cos\theta_3)d\phi_3\,.
\end{eqnarray}
Because of the fibre-coordinate transition maps (\ref{maps}), the
one-form $(d\psi-{\cal A})$ is globally well defined on $Q^{1,1,1}$.
In other words the different one-forms $(d\psi_{\alpha\beta\gamma}-
{\cal A}_{\alpha\beta\gamma})$
defined on the corresponding $H_{\alpha\beta\gamma}$, coincide on the intersections
of the patches.
We can therefore define an $SU(2)^3\times U(1)$-invariant metric on the
total space by:
\begin{equation}
ds^2_{Q^{1,1,1}}=c^2(d\psi-{\cal A})^2+a^2ds^2_{S^2\times S^2\times S^2}\,.
\end{equation}
The Einstein metric of this family is given by
\begin{equation}\label{Qmetric}
ds^2_{Q^{1,1,1}}=\frac{3}{8\Lambda}(d\psi-{\cal A})^2
+\frac{3}{4\Lambda}\sum_{i=1}^3
\left(d\theta_i^2+\sin^2\theta_i\,d\phi_i^2\right)\,,
\label{metq111}
\end{equation}
where $\Lambda$ is the compact space cosmological constant defined in
eq.(\ref{ricciin}). The Einstein metric (\ref{metq111}) was
originally found by D'Auria, Fr\'e and van Nieuwenhuizen
\cite{dafrepvn}, who introduced the family $Q^{p,q,r}$ of D=11
compactifications and found that ${\cal N}=2$ supersymmetry is preserved in
the case $p=q=r$. All the other cosets in the family break
supersymmetry to $N=0$, namely, in mathematical language, are not
Sasakian. In \cite{dafrepvn} the Einstein metric was constructed
using the intrinsic geometry of coset manifolds and using Maurer--Cartan
forms. An explicit form was also given using stereographic
coordinates on the three $S^2$. In the coordinate form of
eq. (\ref{metq111}) the Einstein metric of $Q^{1,1,1}$ was later written
by Page and Pope \cite{pagepopeQ}.
\subsubsection{The baryonic $5$--cycles of $Q^{1,1,1}$ and their volume}
\label{brynQ}
The relevant homology group of $Q^{1,1,1}$ for the
calculation of the baryonic masses is
\begin{equation}
H_5(Q^{1,1,1},{\mathbb R})={\mathbb R}^2\,.
\end{equation}
Three (dependent) five-cycles spanning $H_5(Q^{1,1,1})$ are the restrictions
of the $U(1)$-fibration to the product of two of the three ${\mathbb P}^1$'s.
Using the above metric (\ref{metq111}) one easily computes the volume of these
cycles. For instance
\begin{equation}
{\rm Vol(cycle)}=\oint_{\pi^{-1}({\mathbb P}^1_1\times{\mathbb P}^1_2)}
\left(\frac{3}{8\Lambda}\right)^{5/2}
4\sin\theta_1\,\sin\theta_2\ d\theta_1\,d\theta_2\,d\phi_1\,d\phi_2\,d\psi=
\frac{\pi^3}{4}\left(\frac{6}{\Lambda}\right)^{5/2}\,.
\label{volcycq}
\end{equation}
The volume of the whole space $Q^{1,1,1}$ is
\begin{equation}
{\rm Vol(Q^{1,1,1})}=\oint_{Q^{1,1,1}}\left(\frac{3}{8\Lambda}\right)^{7/2}
8\,\prod_{i=1}^{3}\sin\theta_i\,d\theta_i d\phi_i\,d\psi
=\frac{\pi^4}{8}\left(\frac{6}{\Lambda}\right)^{7/2}\,.
\label{volq111}
\end{equation}
Just as in the $M^{1,1,1}$ case, inserting the above results (\ref{volcycq}, \ref{volq111})
into the general formula (\ref{baryondim}) we obtain the conformal weight of
the  baryon operator corresponding to the five-brane wrapped on this cycle:
\begin{equation}
E_0=\frac{N}{3}\,.
\label{Ccweight}
\end{equation}
\par
The other two cycles can be obtained from this by permuting the role of the three
${\mathbb P}^1$'s and their volume is the same.
This fact agrees with the symmetry which exchanges the
fundamental fields $A$, $B$ and $C$ of the conformal theory, or the
three gauge groups $SU(N)$. Indeed, naming $SU(2)_i$ ($i=1,2,3$) the
three $SU(2)$ factors appearing in the isometry group of $Q^{1,1,1}$,
the stability subgroup of the first of the cycles described above is
\begin{eqnarray}
  H({\cal C}^1) &=& SU(2)_1 \times SU(2)_2 \times U(1)_{B,3}\nonumber\\
  U(1)_{B,3} &\subset & SU(2)_3
\label{hcycq111}
\end{eqnarray}
so that the collective coordinates of the baryon state live on
${\mathbb P}^1 \simeq SU(2)_3/U(1)_{B,3}$. This result is obtained by
an argument completely analogous to that used in the analysis of
$M^{1,1,1}$ $5$--cycles and leads to a completely analogous conclusion.
The baryon state is in the $J_1=0, \, J_2 = 0, \, J_3=N/2$ flavor
representation. In the conformal field theory the corresponding
baryon operator is the chiral field (\ref{q111baryopC})
and the result (\ref{Ccweight}) implies that the conformal weight of
the $C_i$ elementary world--volume field is
\begin{equation}
  h[C_i]=\frac{1}{3}.
\label{hCfield}
\end{equation}
The stability subgroup of the permuted
cycles is obtained permuting the indices $1,2,3$ in eq. (\ref{hcycq111})
and we reach the obvious conclusion
\begin{equation}
  h[A_i]=h[B_j]=h[C_\ell]=\frac{1}{3}.
\label{hABfield}
\end{equation}
This matches with the previous result (\ref{donalfonso}) on the
spectrum of chiral operators, which are predicted of the form
\begin{equation}
  \mbox{chiral operators} = \mbox{Tr} \left(  A_{i_1} \, B_{j_1} \, C_{\ell_1} \,
  \dots \,A_{i_k} \, B_{j_k} \, C_{\ell_k}\right)
\label{chiropq111}
\end{equation}
and should have conformal weight $E=k$. Indeed, we have $k \times (
\frac{1}{3}+\frac{1}{3}+\frac{1}{3}) =k$ !
\section{Conclusions}
We saw, using geometrical intuition,
 that there is  a set of supersingletons fields which are likely to be
the fundamental degrees of freedom of the CFT's corresponding to
$Q^{1,1,1}$ and $M^{1,1,1}$. The entire KK spectrum and
the existence of baryons of given quantum numbers
can be explained in terms of these singletons.

We also proposed candidate three-dimensional gauge theories which should
flow in the IR to the superconformal fixed points dual to the $AdS_4$
compactifications. The singletons are the elementary chiral
multiplets of these gauge theories.
The main problem we did not solve is the existence of chiral operators 
in the gauge theory that have no counterpart in the KK spectrum. These are
the non completely flavor symmetric chiral operators. Their existence
 is due to
the fact that, differently from the case of $T^{1,1}$, we are not able
to write any superpotential of dimension two.  
If the proposed gauge theories are correct, the dynamical
 mechanism responsible for
the disappearing of the non symmetric operators in the IR
has still to be clarified.

It would be quite helpful to have a description of the conifold
as  a deformation of an orbifold singularity \cite{witkleb,morpless}.
It would provide an holographic description of the RG flow between two
different CFT theories and it would also help in checking whether the
proposed gauge theories are correct or require to be slightly
modified by the introduction of new fields.
In general, different orbifold theories can flow to the same
conifold CFT in the IR. 
In the case of $T^{1,1}$,  one can  deform a ${\mathbb Z}_2$ orbifold theory 
with a mass term \cite{witkleb} or a ${\mathbb Z}_2\times {\mathbb Z}_2$ 
orbifold theory with a FI parameter \cite{morpless}.
The mass deformation approach for a ${\mathbb Z}_2\times {\mathbb Z}_2$ 
singularity was attempted for the case of $Q^{1,1,1}$ in \cite{tatar},
where a candidate conifold CFT was written. This theory is deeply different 
from our proposal. It is not obvious to us whether
this theory is compatible or not with the KK expectations.
It also seems that, in the approach followed in \cite{tatar},  
the singletons degrees of freedom needed for
constructing the KK spectrum are not the elementary chiral
fields of the gauge theory but are rather obtained with some change of 
variables which should make sense only in the IR.
The FI approach was pursued in \cite{dallagat}, were
$Q^{1,1,1}$ was identified as a deformation of orbifold singularities
whose associated CFT's can be explicitly written. Unfortunately,
the order of the requested orbifold group and, consequently, the number
of requested gauge factors, make difficult an explicit analysis
of these models and the identification of the
conifold CFT.
It would be quite interesting to investigate the relation between the results 
in \cite{dallagat} and our proposal or to find simpler orbifold singularities
related to $Q^{1,1,1}$ and $M^{1,1,1}$. For the latter one, for the moment,
no candidate orbifold has been proposed. 
  
We did not discuss at all the CFT associated to $V_{5,2}$ and $N^{0,1,0}$.
The absence of a toric description makes more difficult to guess a
gauge theory with the right properties and also to find
associated orbifold models. We leave for the future the investigation
of these interesting models. 
\par
\vskip 0.5cm
\leftline{\bf Acknowledgements} We thank S. Ferrara for many important
and enlightening discussions throughout all the completion of the present
work, and I. Klebanov for very useful and interesting conversations.
We are also indebted to C. Procesi for deep hints and valuable discussions
on the geometric side of representation theory.
\appendix
\section{The scalar potential}
\label{scalapot}
Let us now consider more closely the scalar potential of the ${\cal N}=2$ 
world--volume gauge theories we have conjectured to be associated with the
$Q^{1,1,1}$ and $M^{1,1,1}$ compactifications.
\par
In complete generality, the scalar potential of a three dimensional
${\cal N}=2$
gauge theory with an arbitrary gauge group and an arbitrary number of
chiral multiplets in generic representations of the gauge--group was
written in  eq. (5.46) of \cite{superfieldsN2D3}. It has the following
form:
\begin{eqnarray}
U(z,\overline z,M)&=&
\partial_i W(z)\eta^{ij^*}\partial_{j^*}\overline W(\overline z)\nonumber\\
&&+\ft{1}{2}g^{IJ}\left(\overline z^{i^*}(T_I)_{i^*j}z^j\right)
\left(\overline z^{k^*}(T_J)_{k^*l}z^l\right)\nonumber\\
&&+\overline z^{i^*}M^I(T_I)_{i^*j}\eta^{jk^*}M^J(T_J)_{k^*l}
z^l\nonumber\\
&&-2\a^2g_{IJ}M^IM^J-2\a\zeta^{\tilde I}{\cal C}_{\tilde I}^{\ I}g_{IJ}M^J
-\ft{1}{2}\zeta^{\tilde I}{\cal C}_{\tilde I}^{\ I}g_{IJ}
\zeta^{\tilde J}{\cal C}_{\tilde J}^{\ J}\nonumber\\
&&-2\a M^I\left(\overline z^{i^*}(T_I)_{i^*j}z^j\right)
-\zeta^{\tilde I}{\cal C}_{\tilde I}^{\ I}
\left(\overline z^{i^*}(T_I)_{i^*j}z^j\right)\,,
\end{eqnarray}
where:
\begin{enumerate}
  \item $z^i$ are the {\it complex scalar fields} belonging to the chiral
  multiplets,
  \item $W(z)$ is the holomorphic superpotential,
  \item the hermitian matrices $(T_I)_{i^*j}$  ($I=1,\dots , \mbox{dim} {\cal G}$)
  are the generators of the gauge group $\cal G$ in the (in general reducible)
  representation $\cal R$ supported by the chiral multiplets,
  \item $\eta^{ij^\star}$ is the ${\cal G}$ invariant metric,
  \item $g^{IJ}$ is the Killing metric of $\cal G$,
  \item $M_I$ are the {\it real} scalar fields belonging to the vector multiplets
  that obviously transform in  $ad \, ({\cal G} ) \,$,
  \item $\alpha$ is the coefficient of the Chern--Simons term, if
  present,
  \item $\zeta^{\tilde J}$ are the coefficients of the Fayet
  Iliopoulos terms that take values in the center of the gauge Lie
  algebra $\zeta^{\tilde J} \in Z({\cal G})$.
\end{enumerate}
If we put the Chern Simons and the Fayet Iliopoulos terms to zero
$\alpha =  \zeta^{\tilde J} =0$, the scalar potential
becomes the sum of three quadratic forms:
\begin{equation}
 U(z,\overline z,M)= |\partial W (z) |^2 + \ft{1}{2}g^{IJ}\, D_I(z,\overline z
 \,)\, D_J(z,\overline z) + M^I \, M^J \, K_{IJ}(z,\overline z),
\label{trequadra}
\end{equation}
where the real functions
\begin{equation}
D^I(z,\overline z)=-\overline z^{i^*}(T_I)_{i^*j}z^j
\end{equation}
are the $D$--terms, namely the on--shell values of the vector multiplet auxiliary
fields, while by definition we have put
\begin{equation}
   K_{IJ}(z,\overline z) \stackrel{\mbox{\scriptsize def}}{=}
   \overline z^{i^*}(T_I)_{i^*j}\eta^{jk^*}(T_J)_{k^*l}z^l.
\label{KIJ}
\end{equation}
If the quadratic form $M_I \, M_J \, K_{IJ}(z , \bar z)  $ is
positive definite, then the vacua of the gauge theory are singled out
by the three conditions
\begin{eqnarray}
\frac{\partial W}{\partial z^i} & = & 0, \label{suppoteq}\\
D^I(z,\bar z) & = & 0, \label{Dtermeq} \\
M_I \, M_J \, K_{IJ}(z , \bar z) & = & 0. \label{MMKeq}
\end{eqnarray}
The basic relation between the candidate superconformal gauge theory
$CFT_3$ and the compactifying $7$--manifold $M^7$ that we have used in
eq.s (\ref{toricQ}, \ref{toricM}) is that, in the Higgs branch ($ \langle M_I \rangle =0$),
the space of vacua of $CFT_3$, described by eq.s (\ref{suppoteq}, \ref{Dtermeq}, \ref{MMKeq}), should be equal to
the product of
$N$ copies of $M^7$:
\begin{equation}
  \mbox{vacua of gauge theory} = \underbrace{ M^7 \, \times \, \dots\,\times \, M^7}_{N}/\Sigma_N\,.
\label{NcopieM7}
\end{equation}
Indeed, if there are $N$ M2--branes in the game, each of them can be
placed somewhere in $M^7$ and the vacuum is described by giving all
such locations. In order for this to make sense it is necessary
that
\begin{itemize}
  \item The Higgs branch should be distinct from the Coulomb branch
  \item The vanishing of the D--terms should indeed be a geometric
  description of (\ref{NcopieM7}).
\end{itemize}
Let us apply our general formula to the two cases under consideration and see that
these conditions are indeed verified.
\subsection{The scalar potential in the $Q^{1,1,1}$ case}
Here the gauge group is
\begin{equation}
 {\cal G}=SU(N)_1 \times SU(N)_2 \times SU(N)_{3}
\label{Unkvolte}
\end{equation}
in the non--abelian case $N>1$ and
\begin{equation}
 {\cal G}=U(1)_1 \times U(1)_2 \times U(1)_{3}
\label{U1kvolte}
\end{equation}
in the abelian case $N=1$. The chiral fields $A_i, B_j , C_\ell$
are in the $SU(2)^3$ flavor
representations $({\bf 2},{\bf 1},{\bf 1})$, $({\bf 1},{\bf 2},{\bf 1})$,
$({\bf 1},{\bf 1},{\bf 2})$  and in
the color $SU(N)^3$ representations $({\bf N},\bar {\bf N},{\bf 1})$,
$({\bf 1,N},\bar {\bf N})$,
$(\bar {\bf N},{\bf 1,N})$, respectively (see fig.\ref{ABCcolor}).
We can arrange the chiral fields into a column vector:
\begin{equation}
  {\vec z}  \, = \, \left( \matrix{A_i\cr B_j \cr C_\ell \cr } \right).
\label{zabc}
\end{equation}
Naming $(t_I)_{\phantom{\Lambda}\Sigma}^{\Lambda}$ the $N \times N$ hermitian
matrices such that $\mbox{i}\, t_I$ span
the $SU(N)$ Lie algebra ($I=1,\dots,N^2-1$), the generators of the gauge
group acting on the chiral fields can be written as follows:
\begin{eqnarray}
  T^{[1]}_I &=& \left( \begin{array}{ccc}
    t_I \otimes {\bf 1} & 0 & 0 \\
    0 & 0 & 0 \\
    0 & 0 & -{\bf 1} \otimes t_I \
  \end{array}\right) , \quad T^{[2]}_I = \left( \begin{array}{ccc}
    -{\bf 1} \otimes t_I & 0 & 0 \\
    0 & t_I \otimes {\bf 1}& 0 \\
    0 & 0 & 0 \
  \end{array}\right), \nonumber\\
  \quad  T^{[3]}_I &=& \left( \begin{array}{ccc}
    0 & 0 & 0 \\
    0 & -{\bf 1} \otimes t_I & 0 \\
    0 & 0 & t_I \otimes {\bf 1} \
  \end{array}\right).
\label{Un3gene}
\end{eqnarray}
Then the $D^2$--terms appearing in the scalar potential take the
following form:
\begin{eqnarray}
 \mbox{$D^2$-terms}&=& \ft{1}{2}\,\Bigl [ \sum_{I=1}^{N^2-1} \, \left(
 {\bar A}^i \,\left(  t_I \otimes {\bf 1}\right)  \,
 A_i - {\bar C}^i \, \left( {\bf 1} \otimes t_I \right) \, C_i \right)
 ^2\nonumber\\
 &&
 +\sum_{I=1}^{N^2-1} \, \left( {\bar B}^i \,\left(  t_I\otimes {\bf 1}\right)  \,
 B_i - {\bar A}^i \, \left( {\bf 1} \otimes t_I \right) \, A_i \right) ^2\nonumber\\
 &&+\sum_{I=1}^{N^2-1} \, \left( {\bar C}^i \, \left( t_I \otimes {\bf 1}\right)
 C_i - {\bar B}^i \, \left( {\bf 1} \otimes t_I \right) \, B_i \right) ^2 \Bigr ].
\label{Dterm}
\end{eqnarray}
The part of the scalar potential involving the gauge multiplet
scalars is instead given by:
\begin{eqnarray}
\mbox{$M^2$--terms} & = & M_1^I \, M_1^J \, \left(
 {\bar A}^i \,\left(  t_I t_J \otimes {\bf 1}\right)  \,
 A_i + {\bar C}^i \, \left( {\bf 1} \otimes t_I t_J \right) \, C_i
 \right)\nonumber\\
 &&+   M_2^I \, M_2^J \,\left( {\bar B}^i \,\left(  t_I t_J\otimes {\bf 1}\right)  \,
 B_i + {\bar A}^i \, \left( {\bf 1} \otimes t_I t_J \right) \, A_i \right) \nonumber\\
&& + M_3^I \, M_3^J \,  \left( {\bar C}^i \, \left( t_I t_J \otimes {\bf 1}\right)
 C_i + {\bar B}^i \, \left( {\bf 1} \otimes t_I t_J \right) \, B_i
 \right)\nonumber\\
 && - \, 2 \,M_1^I \, M_2^J \, {\bar A}^i \, \left( t_I  \otimes t_J \right)
 A_i - \, 2 \, M_2^I \, M_3^J \, {\bar B}^i \, \left( t_I  \otimes t_J \right)
 B_i \nonumber\\
 && - \, 2 \,M_3^I \, M_1^J \, {\bar C}^i \, \left( t_I  \otimes t_J \right)
 C_i.
\label{mmterm}
\end{eqnarray}
In the abelian case we simply get:
\begin{eqnarray}
 \mbox{$D^2$-terms}&=& \ft{1}{2}\,\Bigl [ \left (
\vert A_1 \vert^2 + \vert A_2 \vert^2- \vert  C_1 \vert^2 - \vert  C_2 \vert^2
\right)^2
 \nonumber\\
 &&
 +\left (
\vert B_1 \vert^2 + \vert B_2 \vert^2- \vert  A_1 \vert^2 - \vert  A_2 \vert^2
\right)^2\nonumber\\
 &&+\left (
\vert C_1 \vert^2 + \vert C_2 \vert^2- \vert  B_1 \vert^2 - \vert  B_1 \vert^2
\right)^2 \Bigr ],
\label{Dtermab}
\end{eqnarray}
\begin{eqnarray}
 \mbox{$M^2$-terms}&=&  \Bigl [ \left (
\vert A_1 \vert^2 + \vert A_2 \vert^2\right)  (M_1 - M_2)^2
\nonumber\\
&&+\left (
\vert B_1 \vert^2 + \vert B_2 \vert^2\right)  (M_2 - M_3)^2
\nonumber\\
&&+ \left( \vert  C_1 \vert^2 + \vert  C_2 \vert^2\right)  (M_3 -M_1)^2
 \Bigr ].
\label{mmtermab}
\end{eqnarray}
Eq.s (\ref{Dtermab}) and (\ref{mmtermab}) are what we have used in our
toric description of  $Q^{1,1,1}$ as the manifold of gauge--theory
vacua in the Higgs branch. Indeed it is evident from
eq. (\ref{mmtermab}) that if  we give non vanishing {\it vev} to the
chiral fields, then we are forced to put \linebreak $<M_1>=<M_2>=<M_3>=m$.
Alternatively, if we give non trivial {\it vevs} to the vector
multiplet scalars $M_i$, then we are forced to put
$<A_i>=<B_j>=<C_\ell>=0$ which confirms that the Coulomb branch is
separated from the Higgs branch.
\par
Finally, from eq.s (\ref{Dterm}, \ref{mmterm}) we can retrieve
the vacua describing  $N$ separated branes. Each chiral field has
two color indices and is actually a matrix. Setting
\begin{eqnarray}
 < A_{i\vert \Sigma}^{\phantom{i\vert
  \Sigma}\Lambda}>&=&\delta^{\Lambda}_\Sigma \, a_i^\Lambda,
  \nonumber\\
 < B_{i\vert \Sigma}^{\phantom{i\vert
  \Sigma}\Lambda}>&=&\delta^{\Lambda}_\Sigma \, b_i^\Lambda,
  \nonumber\\
  <C_{i\vert \Sigma}^{\phantom{i\vert
  \Sigma}\Lambda}>&=&\delta^{\Lambda}_\Sigma \, c_i^\Lambda,
\label{NbraneQ}
\end{eqnarray}
a little work shows that the potential (\ref{Dterm}) vanishes if each of the
$N$--triplets
$a_i^\Lambda, b_j^\Lambda, c_\ell^\Lambda$ separately satisfies the
$D$--term equations, yielding the toric description of a $Q^{1,1,1}$
manifold (\ref{toricQ}). Similarly, for each abelian generator
belonging to the Cartan subalgebra of $U_i(N)$ and having a non trivial action
on $a_i^\Lambda, b_j^\Lambda, c_\ell^\Lambda$  we have
$<M_1^\Lambda>=<M_2^\Lambda>=<M_3^\Lambda>=m^\Lambda$.
\subsection{The scalar potential in the $M^{1,1,1}$ case}
Here the gauge group is
\begin{equation}
 {\cal G}=SU(N)_1 \times SU(N)_2
\label{SUn2volte}
\end{equation}
in the non--abelian case $N>1$ and
\begin{equation}
 {\cal G}=U(1)_1 \times U(1)_2
\label{U12volte}
\end{equation}
in the abelian case $N=1$. The chiral fields $U_i, V_A$
are in the $SU(3) \times SU(2)$ flavor
representations $({\bf 3,1})$, $({\bf 1,2})$ respectively. As for color,
they are in the $SU(N)^2$ representations
$Sym^2({\mathbb C}^N)\otimes Sym^2({\mathbb C}^{N*})$,
$Sym^3({\mathbb C}^{N*})\otimes Sym^3({\mathbb C}^N)$
respectively (see fig. \ref{UVcolor}).
As before, we can arrange the chiral fields into a column vector:
\begin{equation}
  {\vec z}  \, = \, \left( \matrix{U_i\cr V_A \cr } \right).
\label{zuuuvv}
\end{equation}
Naming $(t^{[3]}_I)_{\phantom{\Lambda\Sigma\Gamma}
\Xi\Delta\Theta}^{\Lambda\Sigma\Gamma}$ the   hermitian
matrices generating $SU(N)$ in the three--times symmetric
representation  and  $(t^{[2]}_I)_{\phantom{\Lambda\Sigma}
\Xi\Delta}^{\Lambda\Sigma}$ the same generators
in the two--times symmetric representation, the generators of the gauge
group acting on the chiral fields can be written as follows:
\begin{eqnarray}
  T^{[1]}_I &=& \left( \begin{array}{cc}
    t^{[2]}_I \otimes {\bf 1} & 0  \\
 0 & -{\bf 1} \otimes t^{[3]}_I \
  \end{array}\right) ,
  \quad T^{[2]}_I = \left( \begin{array}{cc}
    -{\bf 1} \otimes  t^{[2]}_I & 0  \\
    0 & t^{[3]}_I \otimes {\bf 1} \\
\end{array}\right).
\label{SUn2gene}
\end{eqnarray}
Then the $D^2$--terms appearing in the scalar potential take the
following form:
\begin{eqnarray}
 \mbox{$D^2$-terms}&=& \ft{1}{2}\,\Bigl [ \sum_{I=1}^{N^2-1} \, \left(
 {\bar U}^i \,\left(  t^{[2]}_I \otimes {\bf 1}\right)  \,
 U_i - {\bar V}^A \, \left( {\bf 1} \otimes t^{[3]}_I \right) \, V_A \right)
 ^2\nonumber\\
 &&
 +\sum_{I=1}^{N^2-1} \, \left(
 {\bar U}^i \,\left( {\bf 1} \otimes t^{[2]}_I  \right)  \,
 U_i - {\bar V}^A \, \left(  t^{[3]}_I \otimes {\bf 1}  \right) \, V_A \right)
 ^2 \Bigr ],
\label{DtermM}
\end{eqnarray}
while the part of the scalar potential involving the gauge multiplet
scalars is given by
\begin{eqnarray}
\mbox{$M^2$--terms} & = & M_1^I \, M_1^J \, \left(
 {\bar U}^i \,\left(  t^{[2]}_I t^{[2]}_J \otimes {\bf 1}\right)  \,
 U_i + {\bar V}^A \, \left( {\bf 1} \otimes  t^{[3]}_I t^{[3]}_J \right) \,
 V_A
 \right)\nonumber\\
 &&+   M_2^I \, M_2^J \, \left(
 {\bar U}^i \,\left( {\bf 1} \otimes t^{[2]}_I t^{[2]}_J  \right)  \,
 U_i + {\bar V}^A \, \left( t^{[3]}_I t^{[3]}_J \otimes {\bf 1}    \right) \,
 V_A \right) \nonumber\\
 && - \, 2 \,M_1^I \, M_2^J \, {\bar U}^i \, \left( t^{[2]}_I  \otimes t^{[2]}_J \right)
 U_i - \, 2 \, M_2^I \, M_1^J \, {\bar V}^A \, \left( t^{[3]}_I  \otimes t^{[3]}_J \right)
 V_A.
\label{mmtermM}
\end{eqnarray}
In the abelian case we simply get
\begin{eqnarray}
 \mbox{$D^2$-terms}&=& \ft{1}{2}\,\Bigl \{ \left [ 2
\left( \vert U_1 \vert^2 + \vert U_2 \vert^2+ \vert  U_3 \vert^2\right)  -
3 \left( \vert  V_1 \vert^2 + \vert  V_1 \vert^2\right)
\right]^2
 \nonumber\\
 &&
 +\left [ 2
\left( \vert U_1 \vert^2 + \vert U_2 \vert^2+ \vert  U_3 \vert^2\right)  -
3 \left( \vert  V_1 \vert^2 + \vert  V_2 \vert^2\right)
\right]^2 \Bigr \},
\label{DtermabM}
\end{eqnarray}
\begin{eqnarray}
 \mbox{$M^2$-terms}&=&   \left [
4 \left( \vert U_1 \vert^2 + \vert U_2 \vert^2 +\vert U_3 \vert^2\right)
+9 \left( \vert V_1 \vert^2 + \vert V_2 \vert^2 \right) \right] (M_1 - M_2)^2.
\label{mmtermabM}
\end{eqnarray}
Once again from eq.s (\ref{DtermabM}) and (\ref{mmtermabM}) we see
that the Higgs and Coulomb branches are separated.
Furthermore, in eq. (\ref{DtermabM}) we recognize  the
toric description of  $M^{1,1,1}$ as the manifold of gauge--theory
vacua in the Higgs branch (see eq. (\ref{toricM})).
\par
As before, from eq.s (\ref{Dterm}, \ref{mmterm}) we can retrieve
the vacua describing  $N$ separated branes. In this case the color
index structure is more involved and we must set
\begin{eqnarray}
 < U_{i\vert \Lambda\Lambda}^{\phantom{i\vert
  \Lambda\Lambda}\Lambda\Lambda}>&=& \, u_i^\Lambda,
  \nonumber\\
 < V_{A\vert \Lambda\Lambda\Lambda}^{\phantom{i\vert
  \Lambda\Lambda\Lambda}\Lambda\Lambda\Lambda}>&=& \, v_A^\Lambda.
\label{NbraneM}
\end{eqnarray}
A little work shows that the potential (\ref{Dterm}) vanishes if each of the
$N$--doublets
$u_i^\Lambda, v_A^\Lambda $ separately satisfies the
$D$--term equations yielding the toric description of a $M^{1,1,1}$
manifold (\ref{toricM}). Similarly, for each abelian generator
belonging to the Cartan subalgebra of $U_i(N)$ and having a non trivial action
on $u_i^\Lambda, v_A^\Lambda $  we have
$<M_1^\Lambda>=<M_2^\Lambda>=m^\Lambda$.
\section{The other homogeneous Sasakian 7-manifolds}\label{AppB}
In this Appendix we briefly discuss the other three homogeneous Sasakians
of dimension 7, giving a description of their realizations as circle bundles,
the corresponding embeddings of the base manifolds and computing their cohomology.
\subsection{The manifold $N^{0,1,0}$}
Next we have $G=SU(3)$, $H=U(1)$ is generated by $h_1+2h_2$ and
$\tilde H = U(1)\times U(1)$ is the
maximal torus. Accordingly
\begin{equation}
M_a = {\mathbb F}(1,2;3)
\label{n010base}
\end{equation}
is the complete flag variety of lines inside planes in ${\mathbb C}^3$.
A realization of this variety is given by parametrizing
separately the lines and the planes by ${\mathbb P}^2\times{\mathbb P}^{2*}$
and then imposing the incidence relation
\begin{equation}
\Sigma_k \alpha^kz_k=0,
\label{n010eq}
\end{equation}
where $z_i$ and $\alpha_i$ are homogeneous coordinates on
${\mathbb P}^2$ and ${\mathbb P}^{2*}$.
Notice that this relation is the singleton in the tensor product
${\mathbb C}^3\otimes{\mathbb C}^{3*}$.

The generator of the fibre is $h_2$, so
\begin{eqnarray}
\nonumber&&\exp(i\theta h_1)\cdot eH = \exp(-2i\theta h_2)H,\\
\nonumber &&\exp(i\theta h_2)\cdot eH = \exp(i\theta h_2)H,
\end{eqnarray}
showing that $N^{0,1,0}$ is the circle bundle inside ${\cal O}(1,1)$ over the
flag variety ${\mathbb F}(1,2;3)$.

This time $\dim H^0 (M_a, L) = 8$ and the embedding space is ${\mathbb P}^7$;
the ideal of the image is generated by $36-27=9$ equations.

We now list the cohomology groups.
Since ${\mathbb F}(1,2;3)$ is a
${\mathbb P}^1$-bundle over
${\mathbb P}^{2*}$, we can again apply the Gysin sequence to this $S^2$
fibration to compute its cohomology.
This turns out to be $\zz[\omega_1, \omega_2]/
\langle\omega_1^3, \omega_2^2\rangle$; the Chern class of $L$ is $c_1=
\omega_1 +\omega_2$.
We can now apply
the Gysin sequence to the Sasakian fibration, getting
\begin{eqnarray}
H^1(N^{0,1,0}, \zz) &=&
H^6(N^{0,1,0}, \zz) = 0,
 \nonumber\\
H^3(N^{0,1,0}, \zz) &=&
H^4(N^{0,1,0}, \zz) = 0,
 \nonumber\\
H^2(N^{0,1,0}, \zz) &=& \zz \cdot \omega_1,
 \nonumber\\
H^5(N^{0,1,0}, \zz) &=& \zz \cdot \alpha, 
\label{n010gys}
\end{eqnarray}
where $\pi_* \alpha = \omega_1^2 - \omega_1\omega_2$, and the pullbacks
are left implicit.

\subsection{The manifold $V_{5,2}$}

The last Sasakian is $V_{5,2}=SO(5)/SO(3)$, where $SO(3)$ acts on the first
three basis vectors of ${\mathbb R}^5$.
Here  $\tilde H$ is $SO(3)\times SO(2)$;
so $M_a$ is the homogeneous space $SO(5)/SO(3)\times SO(2)$, which is
actually a quadric in ${\mathbb P}^4$. To see this, recall the isomorphism
$Spin(5) \simeq Sp(2,{\mathbb H})$, the compact form
of $Sp(4,{\mathbb C)}$. This last group is of rank $2$ and has two
maximal parabolic subgroups.
The two simple roots can be chosen as $L_1-L_2$ and
$2L_2$ \cite{FH}. The parabolic subgroup we are interested in is given by
"marking" the long root $2L_2$, i.e. by adding to the Borel subalgebra
the vector $Y_{1,2}$ with root $-L_1+L_2$. A little calculation
shows that the parabolic subalgebra we get in this way is the span
of the Cartan subalgebra and the root vectors with roots
$2L_1,2L_2,L_1+L_2,L_1-L_2,-L_1+L_2$. The corresponding matrices
have the block form
\begin{equation}
\left( \begin{array}{cc}
A & B \\
0 &-A^t
\end{array}
\right),
\label{v52triang}
\end{equation}
with $A$ generic and $B$ symmetric $2\times 2$ matrices. As such it is
clear that it stabilizes a $2$-plane in ${\mathbb C}^4$.
On the other hand, $Sp(4, {\mathbb C})$ acts on
$\wedge ^2{\mathbb C}^4\simeq {\mathbb C}^6$,
with Pl\"ucker coordinates $p_{ij}=a_ib_j-a_jb_i,\;(i<j)$,
preserving the standard
symplectic form $M=p_{13}+p_{24}$.
Summing up, this action preserves the intersection  of the Pl\"ucker quadric
$p_{12}p_{34}-p_{13}p_{24}+p_{14}p_{23}=0$ with the
hyperplane $p_{13}+p_{24}=0$. Therefore $M_a$ is a quadric in ${\mathbb
P}^4$.

The character of $\tilde H$ is simply the projection on  $SO(2)$, which
is the fundamental character
associated to the parabolic subgroup above. Hence, $V_{5,2}$ is the circle
bundle inside the restriction of
${\cal O}(1)$ over ${\mathbb P}^4$ to the quadric $M_a$; the embedding is the
trivial one, and there is just $15-14=1$ equation (the quadric itself).

In this case it is more direct to observe that $V_{5,2}$ is an
$S^3$-bundle over $S^4$. In fact it is the cone over the quadric in
${\mathbb P}^4$ intersected with $S^{10}$ and this is in turn isomorphic to
the unit sphere bundle in the tangent bundle of $S^4$.
The Gysin sequence gives that the only non-vanishing cohomology groups
are
$ H^0(V_{5,2}, \zz)=
 H^7(V_{5,2}, \zz) = \zz$
and possibly
$ H^4(V_{5,2}, \zz)$ which is torsion.

\subsection{Sasakian fibrations over ${\mathbb P}^3$}

Recall \cite{BG} that every homogeneous Sasakian-Einstein
$7$-manifolds  is a circle bundle over an
algebraic homogeneous space of complex dimension $3$.
There is one missing in the list above, namely ${\mathbb P}^3$.
As we already mentioned, there is another maximal
parabolic subgroup  $P\subset Sp(4,{\mathbb C})$ given by marking the
short root $L_1-L_2$. After a suitable Weyl action,
the compact form $U(1)\times SU(2)$ of $P$ is
embedded in $Sp(2,{\mathbb H})$ as
\begin{equation}
\left( \matrix{e^{i\theta} &0 &0\cr
                 0           &U &0\cr
                 0           &0 &e^{-i\theta} }\right),
\label{p3embed}
\end{equation}
where $U$ is in $SU(2)$.
As such it stabilizes a line in a $3$-plane in ${\mathbb C}^4$.
If we look at the fibration $p:{\mathbb F}(1,3;4)\rightarrow {\mathbb P}^3$
given by forgetting the second element of the flags $V_1\subset V_3$,
we see that the map $V_1\mapsto V_1\subset Ker M(V_1,\cdot )$ is a section
of $p$ which is $Sp(2,{\mathbb H})$ invariant. $M_a$ is the image
of this section and hence
\begin{equation}
Sp(2,{\mathbb H})/U(1)\times SU(2) \simeq {\mathbb P}^3.
\label{p3base}
\end{equation}
It is clear that the Sasakian fibration is $M^7=Sp(2,{\mathbb H})/SU(2)\simeq
S^7$ and obviously $S^7/U(1)={\mathbb P}^3$. Notice that
$S^7/SU(2)= {\mathbb P}^1({\mathbb H})=S^4$ and we have a commutative diagram
\begin{equation}
\begin{array}{ccc}
S^7&\buildrel {id}\over \longrightarrow & S^7\\
U(1)\downarrow\phantom{U(1)}&         &\phantom{SU(2)}\downarrow SU(2)\\
{\mathbb P}^3&\buildrel {S^2}\over\longrightarrow& S^4
\end{array}\,,
\label{p3commut}
\end{equation}
where the action of $Sp(2,{\mathbb H})$ on ${\mathbb P}^3$ preserves the
fibration given by the bottom line.
If we forget about this fibration, ${\mathbb P}^3$ can be considered a
homogeneous space of $SU(4)$ and again the Sasakian fibration over it
is $S^7$.
There are two more ways of getting $S^7$ as a homogeneous space (namely
$SO(8)/SO(7)$ and $SO(7)/G_2$), but the action of the group does not
preserve the $U(1)$ fibration over ${\mathbb P}^3$.
%
%
\section{The necessary symmetrization of color indices}\label{AppC}
In this Appendix we show that the symmetrization of {\sl flavor}
indices for the chiral operators of the $M^{1,1,1}$ theory implies
the symmetrization of the color indices. For this purpose,
let us concentrate on the indices of one of the two $SU(N)$ {\sl color}
groups.

The problem is to construct uncolored fields polynomially depending 
(meaning totally symmetric) on the 
$U^i_{\alpha\beta}$ and $V_A^{\alpha\beta\gamma}$. Since these fields 
belong to the {\it irreps} $Sym^2({\mathbb C}^N)$ and
$Sym^3({\mathbb C}^{N*})$ respectively, we need $3k$ $U$'s and $2k$ $V$'s
to have the right number of indices to saturate.
Hence, we have to find dual irreducible subrepresentations in the
decompositions  
$$Sym^{3k}(Sym^2({\mathbb C}^N)) = Sym^{6k}({\mathbb C}^N) \oplus_{\lambda} 
W_{\lambda}, $$
$$ Sym^{2k}(Sym^3({\mathbb C}^{N*})) = Sym^{6k}({\mathbb C}^{N*}) \oplus_{\mu} 
 W_{\mu}.$$
The first two terms in these decompositions are obviously paired. 
We claim that these are the only ones.
If there is another pair of {\it irreps} $W_\lambda$, $W_\mu$ which are dual, 
then each must be invariant under both the permutation subgroups  
$H_1 = \Sigma_{2k}\times \Sigma_3$ and $H_2 = \Sigma_{3k}\times \Sigma_2$ of 
$\Sigma_{6k}$. 
So we have only to show that these subgroups generate the 
whole $\Sigma_{6k}$.

First observe that we can order $6k$ letters in $2k$ 
triples in such a way that $H_1$ acts with $\Sigma_3$ permuting letters
of the first triple, and $\Sigma_{2k}$ permuting the triples.
The action of $H_2$ is faithful but otherwise arbitrary. 
To these actions we can associate a graph whose vertices are the 
triples and whose links connect the triples which contain letters 
permuted by $\Sigma_2$; $\Sigma_{3k}$ permutes the links, while 
$\Sigma_{2k}$ permutes vertices. Notice that each link connects in fact 
two precise letters within triples; it doesn't matter which ones, since 
there is symmetry within each triple.
 
If two vertices are connected by a link we can permute any two letters in 
these vertices, using $\Sigma_3$ within every single vertex and
$\Sigma_2$ exchanging the letters at the endpoints of the links. 
Thus, we have the action of the full symmetric group of the letters 
belonging to every connected component of the graph.

If there are two disconnected components, we can permute a letter in one 
component with a letter in the other as follows. First permute the 
nodes to which they belong by the action of $\Sigma_{2k}$; then use the 
symmetric group of each component to put the extra two letters of 
each involved triple at the endpoints of a link. 
Next exchange the couple of links got in this way by an action of
$\Sigma_{3k}$.
Finally use again the symmetric group of each component to restore the 
sequence of letters we started from, except for the two which have been
exchanged.

Summing up this proves that, if we consider combinations of the
$U$'s and $V$'s completely symmetric on {\sl flavor} indices,
the structure of the saturation of the {\sl color} indices is unique:
the totally symmetric one saturated with its dual. 
%
%
\end{document}